\documentclass[review]{elsarticle}

\usepackage{lineno,hyperref}

\usepackage{multirow}
\usepackage{placeins}
\usepackage{graphicx}
\usepackage{subcaption}
\usepackage{comment}
\usepackage{xcolor}
\graphicspath{{figures/}}
\DeclareGraphicsExtensions{.pdf,.jpeg,.png,.jpg}

\modulolinenumbers[5]

\journal{Nuclear Instruments and Methods in Physics Research Section A}

\begin{document}

\begin{frontmatter}

\title{Performance evaluation of the Boron Coated Straws detector with Geant4}



\author[mymainaddress,mysecondaryaddress,mytertiaryaddress]{M.~Klausz\corref{mycorrespondingauthor}}
\cortext[mycorrespondingauthor]{Corresponding author}
\ead{milan.klausz@energia.mta.hu}
\author[mysecondaryaddress]{K.~Kanaki}
\author[mymainaddress,mytertiaryaddress]{P.~Zagyvai}
\author[mysecondaryaddress,myquinternaryaddress]{R.J.~Hall-Wilton}

\address[mymainaddress]{Hungarian Academy of Sciences, Centre for Energy Research, 1525 Budapest 114., P.O. Box 49., Hungary}
\address[mysecondaryaddress]{European Spallation Source ESS ERIC, P.O Box 176, SE-221 00 Lund, Sweden}
\address[mytertiaryaddress]{Budapest University of Technology and Economics, Institute of Nuclear Techniques, 1111 Budapest, M\H uegyetem rakpart 9., Hungary}
%\address[myquaternaryaddress]{Oak Ridge National Lab, Neutron Technologies Division, Oak Ridge, TN 37831-6475, USA}
\address[myquinternaryaddress]{Università degli Studi di Milano-Bicocca, Piazza della Scienza 3, 20126 Milano, Italy}

\begin{abstract}
The last decade has witnessed the development of several alternative neutron detector technologies, as a consequence of upcoming neutron sources and upgrades, as well the world-wide shortage of $^3$He. One branch of development is the family of $^{10}$B-based gaseous detectors. This work focuses on the boron coated straws (BCS) by Proportional Technologies Inc., a commercial solution designed for use in homeland security and neutron science. A detailed Geant4 simulation study of the BCS is presented, which investigates various aspects of the detector performance, e.g.\,efficiency, activation, absorption and the impact of scattering on the measured signal. The suitability of the BCS detector for Small Angle Neutron Scattering (SANS) and direct chopper spectrometry is discussed.
  
\end{abstract}


\end{frontmatter}

%\linenumbers

\section{Introduction}

For many years $^3$He-based detectors have been dominant in the field of
neutron scattering science, as they satisfied scientific requirements and
$^3$He was available in sufficient quantities at an affordable price. 
The situation has changed in recent years due to the worldwide $^3$He crisis~\cite{he3crisis1, he3crisis2} that necessitated the development
of alternative neutron detector technologies
%~\cite{zeitelhack2012}
based on $^{10}$B$_4$C~\cite{kirstein2014,multiwire_2013,bcs_evolution,mgcncs,bandgem4,rpc,rpc_arxiv},
$^6$LiF~\cite{lif,ear1,sintef} and
scintillators~\cite{isis1,isis2,wang11,nop_sonde,sonde_arxiv,SCI_katagari,jparc,SCI_bell}.

%Additionally, it is already proven that $^{10}$B$_4$C-based detectors are capable of outperforming $^{3}$He detectors in terms of spurious scattering of neutrons~\cite{estia_prop, MB2017}. Several studies demonstrate the performance of this detector type and its suitability for neutron scattering experiments~\cite{MB2017,crisp,mauri2,phd_piscitelli}.

More importantly though, these new technologies are required to exceed the scientific capabilities of previous detectors, as imposed by
future instrument upgrades and upcoming research facilities.
Such a facility is the European Spallation Source
(ESS) ERIC~\cite{esstdr, ess2018} that aspires to lead neutron scattering research.
The cutting edge neutron scattering instruments set high requirements
for the detectors, that could otherwise become the bottleneck of the
instrument's scientific performance.
%The fulfillment of the performance criteria must be proven prior to the installation so the characteristics of the detectors have to be well known beforehand. 
It is therefore important to understand every aspect of detector
performance before qualifying it for a particular neutron technique.

Monte Carlo simulations could and should play a key role in
the development and characterisation of detectors as a reliable, cheap and versatile tool~\cite{KANAKI2018386}. 
Simulations not only make it easier to analyse and compare detectors
and detector arrangements without building a physical prototype every time, but also enable the quantification of
otherwise unmeasurable properties. The primary goal of this work is to
perform a comprehensive characterisation of the BCS detector -- a
promising $^3$He detector replacement technology developed initially for homeland security applications --, using Monte Carlo simulations.
The study includes the efficiency of the detectors, the absorption and
activation of the materials thereof, and the impact of the material
budget on scattering. %The simulations were done with different neutron wavelengths and detector arrangements.
A generic detector geometry is implemented in Geant4~\cite{geant4a,geant4b,geant4c_inpresscorrectedproof}, which can
serve the needs of various neutron scattering techniques. The
following sections introduce the detector specifics and the respective model,
define appropriate figures of merit and discuss the evaluation of the BCS performance.

%The Monte Carlo simulations are carried out using Geant4~\cite{geant4a,geant4b,geant4c_inpresscorrectedproof}. 
%The boron coated straw detector technology and the Geant4 model in use is introduced in the next section.

\section{Detector technology and simulation model}

\subsection{The BCS detector}

%NOTES: introduce BCS technology with the multiple applications based on  LACY et al.: EVOLUTION OF NEUTRON STRAW DETECTOR APPLICATIONS IN HOMELAND SECURITY
% Different desings (different straw diameters, lengths, AND "The figure also shows two alternate designs, whereby the round straws have been replaced with starshaped
% straws, refered to as Star1 and Star2, in order to increase the boron-coated area by a factor of 1.30 and 1.98, respectively."//[EVOLUTION of....])

\begin{figure}[!h]  
  \centering
  \begin{subfigure}{.7\textwidth} %{8cm}
      \includegraphics[width=\textwidth]{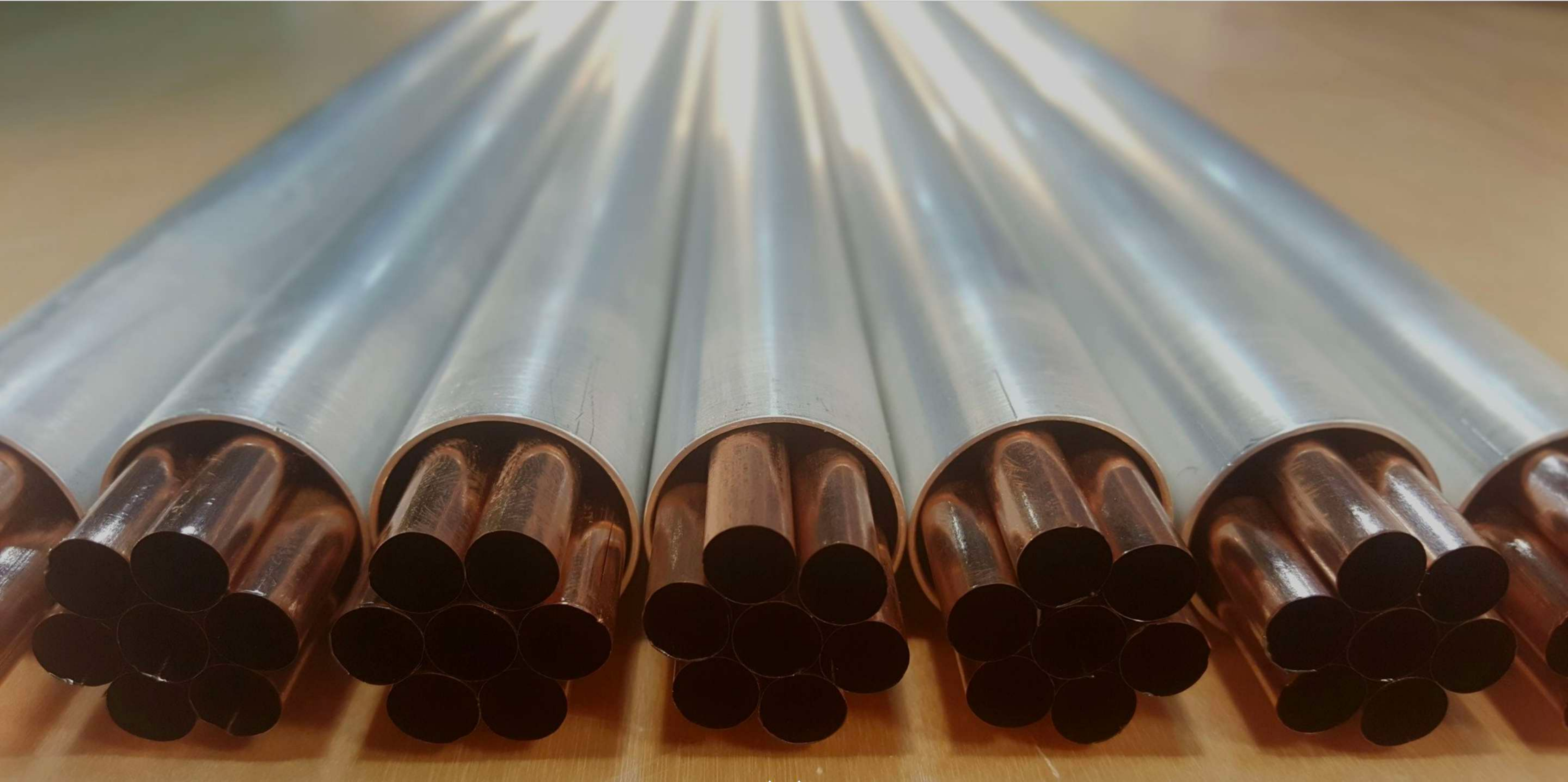}
     \end{subfigure}
   \caption{\footnotesize Cross-sectional view of the BCS
     detector~\cite{bcs_propTech}. The copper straws inside the
     aluminium tubes are exposed.}
  \label{bcs_propTech}
\end{figure}

BCS is a position sensitive $^{10}$B-based gaseous neutron detector developed by Proportional Technologies, Inc.\,(PTI)~\cite{bcs_propTech}. 
It consists of a long thin-walled aluminium tube, containing seven
copper straws arranged hexagonally with one in the centre (see Fig.~\ref{bcs_propTech}).
The inner wall of the straws is coated with a 1~$\mu$m thin B$_4$C
converter layer enriched in $^{10}$B by 95\%. The straws are filled with an Ar/CO$_2$ mixture (90/10 by volume) at 0.7~atm.
A bias voltage is applied between the tube and resistive Stablohm wires tensioned in the center of each straw as anodes, making them work in proportional mode~\cite{sauli14}. 
The charge is read out at both ends of the detector using charge division to acquire longitudinal position information along the straws.

The length and diameter of the straws, and therefore of the tubes, can
vary depending on the application. It is claimed that a straw diameter
of 2~mm up to~15 mm or even more can be achieved~\cite{bcs_evolution},
but in most publications either 4~mm or 7.5~mm is used
~\cite{bcs_experimental,bcs_iniPerformance,bcs_multiplex}. Generally,
several BCS tubes are placed behind each other in successive layers in order to achieve the desired coverage, uniformity and detection efficiency.
The main application field of the BCS detectors is homeland security but they have the potential to be used as large area position-sensitive
detectors for SANS and chopper spectrometers.

%~\cite{bcs_evolution} "Using this methodology, straw diameters can be readily achieved over the range 2 mm up to 15 mm or more"

% My Talks
%A detector consists of an aluminum tube with a diameter of one inch which contains seven copper straws. The straws are configured hexagonally in a tube and their inner wall is coated with one micron of B4C as a neutron converter layer.  They are filled with a gas mixture of argon and carbon-dioxide on sub-atmospheric pressure and a resistive wire is tensioned in the center of each of them which serves as anode. A bias voltage is applied between the anode wire and cathode tube, which makes the straws work in proportional mode and as the charges are read out in both end applying charge division method makes them a position sensitive detector.

% Rates Paper:
% A BCS detector consists of an aluminium tube, containing seven
% copper straws arranged hexagonally (see Fig.~\ref{bcs_zoom}). The tubes are 1~m long with a
% diameter of 2.54~cm. The straw inner wall is
% coated with a 1~$\mu$m thin B$_4$C converter layer 
% enriched in $^{10}$B by 95\%. They are filled with an Ar/CO$_2$
% mixture (90/10 by volume) at 0.7~atm.
% A bias voltage is applied between the tube and resistive
% anode wires, which are tensioned in the center of each straw.
% This makes the straws work in proportional mode. The charge is read out at both ends
% of the detector using charge division to acquire the position information. The
% tubes can be arranged in successive layers in order to achieve the
% desired coverage, uniformity and detection efficiency.

\FloatBarrier
\subsection{Geant4 model}
%{\color{red} fix the order: first geometry, materials,  generator description, last is the hit definition, thresholds etc.}

\begin{figure}[!h]  
  \centering
  \begin{subfigure}{0.40\textwidth} %{8cm}
    \includegraphics[width=0.9\textwidth]{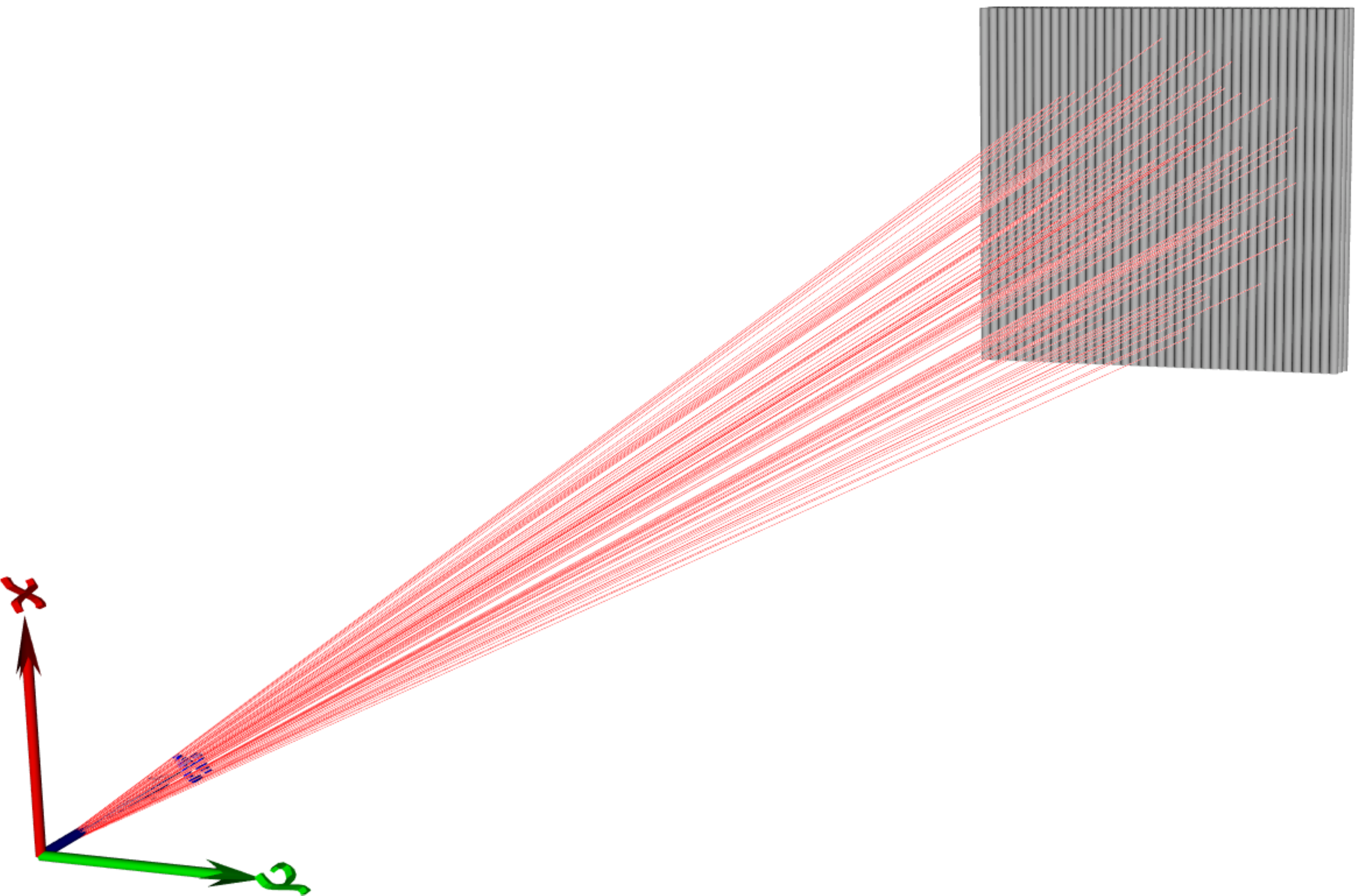}
    \caption{\footnotesize The primary neutrons (in red) hit
      the detectors 5~m away from the source position.}
    \label{bcsGeo_neutrons}    
  \end{subfigure}
  \begin{subfigure}{0.59\textwidth} %{8cm}
    \centering
    %\raggedleft
    %\includegraphics[width=7.5cm]{bcs_zoom2}
    \includegraphics[width=\textwidth]{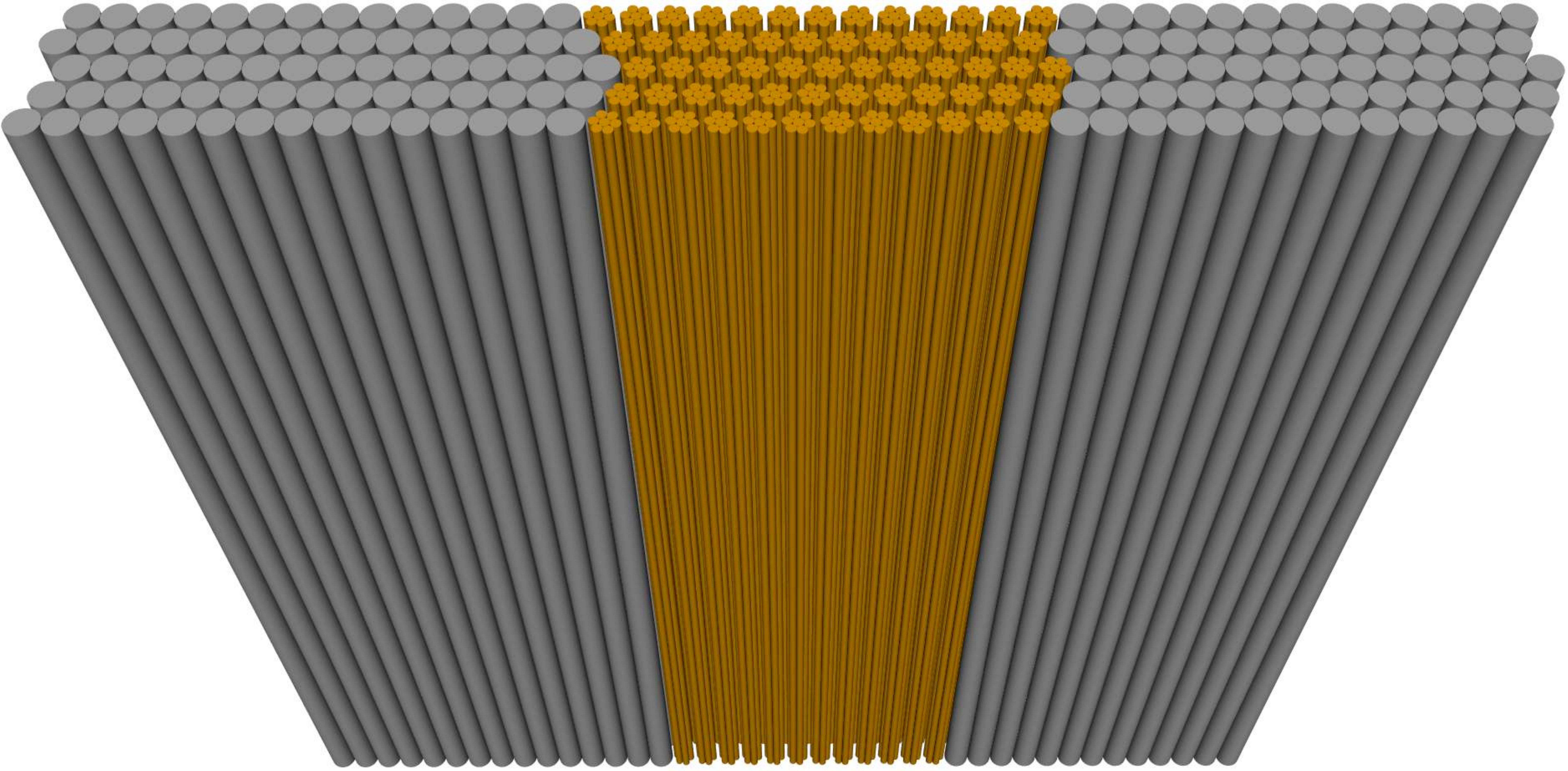}
    \caption{\footnotesize Enlarged view of the 5 consecutive BCS tube
      panels with straws exposed in the middle section for illustration. Each aluminium tube (grey) contains 7 copper straws (orange) arranged hexagonally.}
    \label{bcs_zoom}
  \end{subfigure}
  \caption{\footnotesize The simulation arrangement containing the neutron generator %placed at the beginning of the coordinate system
  and 5 panels ($\times$~40~tubes/panel) of BCS tubes covering 1~m~$\times$~1~m. % at 5~m downstream along the z-axis. 
  The figures are taken from previous study of the detectors in the same arrangement~\cite{ratesPaper}.}
  \label{bcsGeo}
\end{figure}

\begin{figure}[!h]  
  \centering
  \begin{subfigure}{0.49\textwidth} %{8cm}
    \includegraphics[width=\textwidth]{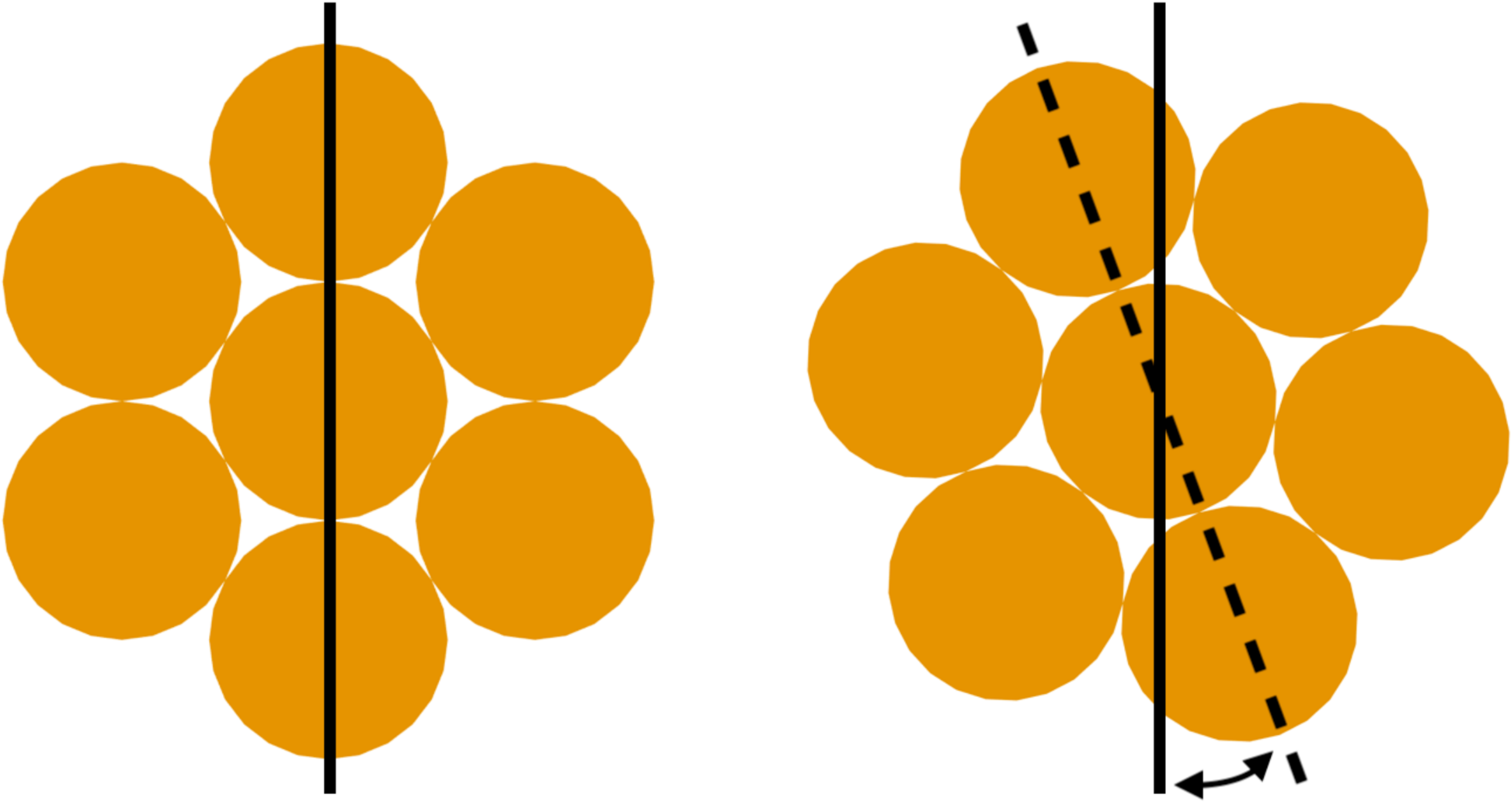}
    \caption{\footnotesize A rotation of 20$^\circ$ around its cylindrical axis is applied for every detector tube.}
    \label{bcs_rotation}    
  \end{subfigure}
  \begin{subfigure}{0.49\textwidth} %{8cm}
    \centering
    \includegraphics[width=.9\textwidth]{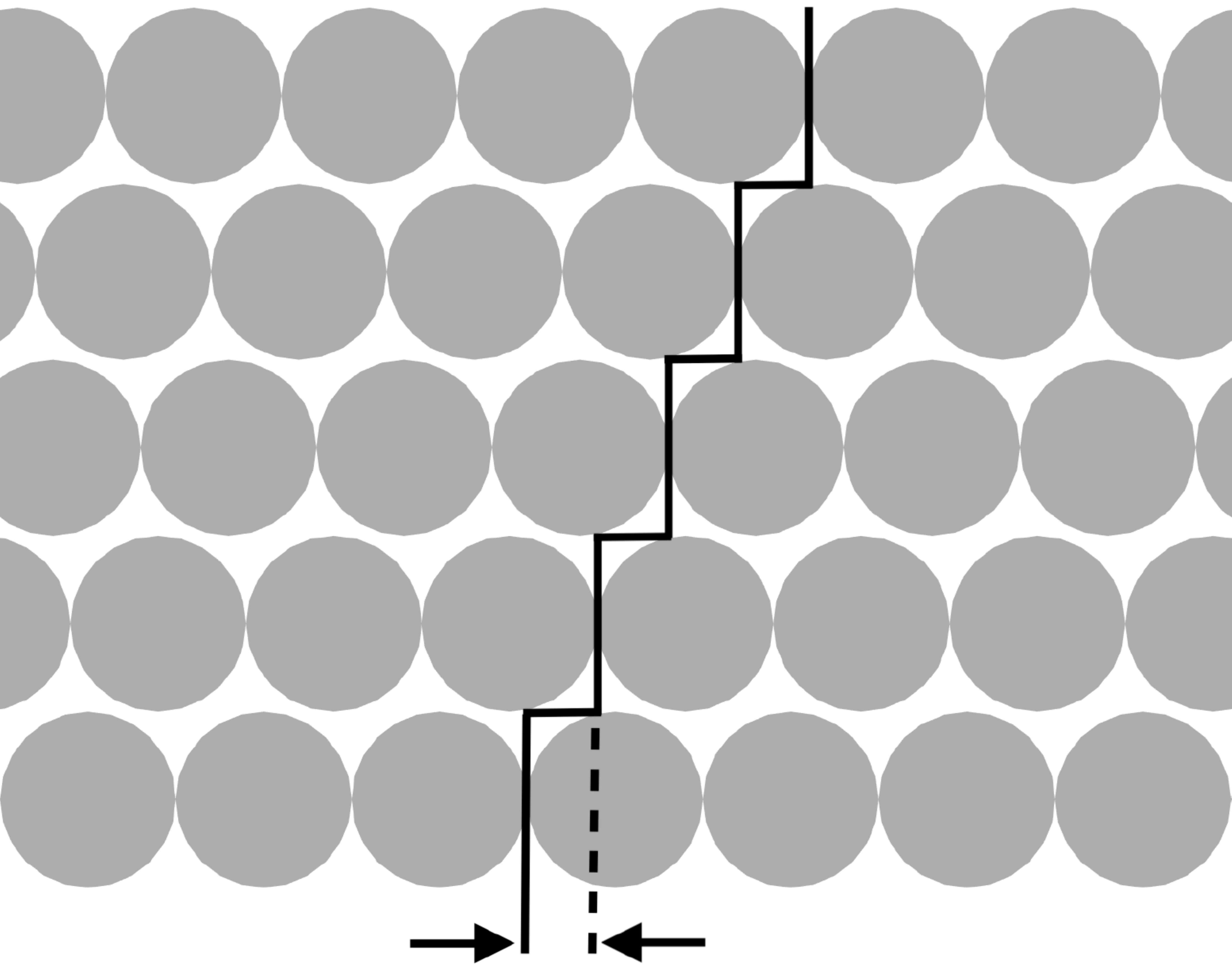}
    \caption{\footnotesize A horizontal shift of 10.16~mm is used for adjacent panels.}
    \label{bcs_offset}
  \end{subfigure}
  \caption{\footnotesize Rotation and translation of the detector tubes in order to increase the uniformity of the conversion efficiency for neutron entering the detector arrangement in different positions and angles.}
  \label{bcs_RotationOffset}
\end{figure}

The Geant4 geometry model implemented for this study is a generic
1~m$^2$ detector arrangement that  % for many SANS instruments around the world 
consists of 5 consecutive detector panels with 40 tubes in each (see Fig.~\ref{bcsGeo}).  
For higher uniformity in the conversion efficiency, the tubes are rotated by 20$^\circ$ around their cylindrical axes and a relative horizontal shift of 10.16~mm is applied for adjacent panels (see Fig.~\ref{bcs_RotationOffset}) to avoid high differences in the path length in the B$_4$C converter layer~\cite{bcs_Davidetalk}.

The model of a single BCS detector consists of the aluminium tube that
contains the 7 copper straws arranged hexagonally with 1~$\mu m$
enriched B$_4$C layers on their inner surface. The void inside a
tube is filled with the Ar/CO$_2$ mixture (90/10 by volume) at
0.7~atm. The exact parameters of the full scale model are listed in
table~\ref{tab:geom}. 
\begin{table}[htp]
\begin{center}
\begin{tabular}{|c c @{ [} l |}
\hline
%Parameter & value \\\hline\hline

Tube diameter & 25.4 &mm]\\ %\hline
Tube thickness & 0.94 &mm]\\ %\hline
Straw diameter & 7.5 &mm]\\ %\hline
Straw thickness & 25 &$\mu$m]\\ %\hline
B$_4$C thickness & 1 &$\mu$m]\\  %\hline
Tube\&straw length & 1 &m]\\
\hline
\end{tabular}
\end{center}
\caption{Geometry parameters in the model of the BCS detectors.}
\label{tab:geom}
\end{table}

All materials are selected from the Geant4 database of NIST materials,
except for Al and Cu. The latter are described with the use of the
NCrystal library~\cite{ncrystalArxiv,ncrystal,KANAKI2018386}, as
their crystalline structure is important for the correct treatment of
their interaction with neutrons. 
The density of the NCrystal aluminium and copper are $\rho_{Al}$=2.70~g/cm$^3$ and $\rho_{Cu}$=8.93~g/cm$^3$ respectively. The Geant4 physics list used is QGSP\_BIC\_HP. 
%The density of the NCrystal aluminium and copper are $\rho_{Al}$=2.69865~g/cm$^3$ and $\rho_{Cu}$=8.93484~g/cm$^3$ respectively. The Geant4 physics list used is QGSP\_BIC\_HP.%~\cite{dgcodechep2013}. 

The detector system is illuminated with neutrons from a point source at a 5~m distance from the centre of the geometry in vacuum.
The neutron source is an isotropic, monochromatic conical beam with an opening angle of 10.6$^\circ$; this ensures that the direct beam crosses all 5 panels to minimise edge effects. 

% mention all the approximations used in this paper
The model does not contain the anode wires in the straws and therefore neither the charge collection nor the readout are simulated. A detection event is recorded if a neutron's conversion products deposit more energy in the counting gas than a preset threshold applied to mimic discrimination of the gamma background.
In different studies of BCS detectors, different thresholds are reported, e.g.\,30~keV~\cite{bcs_highRateHighRes}, 73~keV~\cite{bcs_evolution} and 200~keV~\cite{bcs_experimental}. %depending on the application. 
Based on these values, in the simulations of this study an energy
threshold of 120~keV is applied, which corresponds to an appropriate threshold for $^{10}$B detectors to achieve sufficient $\gamma$/thermal neutron discrimination for neutron scattering applications~\cite{Croci_2013,khaplanov2013,mauri2}. % on the energy deposition of the charged conversion products.

The simulation of each neutron is completely independent so pile-up is
not possible. If a neutron enters the copper straw in a BCS detector,
it is counted as incident for that straw. If the neutron is converted
in the B$_4$C layer, it is counted as converted. The detection event's
two coordinates perpendicular to the straw are defined by the virtual
position of the wire in the center of the straw. In order to get the
third, longitudinal coordinate, first the weighted average of the
deposited energy by the conversion products is determined, then a
Gaussian distribution with that mean value is sampled. The full width
at half maximum (FWHM) for this smearing is set to 0.6~cm based on
experimental results~\cite{bcs_experimental,bcs_Davidetalk}. Although
the longitudinal resolution for a tube detector depends on the
position along its length (higher in the centre than closer to the
ends~\cite{bcs_multiplex}), and is very much a function of the analogue quality and signal treatment in the electronic readout, this study assumes uniform resolution.

% PLOTS: 
% introduce the sans geometry (generic geometry, not sans) 
% // 5-10 meters maybe show that it doesnt matter 1m$^2$ detector.

%we do not take into account the spatial resolution in this study. Or do we? This will be the LAST thing to do

% In the Geant4 model of the current study 5 layers of BCS detectors, referred to as
% panels, consisting of 40 tubes each cover the same 1~m~$\times$~1~m area as the $^3$He detector model. Each tube
% is rotated by an angle of 10$^{\circ}$ around its cylindrical axis and the adjacent panels are
% positioned behind each other with a relative shift of 1.016~cm along
% the horizontal axis (see
% Fig.~\ref{bcs_zoom}) for performance optimisation reasons. 
% giving the optimal layout based on both simulations and measurements~\cite{lacy_icns17}.
 
\FloatBarrier
\section{Detector efficiency}

Detection efficiency is one of the key performance parameters of a
detector. With new and stronger sources coming up it is important to
fully exploit the high brilliance of a neutron pulse and accommodate a
larger number of users. This aspect of neutron detectors has come to
focus with the replacement detector technologies. A simulation tool
like Geant4 can shed a lot of light in the response of a complex
geometry like that of BCS.

%New facilities are built with stronger and stronger sources in order
%to increase the intensity of neutron signal, so it's a natural requirement to get use of it by having the detection efficiency as high as possible. is needed, that can be achieved primarily with the most detected neutrons.
%New facilities are built with stronger and stronger sources in order to increase the intensity of neutron signal, so it's a natural requirement to get use of it by having the detection efficiency as high as possible. 
%{\color{red} a bit of an abrupt start, say a few things about why this property is important. we produce lots of neutrons and we don't want them to be wasted, so this requirement being as high as possible ensures that the pulse will be properly exploited.}

The functional unit of a BCS detector is a single straw. The straw detection efficiency can be expressed in several valid ways. In this
work the following definition is used:
\begin{itemize}
%\item \textbf{Detected efficiency} is the number of neutrons detected in a straw over the number of source neutrons that have an initial direction toward that straw. 
%{\color{gray}
%\item \textbf{Measureable detection efficiency} is the
%    number of neutrons detected in a straw over the number of source
%    neutrons contained in the solid angle subtended to the straw.
%\begin{equation} 
%\epsilon_{det}=\frac{\textrm{\# of detected neutrons in the straw}}{\textrm{\# of neutrons in the direction of the straw}}.
%\end{equation}
%}
%\item \textbf{Real detection efficiency} is the number of neutrons detected in a
%  straw over the number of incident neutrons in that straw
%  from every direction. A neutron is counted as incident every time it enters the copper of the straw from the outside.
\item \textbf{Detection efficiency} is the number of neutrons detected in a
  straw divided by the number of incident neutrons in that straw
  from every direction. A neutron is counted as incident every time it enters the copper of the straw from the outside.
\begin{equation} 
\epsilon=\frac{\textrm{\# of detected neutrons  in the straw}}{\textrm{\# of incident neutrons in the straw}}.
\end{equation}
\end{itemize}
%The replacement of the number of detected neutrons with the number of
%converted neutrons in the nominators results in the respective conversion efficiencies. 
The replacement of the number of detected neutrons with the number of
converted neutrons in the numerator results in the respective conversion efficiencies.

%{\color{red} this ratio is not necessary relating conversion and detection efficiency. It is the fraction of detected neutrons over
%  converted ones per wavelength bin. You'll have to change your legend in (a). }
  
Another relevant quantity is the detection to conversion ratio (DCR), that is the fraction of detected neutrons over converted ones.
As previously mentioned, to get a detection event after a conversion, at least one of the conversion products has to leave the boron carbide layer and deposit enough energy in the counting gas to overcome the preset threshold.
Detailed calculations of this exist~\cite{piscitelli2013,phd_piscitelli, Basanez:2018nis}
For BCS it is claimed that for 1~$\mu$m of B$_4$C, one of the two charged
conversion products has a 78\%
probability to escape the converter and ionise the counting gas in the straw~\cite{bcs_iniPerformance}. 
This is the theoretical maximum for DCR, with no energy threshold.
With the applied threshold of 120~keV, the simulated DCR is 70\%, regardless of the wavelength of the converted neutrons, due to the small thickness of the B$_4$C layer.
This value also gives an upper limit for the detection efficiency, as it is the convolution of the conversion efficiency and the DCR.
The higher the threshold is, the lower the DCR and therefore the lower the detection efficiency will be.
The optimal value depends on the gamma background of the measurement and ought to be chosen carefully.

\begin{figure}[!h] 
  \centering
  \begin{subfigure}{1.0\textwidth}
    \centering
    \includegraphics[width=\textwidth]{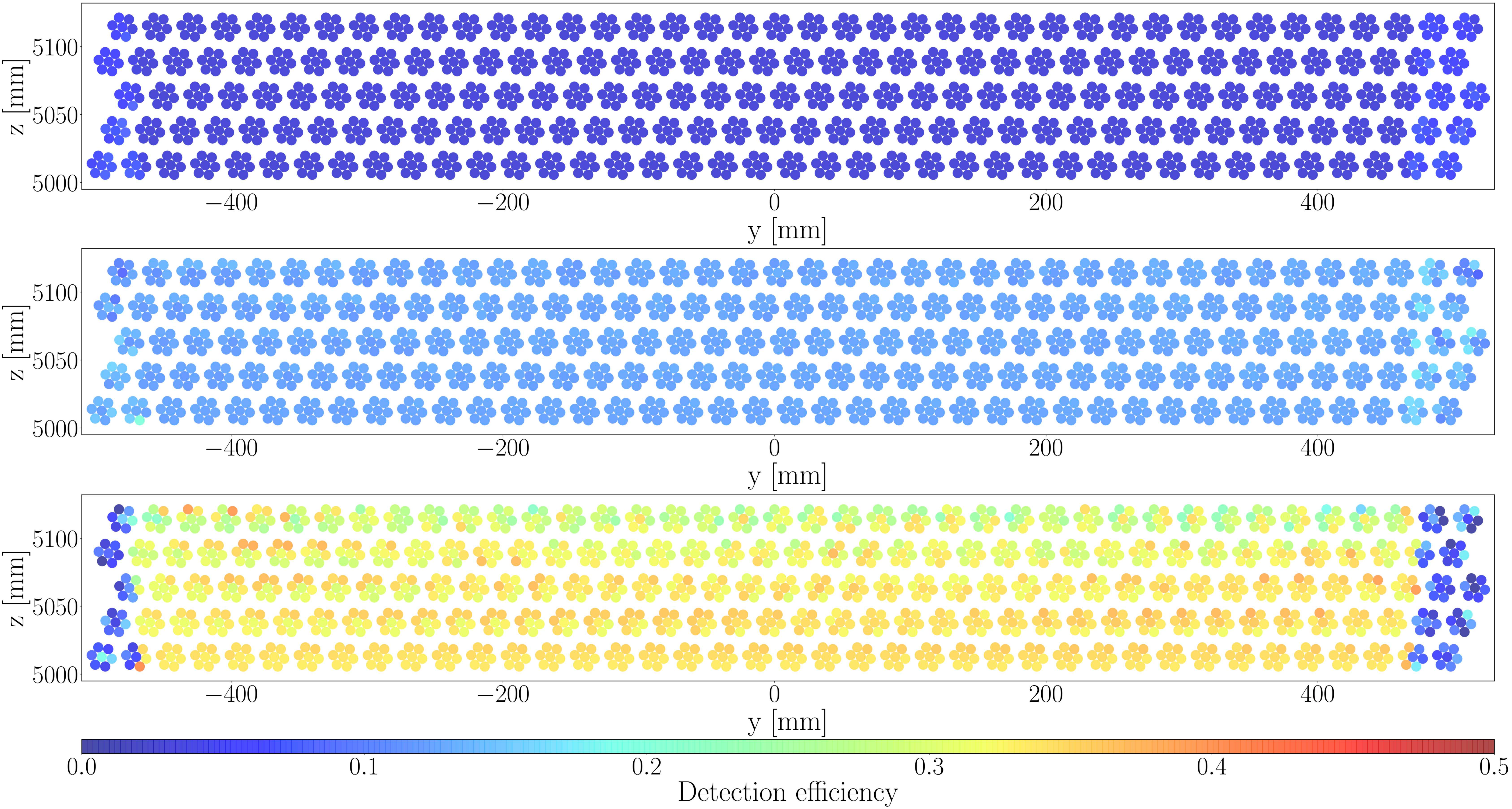}
    %\caption{\footnotesize $\lambda$ = 0.6~{\AA}.}
    %\label{realDetEff_ang0p6}
  \end{subfigure}
  
%  \centering
%  \begin{subfigure}{1.0\textwidth}
%    \centering
%    \includegraphics[width=\textwidth]{Efficiency_Average_Detection_ang3_8e6.pdf}
%    \caption{\footnotesize  $\lambda$ = 3~{\AA}.} 
%    \label{realDetEff_ang3}
%  \end{subfigure}%q
%  
%   \centering
%  \begin{subfigure}{1.0\textwidth}
%    \centering
%    \includegraphics[width=\textwidth]{Efficiency_Average_Detection_ang11_8e6.pdf}
%    \caption{\footnotesize $\lambda$ = 11~{\AA}.}
%    \label{realDetEff_ang11}
%  \end{subfigure}%
  \caption{\footnotesize {Detection efficiency of straws with
      monoenergetic incident neutrons. The results are visualised using the cross-section image of the 5 panel detector system. The colors represent the detection efficiency of the straws from simulations with three neutron wavelengths: 0.6~{\AA} (top), 3~{\AA} (middle) and 11~{\AA} (bottom). The conical neutron source is placed at the origin, some of the straws at the edge of the panels are not illuminated directly to minimise edge effects. }}  %\color{red}(TODO: Make it ONE figure! Also indicate direction of incoming beam in the caption, although it's rather obvious... 'directly illuminated+where the beam comes'}
  \label{figStrawRealEfficiency}
\end{figure}

The detection efficiency for each straw with monochromatic
0.6~{\AA}, 3~{\AA} and 11~{\AA} neutrons is depicted in
Fig.~\ref{figStrawRealEfficiency}. The efficiency of a single straw is
quite low with an average of 3\%, 12\% and 31\% respectively. 
This is why 7 of them are packed together in a BCS tube and this is
why employing even more overlapping straws in consecutive panels of
detectors is necessary for most applications.
The detection efficiency of the straws is quite uniform for a particular wavelength across all panels.  
This is because of the monoenergetic neutron sources, but a previous study with polyenergetic neutrons demonstrated significant differences between the panels, attributed to the hardening of the neutron spectra and the thermalisation of the neutrons via scattering~\cite{ratesPaper}.
The slightly lower efficiencies in the last panels for 11~{\AA} indicate that the thermalisation through scattering occurs indeed, and leads to neutrons with lower wavelength in the neutron spectrum and therefore lower average efficiency for the straws in the back, but also show that it is a minor effect compared to the hardening of the neutron spectra.

%{\color{red} As previously mentioned, the detection efficiency is a
%  convolution of the conversion efficiency and the DCR.}
%The conversion efficiency could be easily increased by using a thicker
%conversion layer but that would significantly lower the escape
%probability of the conversion products 
%%{\color{red} not necessarily true. It depends whether you are detecting in transmission or backscattering. 
%($\alpha$ and $^7$Li particles)
%and consequently the DCR. Therefore, the only straightforward way to
%increase the detection efficiency is by employing more overlapping
%straws in consecutive panels of detectors.

\begin{figure}[!h]  
  \centering
  \begin{subfigure}{0.49\textwidth} %{8cm}
      \includegraphics[width=\textwidth]{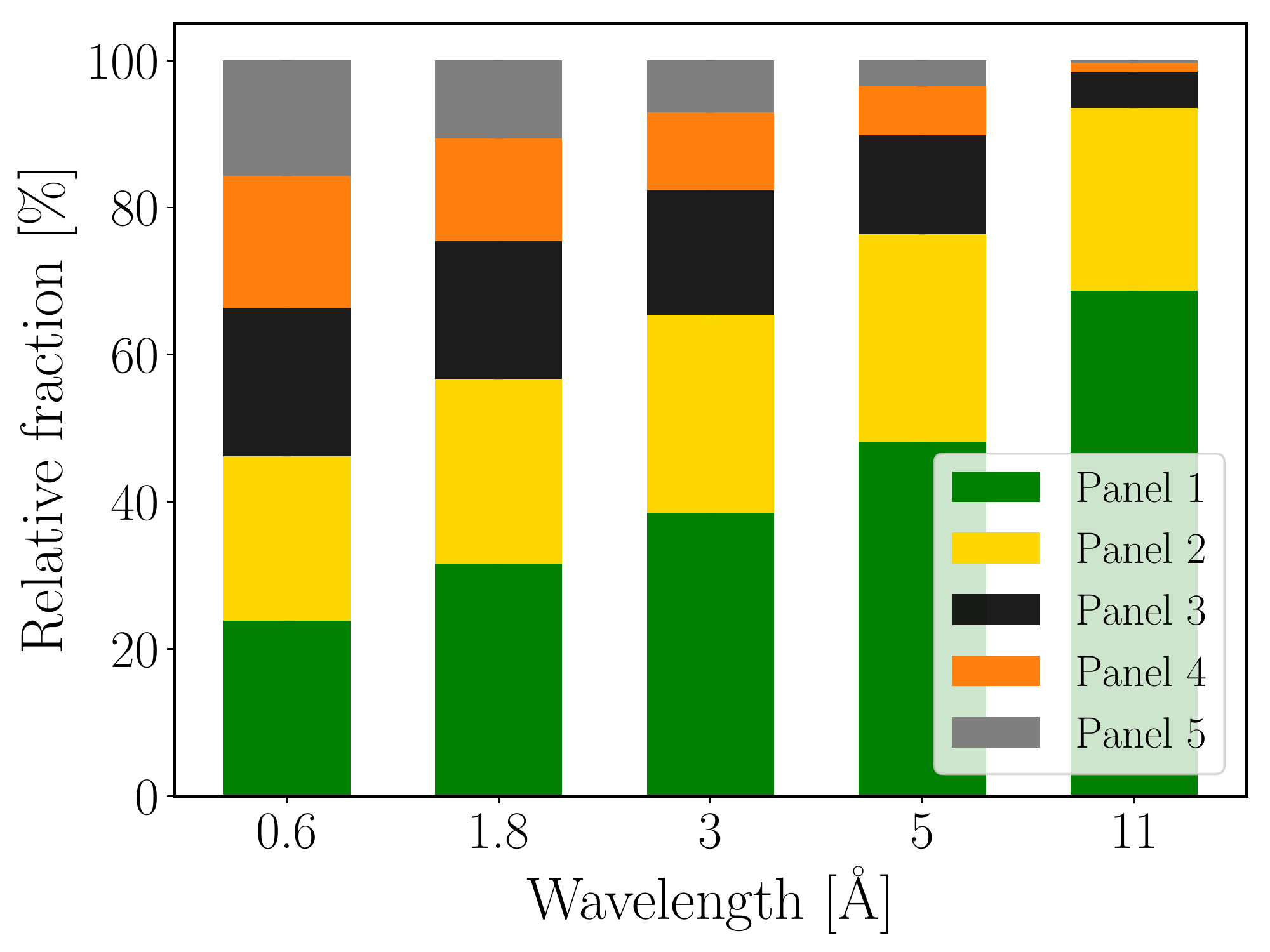}
     \end{subfigure}
     \begin{subfigure}{0.49\textwidth} %{8cm}
      \includegraphics[width=\textwidth]{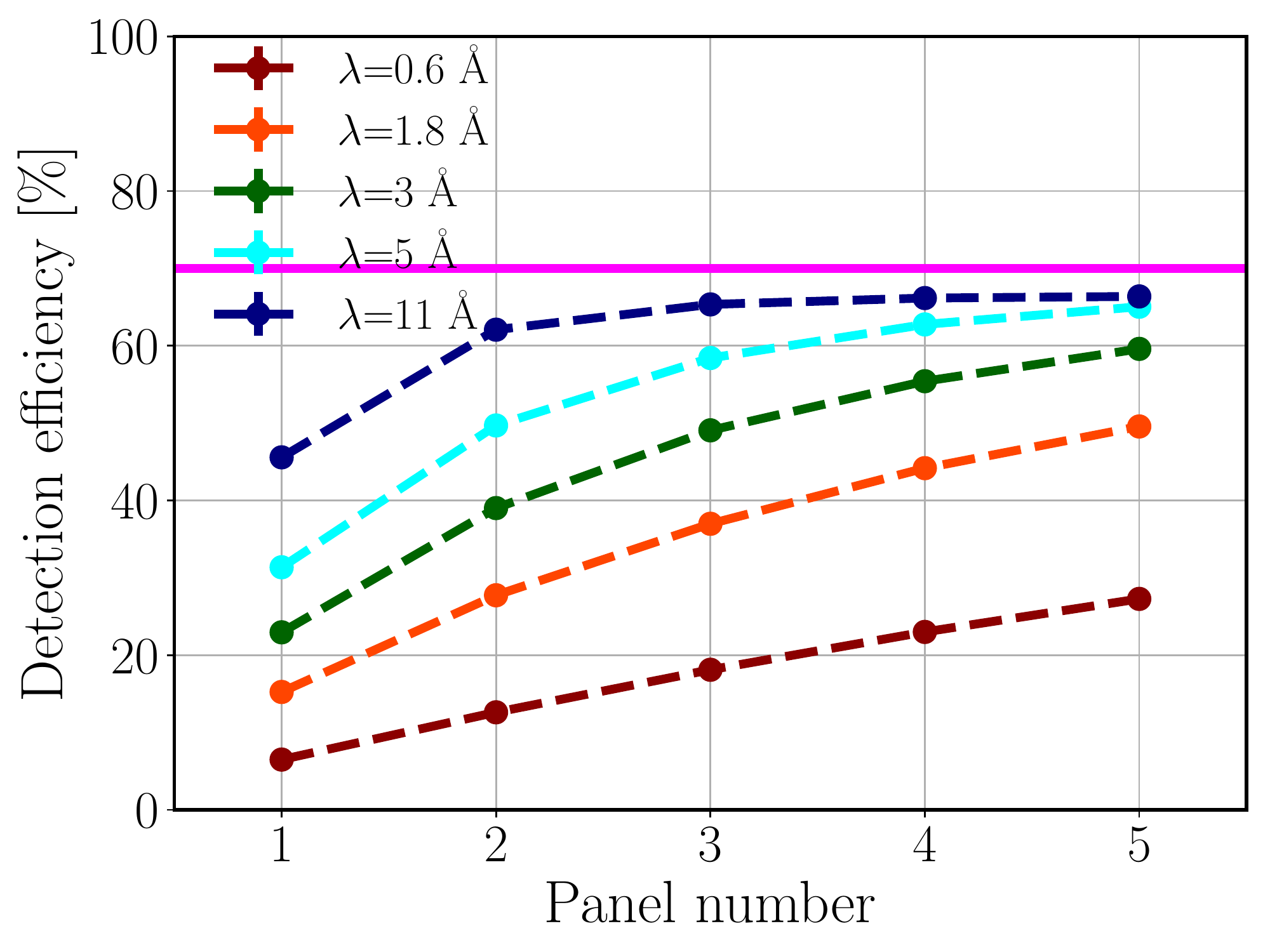}
     \end{subfigure}
   \caption{\footnotesize Contribution of each panel to the total
     number of detected neutrons (on the left) and the resulting
     global detection efficiency (on the right) for different
     wavelengths. The purple line marks the highest possible global
     detection efficiency defined by the DCR.}
  \label{fig:PanelContribution}
\end{figure}

Fig.~\ref{fig:PanelContribution} depicts the contribution of each
panel to the total number of detection events and the resulting global detection efficiency.
For all wavelengths the first panel registers  most  of the detection events and the 
impact of the additional panel becomes lower and lower. 
This is more noticeable for high wavelengths where the efficiency saturates 4-5\% below the maximum of 70\% defined by the DCR. 
The detectors in the fifth panel contribute only a 0.21--4.27\% to the global detection efficiency for the simulated neutron wavelengths. 
These results imply that depending on the application and the optimum
wavelength thereof, using 4~panels might be a more cost efficient solution. 

The results presented are derived from simulations with all 5 panels in place. 
This means that efficiencies with fewer than 5 panels are somewhat overestimated because of the back-scattered neutrons from later panels, but the simulations with fewer panels showed that this effect means a $<$0.5\% difference only.
%{\color{red} explain whether you add the panels one-by-one to the model or if this
%is a cumulative approach (the increase of efficiency of a panel is less then 0.5\% due to the backscattering). Do these figures actually show different things? Do
%you need them both?}
%be an optimal choice taking into account the financial aspects. (Cost-efficient)

%Due to the complex detector geometry, regarding the geometric path
%length differences in the converter layers in different directions,
%the number of detector panels has an impact on the uniformity of the
%efficiencies but that is out of the scope of this study. {\color{red} remove}

It is worth mentioning that the thickness of the converter layer has an impact on the detection efficiency.
As previously said, the latter results from the convolution of the conversion efficiency and the DCR.
The conversion efficiency could be increased by using a thicker
conversion layer but that would lower the escape probability of the conversion products 
%($\alpha$ and $^7$Li particles) 
and consequently the DCR~\cite{bcs_evolution,mgcncs,piscitelli2013}.
The cumulative effect of the converter layer thickness is wavelength dependent and not straightforward. 
The simulations show that for high wavelengths %such as 5~{\AA} and 11~{\AA}, 
the global detection efficiency could be increased with lower B$_4$C thickness due to the increase of DCR but for lower wavelengths the decrease of the conversion efficiency overrules it and the detection efficiency decreases.
In the other direction, the efficiency for low wavelengths can slightly benefit from thicker converter layers but for higher wavelengths where the conversion efficiency is already high, the lower DCR lowers the detection efficiency.
In this study the commercial converter layer thickness of 1~$\mu$m is
used. A more detailed investigation of converter thicknesses and efficiency optimisation is out of scope here.
%Therefore, the only straightforward way to increase the detection efficiency is by employing more overlapping
%straws in consecutive panels of detectors.

\FloatBarrier
 \section{Absorption in detector components}

The previous section demonstrates that for achieving a higher
detection efficiency it is necessary to use multiple panels of
detectors. 
%{\color{red} comment for previous section: you should mention that there are several parameters entering the eff optimisation, your next sentence is valid assuming some optimisation has taken place a priori.}
This does not only increase the number of conversion and detection events but as a consequence the undesired absorption in the non-converting materials of the detector, namely aluminium and copper also increases.
However, it is not only these two materials that can absorb neutrons without
leading to a detection but B$_4$C too.
As mentioned before, not all conversion events result in a detection event because in some cases the conversion products do not exit the converter layer or they do not deposit sufficient energy in the detector gas to overcome the applied threshold.
%A neutron is detected only if one of the conversion products exits the converter layer and deposits sufficient energy in the detector gas to overcome the applied energy threshold.
In addition, there is a rather small amount of neutron absorption in carbon and $^{11}$B without conversion products to trigger a detection event. 
These two event classes together are hereinafter referred to as absorption in B$_4$C.
 
\begin{figure}[!h]  
  \centering
  \begin{subfigure}{0.5\textwidth}
    \centering
    \includegraphics[width=\textwidth]{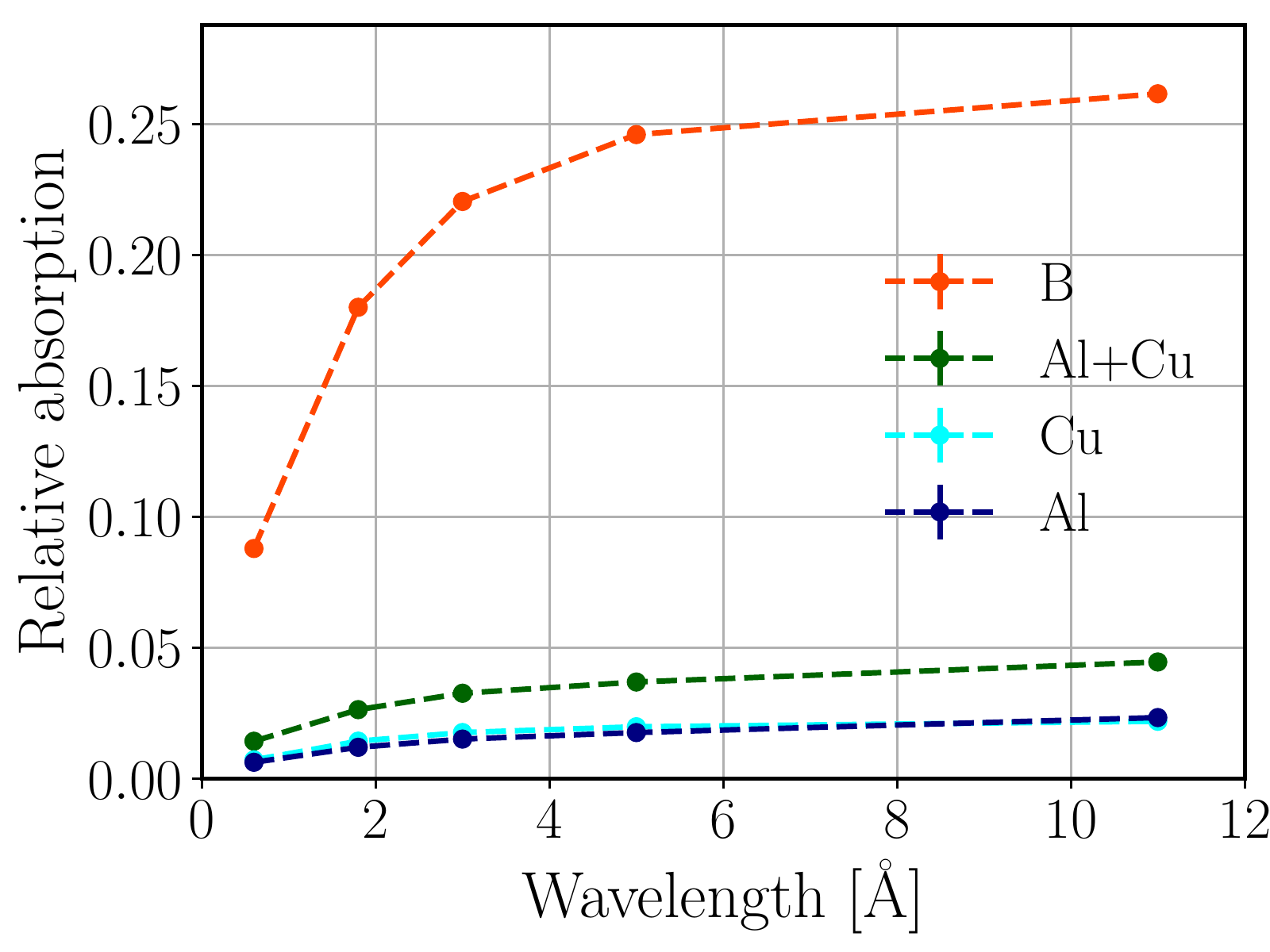}
      \end{subfigure}%
 % \begin{subfigure}{0.5\textwidth}
 % \centering
 % \includegraphics[width=\textwidth]{materialAbsorption_noJustBoron}
 %  \end{subfigure}
  \caption{\footnotesize Relative absorption in the BCS detector
    materials as a function of incident neutron wavelength. `B'
    (orange) represents the absorption in the B$_4$C layer. `Cu'
    (cyan) and `Al' (blue) are the relative absorption in copper and
    aluminium separately. `Al+Cu' (green) is the sum of the latter two, giving the absorption in
    aluminium and copper together. The dashed line only connects the
    markers.}
  \label{materialAbsorption}
\end{figure}

 In a single Geant4 simulation it is possible to register the number of
 neutrons absorbed in aluminium, copper and B$_4$C separately and
 compare the effect of these materials. 
 The latter could also be done with multiple simulations using different models with the materials out of focus replaced with vacuum to eliminate their effect on each other,
 but these effects change the results by $<$3\% so we present the numbers
 using the model with all materials in place. 
 %{\color{red} mention you only run a single simulation with all materials active. You can also
 %mention that the results are similar if your activate them one-by-one.} 
In Fig.~\ref{materialAbsorption} the relative absorption is depicted for
simulations with five different neutron wavelengths. The relative absorption in any material or materials is defined as the number of neutrons absorbed in that material over the number of incident neutrons for the entire detector system.

 Most of the undesired absorption occurs in the
 converter layer. This is not surprising with 30\% of the converted
 neutrons not triggering a detection event (70\% DCR). The absorption in
 the copper is higher than in aluminium except for the highest
 wavelength (11~{\AA}), but the difference is rather small in every case. This might be unexpected as the absorption cross-section of
 copper is approximately 16 times higher for these wavelengths (see
 Fig.~\ref{ncrystalCrossSections} in appendix), but the total volume of aluminium
 is 17.6 times higher than that of copper, therefore the average path length
 of the neutrons in aluminium is much longer, an effect that compensates the cross-section difference.
 
It is possible to make a `naive' analytical estimation of the absorption in a material using Eq.~\ref{eq:estRelAbs}:
{
\begin{equation}
\textrm{Relative absorption} = 1 - e^{- \Sigma_a \cdot l},
\label{eq:estRelAbs}
\end{equation}
where $\Sigma_a$ is the macroscopic absorption cross section given by Eq.~\ref{eq:Sigma_a} and $l$ is the path length in a material.
}
{
\begin{equation}
\Sigma_a = \sigma_a \cdot \rho_{A} = \sigma_a \cdot \frac{\rho_m \cdot N_A}{M},
\label{eq:Sigma_a}
\end{equation}
where $\sigma_a$ is the microscopic absorption cross section,
$\rho_{A}$ is the atomic density, $\rho_{m}$ is the mass density, $M$ is the molar mass of the material and $N_A$ is the Avogadro number.
}

\begin{figure}[!h]  
  \centering
  \begin{subfigure}{\textwidth} %{8cm}
      \includegraphics[width=\textwidth]{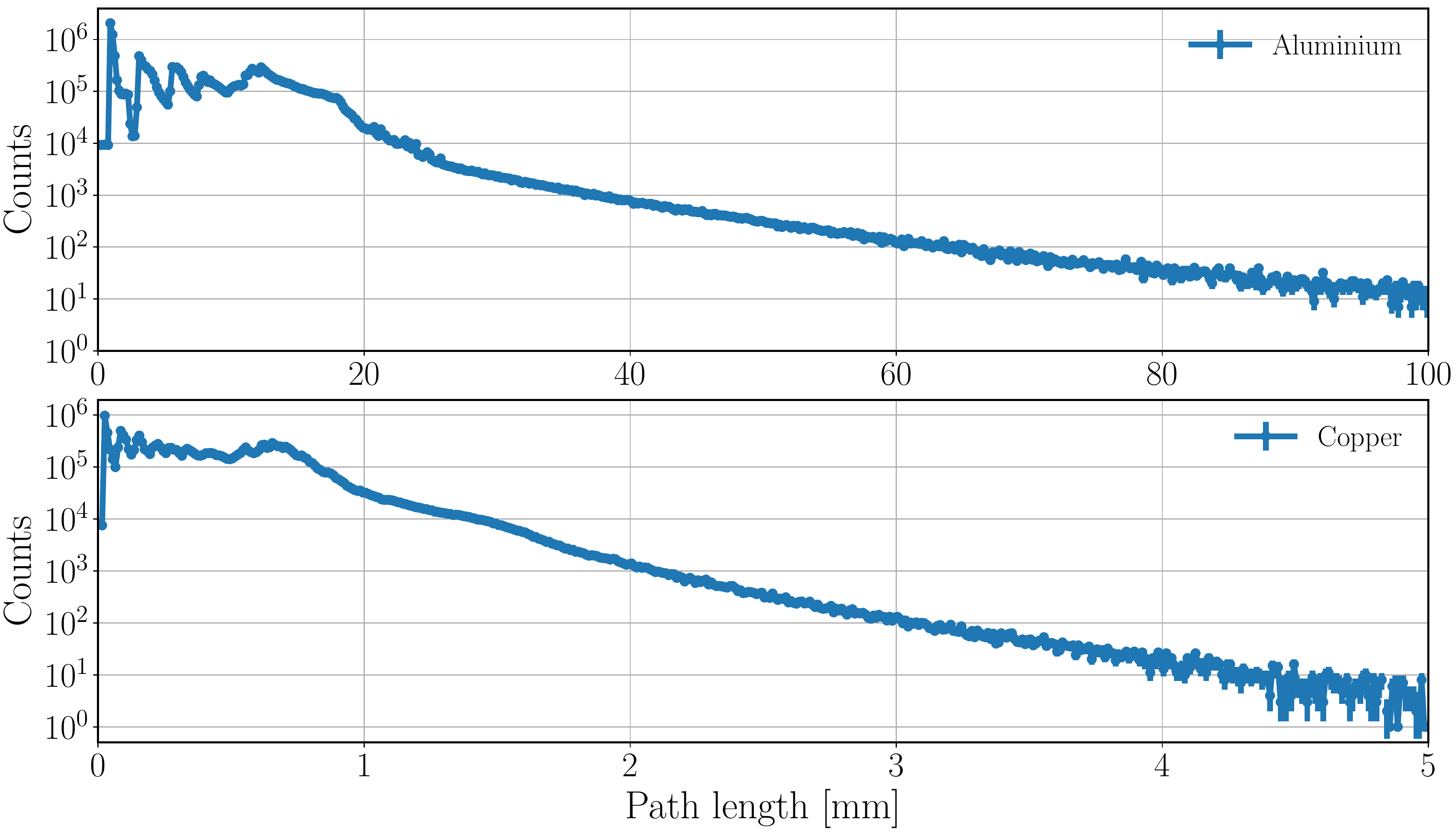}
     \end{subfigure}
   \caption{\footnotesize Neutron path length distribution in
     aluminium (top) and copper (bottom) from simulations with
     $\lambda$=1.8~{\AA} for all neutrons crossing the detector regardless of whether they
     interact with its materials or not. The label `Counts' means the number of neutrons with a total path length in aluminium or copper within the limits of a particular bin of the histogram.}
  \label{pathLength}
\end{figure}

For a specific neutron wavelength each parameter is known except the
path length in the materials. One way to estimate the latter is to
assume that a neutron stops halfway through the detector system
after crossing an aluminium wall 2 times and a copper wall 6 times frontally in each panel. 
Using the wall thicknesses provided in Tab.~\ref{tab:geom} the results are $l_{Al}$=4.7~mm and $l_{Cu}$=0.375~mm.
%Based on the aluminium tube and copper straw thicknesses provided in
%Tab.~\ref{tab:geom} only a naive estimation could be given for the path lengths in 5 panels. Assuming that a neutron goes through the
%aluminium wall 2 times and the copper wall 6 times frontally in each
%panel, the results would be 9.4~mm for the aluminium and 0.75~mm for
%copper.  From the simulations, the distribution of the path length in
%each material can be extracted, as displayed in Fig.~\ref{pathLength}. 
%%As indicated in the legend bar, for 1.8~{\AA} neutrons the average path length in aluminium is {\color{red}5.65}~mm and the same for copper is {\color{red}0.66}~mm.
%%The average path lengths for other wavelengths, along with the absorption cross sections, the average effective thicknesses and the relative absorption in these materials are shown in Tab.~\ref{tab:absorption}.
%With the average path length values an estimation can be given for the relative absorption, using the formula in Eq.~\ref{eq:estRelAbs}.
The path length of the neutrons is of course not constant even for a specific wavelength, as presented in Fig.~\ref{pathLength}. For aluminium, the beginning of the first peak in the histogram corresponds to the wall thickness of a tube, because that is the minimum distance in aluminium that a neutron has to pass to be absorbed in a straw. Due to the circular tube geometry and the conical beam, most of the neutrons do not enter the tube wall perpendicularly, so the neutrons absorbed in the first tube they enter can have a path length in aluminium longer than this minimum, that results in the first peak. The beginning of the second peak corresponds to three times the wall thickness of the tube, as the that is the  minimum distance in aluminium for a neutron that is absorbed in the second tube it enters. The upcoming peaks are more and more blurred as the path length difference in different directions, and the number of scattered neutrons become more and more important. The last broad peak corresponds to the neutrons that pass through all panels without being absorbed and the rest of the histogram contains only scattered neutrons.
Similar effects appear for copper because of the straws, but with more overlapping layers and complex geometry.

%\begin{figure}[!h]  
%  \centering
%  \begin{subfigure}{0.5\textwidth}
%    \centering
%    \includegraphics[width=\textwidth]{SimpleHist1D_Sum_length_of_TubeWall_segments_per_neutron_ang2}
%    %\caption{AluminiumPathLength2Ang}
%  \end{subfigure}%
%  \begin{subfigure}{0.5\textwidth}
%    \centering
%    \includegraphics[width=\textwidth]{SimpleHist1D_Sum_length_of_StrawWall_segments_per_neutron_ang2}
%    %\caption{CopperPathLength2Ang}
%  \end{subfigure}
%  \caption{\footnotesize Path length distribution of the neutrons in aluminium (on the left) and copper (on the right) from simulations with $\lambda$ = 1.8 {\AA}. }
%  \label{pathLength}
%\end{figure}

With the average path length of the neutrons extracted from the
simulations a more accurate estimate can be made proving the relevance of the formula in Eq.~\ref{eq:estRelAbs} and supporting the results. 
The relative absorptions from the two estimation methods and the
results from simulation are presented in
Table~\ref{tab:absorption}.
The estimations using the average path lengths are in very good agreement with the simulations. All results are within 2.5\%, except for the the lowest wavelength but even there the difference is less than 13\%. This shows how well such an easy formula describes the process of absorption in the detectors.
The naive estimation also gives acceptable results, given that it's a rough estimation. 
Some numbers are off by a factor 3.3, but for medium wavelengths the difference is less than 70\%. 
For wavelengths where the average path length is longer than the used
fixed number, the results are underestimated, and the other way around, overestimated path lengths lead to overestimated absorptions.
More accurate estimations could be made with more sophisticated formula but even in this state, both estimations support the simulation results. 

% {\color{red} what did I learn? that the
%  naive estimate is an underestimate but an ok prediction,
%  a factor 1-5 off?  That with average numbers you
%  can do a pretty good job? Do numbers check out? do they make you
%  believe the simulation?}

%NOTE: These estimations don't take into account the change in the wavelength..and the difference pathlengths
%The estimated relative absorption values alongside with the relative absorption directly from the simulations is shown in Table~\ref{tab:absorption}. 
%The results show how the path length difference compensates for the cross section difference that in the end leads to closely the same impact in terms of absorption. 
\begin{table}[htp]
\begin{center}
\begin{tabular}{| c | c c || c c | c c || c c |}
\hline
 \multirow{2}{*}{$\lambda$ [{\AA}]} & \multicolumn{2}{c||}{Rel.~Abs.$_{naive}$ [\%]}  & \multicolumn{2}{c|}{$l_{AVG}$[mm]} & \multicolumn{2}{c||}{Rel.~Abs.$_{est}$ [\%]} & \multicolumn{2}{c|}{Rel.~Abs.$_{sim}$ [\%]}  \\
      & Al & Cu & Al & Cu & Al & Cu & Al & Cu \\\hline
0.6 & 0.23 & 0.40 & 11.90 & 0.66 & 0.57 & 0.71 & 0.65 & 0.79 \\
1.8 & 0.65 & 1.19 & 8.43  & 0.45 & 1.17 & 1.44 & 1.20 & 1.44 \\
3    & 1.13 & 2.00 & 6.38  & 0.33 & 1.52 & 1.78 & 1.51 & 1.76 \\
5    & 1.85 & 3.30 & 4.44  & 0.22 & 1.75 & 1.97 & 1.76 & 1.98 \\
11  & 4.02 & 7.11 & 2.69 & 0.11 & 2.32 & 2.19 & 2.30 & 2.15 \\
\hline
\end{tabular}
\end{center}
\caption{\footnotesize
Estimated and simulated relative absorption in aluminium and
copper. The naive estimations (left) are made with constant path
lengths of $l_{Al}$=4.7~mm and $l_{Cu}$=0.375~mm for all
wavelengths. The more accurate estimations (centre) are made with the
average path lengths ($l_{AVG}$) from the simulations. The right
column contains the relative absorption extracted from simulation alone.
}
\label{tab:absorption}
\end{table}

%\begin{table}[htp]
%\begin{center}
%\begin{tabular}{| c | c c c c c c c c c c |}
%\hline
% \multirow{2}{*}{$\lambda$ [{\AA}]}  & \multicolumn{2}{c}{$\sigma_a$[b]} & \multicolumn{2}{c}{$\Sigma_a$[1/mm]} & \multicolumn{2}{c}{$l_{avg}$[mm]} & \multicolumn{2}{c}{Rel.~Abs.$_{est}$ [\%]} & \multicolumn{2}{c|}{Rel.~Abs.$_{sim}$ [\%]}  \\
%      & Al & Cu & Al & Cu & Al & Cu & Al & Cu & Al & Cu \\\hline
%0.6 & 0.08 & 1.26  & 0.00048 & 0.0108 & 11.90 & 0.659 & 0.57 & 0.71 & 0.65 & 0.79 \\
%1.8 & 0.23 & 3.78  & 0.00139 & 0.0320 & 8.43   & 0.453 & 1.17 & 1.44 & 1.20 & 1.44 \\
%3    & 0.39 & 6.30  & 0.00241 & 0.0539 & 6.38   & 0.332 & 1.52 & 1.78 & 1.51 & 1.76 \\
%5    & 0.64 & 10.51& 0.00398 & 0.0895 & 4.44   & 0.222 & 1.75 & 1.97 & 1.76 & 1.98 \\
%11  & 1.41 & 23.12 & 0.00873 & 0.1968 & 2.69 & 0.113 & 2.32 & 2.19 & 2.30 & 2.15 \\
%\hline
%\end{tabular}
%\end{center}
%\caption{
% \ref{ncrystalCrossSections}
%{\color{green} Should I omit $\sigma_a$ or $\Sigma_a$ from the table?}
%}
%\label{tab:absorption}
%\end{table}

%{\color{green} "Here we can discuss the absorption as well (a function of panels too)} \cite{nxsg4}

\begin{figure}[!h]  
  \centering
  \begin{subfigure}{0.5\textwidth} %{8cm}
      \includegraphics[width=\textwidth]{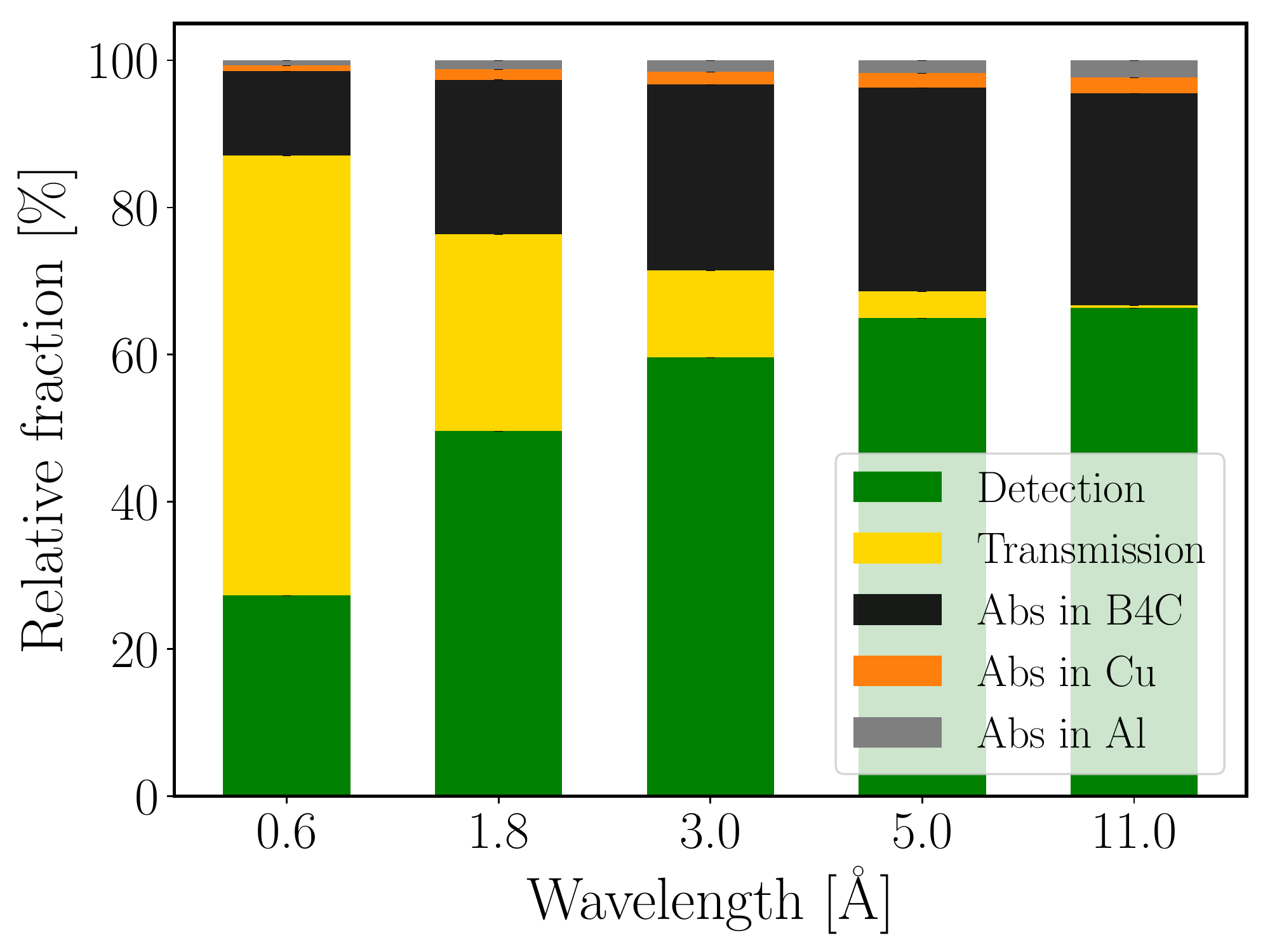}
     \end{subfigure}
   \caption{\footnotesize Proportion of absorption, transmission and detection from simulations with monoenergetic neutrons.}
  \label{fig:absTransDet}
\end{figure}

Fig.~\ref{fig:absTransDet} shows the proportion of absorption in different materials, transmission and detection from simulations. As all neutrons enter the detector system, the proportion of detection events is by definition the global detection efficiency described in the previous section. 
%\begin{table}[htp]
%\begin{center}
%\begin{tabular}{| c | c c c c c |}
%\hline
% \multirow{2}{*}{$\lambda$ [{\AA}]}  & \multicolumn{3}{c}{Absorption [\%]} & \multirow{2}{*}{Transmission [\%]} & \multirow{2}{*}{Detection [\%]}    \\
%& Al & Cu & B$_4$C & &  \\\hline
%0.6 & 0.65	&	0.79	&	11.53	&	59.76	&	27.27 \\
%1.8 & 1.20	&	1.44	&	20.97	&	26.82	&	49.56 \\
%3 & 1.51	&	1.76	&	25.31	&	11.83	&	59.59 \\
%5 & 1.76	&	1.98	&	27.65	&	3.57		&	65.04 \\
%11 & 2.30	&	2.15	&	28.84	&	0.33		&	66.37 \\
%\hline
%\end{tabular}
%\end{center}
%\caption{Proportion of unwanted absorption, transmission and detection from simulations with monoenergetic neutron beams with all material in place. 
%{\color{red}(8e6 neutrons)}
%}
%\label{tab:absTransDet}
%\end{table}
The results indicate that for the lowest wavelengths 60-30~\% of the
neutrons leave the detector system even with 5 panels but for the
highest wavelength this value drops below 0.5\%. 
This high transmission number at lower and medium wavelengths emphasises the need for a shielding layer behind the panels.
Proportion of absorption in aluminium and copper together is approximately 1.5\%
for low wavelengths and stays below 4.5\% even for the highest
wavelength. For the neutron wavelengths that are more relevant to neutron
scattering techniques, the absorption of this scale is acceptable and
justify the use of successive detection panels.

The obtained results correspond to pure unalloyed materials; alloyed materials and impurities may significantly increase the absorption due to the presence of isotopes with high absorption cross-section despite their low concentration. For example, the macroscopic absorption cross section of Al5754~\cite{Al5754}, an aluminium alloy typically used in nuclear science for mechanical structures, can be 18\% higher than the pure aluminium mainly due to its manganese content.

\FloatBarrier
\section{Activation}

%neutron absorption or neutron capture
Neutron absorption in the detector materials potentially has another
negative effect besides lowering the detection efficiency, namely the neutron activation of these materials.
%note: some atoms require more than one neutron capture to become unstable.
Activation might interfere with the normal operation of the detectors in two ways.
Firstly, the gamma rays and particles emitted by the excited nucleus
and the decay products might form a background during measurements in addition to that of prompt gammas.
Secondly, after the measurements, the radiation coming from the radioactive nuclei might 	not allow anyone to get close to the detectors (e.g. for maintenance) owing to the high gamma dose rate.
The purpose of this section is both to determine whether the background from the activation is significant for the measurements, and to find out how much time one has to wait after the measurements to be safe to approach the detectors.

This is intended to be a generic study, therefore the activation is
calculated for pure aluminium and copper instead of a specific alloy. 
The activation of the previously mentioned aluminium alloy is already investigated
in~\cite{ArActivation}. In that work it is also concluded that the
activation of the Ar/CO$_2$ is a minor effect compared to the same of
the aluminium-housing, and the beta radiation is negligible both in
terms of background and radiation protection, so these aspects are not addressed here.
%analytically based on results from the simulations

%NOTE: "Alloy Al5754 [31] has been chosen as a typical alloy used in nuclear science for mechanical structures"

%assuming constant flux of monoenergetic neutron beam, with the use of the relative absorption results from the previous section. %TODO not final..

At spallation sources the detectors are irradiated in pulses, but as the pulses and repetition times are much shorter than activation and decay times of the relevant isotopes, a constant average flux with the same integral can be used to determine activation. 
%The pulse length at ESS is much shorter than activation and decay times of the relevant isotopes, so for the sake of simplicity, the time average of the flux is used. 
The calculations are performed for a constant flux of
10$^9$~n/cm$^2$/s on a 1~cm$^3$ cubic sample, assuming 5\% of neutrons
are scattered toward the detectors. 
These numbers represent a worst case scenario for an intended SANS application~\cite{ratesPaper} but the results to be presented scale linear with the flux, making it is easy to adopt them to any other particular application. 
This assumption gives an incident flux of $5\cdot10^3$~n/cm$^2$/s for the 1~m$^2$ detector system, that means a neutron intensity of $5\cdot10^7$~n/s.
%{\color{red}TODO: ~\ref{activationDiagram}Diagram} 
Using this number as source intensity for the previously introduced simulation arrangement,
the intensity of neutron absorption in each material can be calculated for different monoenergetic beams
using the relative absorption results from Table~\ref{tab:absorption}. 
The results of these monoenergetic beams covering the most relevant wavelength range can be used to give estimation for a particular neutron spectrum by choosing the most representative wavelength or by corresponding segmentation using the upper wavelength for each range.
The estimation presented here is done for 3~{\AA} neutrons, with 1.51\% and 1.76\% of neutrons absorbed in aluminium and copper, respectively.
From the neutron absorption intensity, the activation of an isotope is
calculated for the irradiation duration $t_{irr}$ with Eq.~\ref{eq:activityIrrad},
\begin{equation}
a(t_{irr}) = I_a\cdot\left(1-\exp\left(-\lambda\cdot t_{irr} \right) \right)
\label{eq:activityIrrad}
\end{equation}
where $I_a$ is the neutron absorption intensity and $\lambda$ is the
decay constant of the regarded isotope. 
Pure aluminium contains only the $^{27}$Al isotope, but copper has two
natural isotopes -- $^{63}$Cu and $^{65}$Cu, so the absorption intensity is shared between them with respect to the their natural abundance and absorption cross-section. 
Neutron activation of the activation products (secondary activation) is neglected due to the low probability of the multiple neutron capture by the same nucleus. 
The irradiation time for the calculations is 10$^6$~s ($\approx$11.5 days), that will be approximately an operation cycle for ESS.

The activity of an isotope after irradiation and 
cooling time $t_c$ is given by Eq.~\ref{eq:activityCooling},
\begin{equation}
a(t_c) = a_0\cdot\exp\left(-\lambda\cdot t_c \right)
\label{eq:activityCooling}
\end{equation}
where $a_0$ is the activity reached by the end of the irradiation. 

\begin{figure}[!h]  
  \centering
  \begin{subfigure}{\textwidth} %{8cm}
      \includegraphics[width=\textwidth]{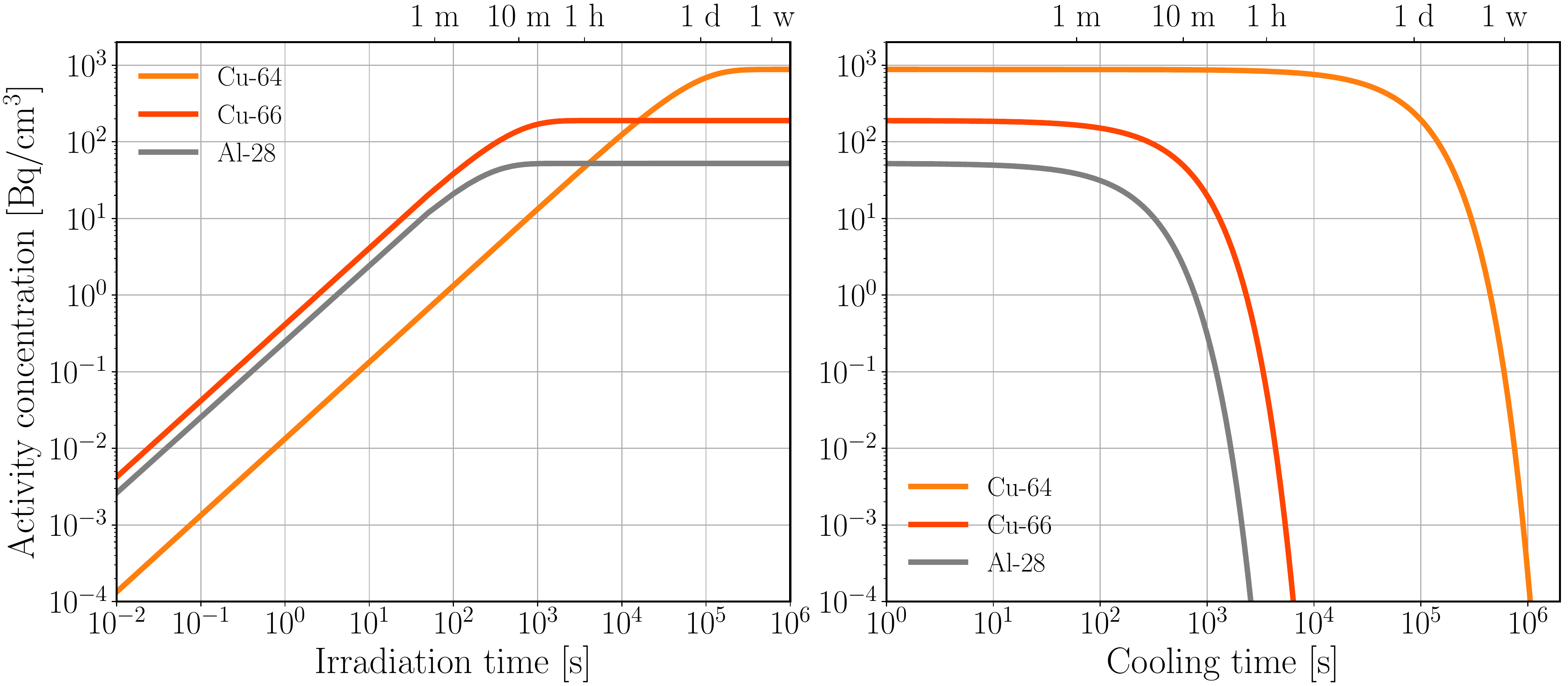}
     \end{subfigure}
   \caption{\footnotesize Activity concentration of the produced radionuclei during 10$^6$~s of irradiation with 3~{\AA} neutrons and during the cooling period.}
  \label{fig:activity}
\end{figure}

In order to express the results in activity concentration, the total activity of the isotopes are normalised with the total volume of the respective material. 
The volume of aluminium and copper for the five detector panels are V$_{Al}$=14447~cm$^3$ and
V$_{Cu}$=822~cm$^3$. The activity concentration of the isotopes of interest during irradiation and cooling are presented in Fig.~\ref{fig:activity}.

%The results show that the activity concentration of the produced radionuclei saturates by the end of irradiation time. These end values are used to calculate the decay gamma emission from a unit volume per second. 
%In this study the gamma efficiency is approximated conservatively with 10$^{-7}$~\cite{bcs_evolution} for all photon energies.
%Due to the constant gamma efficiency, only the total number of photons per decay has to be determined regardless of the photon energies. 
%In order to do that, the yields of the possible decay gamma lines are summed, resulting in a gamma yield of $\gamma_{d,Al-28}$=1.0, $\gamma_{d,Cu-64}$=2.89$\pm$0.06$\cdot$10$^{-3}$, $\gamma_{d,Cu-66}$=9.224$\pm$0.004$\cdot$10$^{-2}$ photons per decay~\cite{nucleardata}. 
%With the saturation activity concentration values, this gives gamma intensities of
%I$_{d,Al-28}$=52.26~s$^{-1}$, I$_{d,Cu-66}$=17.48~s$^{-1}$ and
%I$_{d,Cu-64}$=2.546~s$^{-1}$ for a unit volume. The total decay gamma
%intensity for the whole detector system is
%I$_{\gamma_d}$=7.715$\cdot$10$^5$~s$^{-1}$. 
The results show that the activity concentration of the produced radionuclei saturates by the end of irradiation time. These end values are used to calculate the decay gamma emission of these radionuclei from a unit volume per second. 
In this study the gamma efficiency is approximated conservatively with 10$^{-7}$~\cite{bcs_evolution,khaplanov2013} for all photon energies.
Due to the constant gamma efficiency, only the total number of photons per decay has to be determined regardless of the photon energies. 
In order to do that, the yields of the possible decay gamma lines presented in Table~\ref{tab:decay} are summed, resulting in a gamma yield of $\gamma_{d,Al-28}$=1.0, $\gamma^*_{d,Cu-64}$=4.73$\cdot$10$^{-2}$, $\gamma_{d,Cu-66}$=9.22$\cdot$10$^{-2}$ photons per decay. 
In addition to decay gammas, $^{64}$Cu emits a positron in 17.40\% of the decays, that will produce 2 photons of 511~keV by annihilation. These photons dominate the total gamma yield of $^{64}$Cu, that is $\gamma_{d,Cu-64}$=0.353.
The x-rays from decay and the bremsstrahlung of the emitted electrons are neglected due to their low energy compared to a reasonable pulse height threshold.
\begin{table}[htp]
\begin{center}
\begin{tabular}{| c c c c |}
\hline
Isotope & E [keV] & Yield [\%] & Flux to dose [$\frac{\mu Sv/h}{\gamma /cm^2s}$]\\\hline
Al-28 & 1778.969 & 100 & 0.02473 \\
Cu-64 & 1345.84 & 0.473 & 0.02028\\
Cu-64* & 511 & 34.8 & 0.00907\\
Cu-66 & 833.537 & 0.220 & 0.01387\\
Cu-66 & 1039.231 & 9 & 0.01666\\
Cu-66 & 1333.120 & 0.0037 & 0.02013\\
\hline
\end{tabular}
\end{center}
\caption{\footnotesize
Decay gamma lines of the activated isotopes with their production yield per decay~\cite{nucleardata} and the flux to dose conversion factor corresponding to the photon energy~\cite{ess0019931}. The 511~keV photons of Cu-64 are the results of the annihilation of the emitted positrons.   
}
\label{tab:decay}
\end{table}
With the saturation activity concentration values of A$_{Al-28}$=52.3~Bq, A$_{Cu-64}$=881~Bq 
and A$_{Cu-66}$=190~Bq, the gamma intensities are
I$_{d,Al-28}$=52.3~s$^{-1}$, I$_{d,Cu-64}$=331~s$^{-1}$ and
I$_{d,Cu-66}$=17.5~s$^{-1}$ for a unit volume. 
For the total volume of aluminum and copper, the gamma intensity is 
I$_{\gamma_{Al}}$=7.55$\cdot$10$^5$~s$^{-1}$ and I$_{\gamma_{Cu}}$=2.70$\cdot$10$^5$~s$^{-1}$.
The total decay gamma intensity for the whole detector system is
I$_{\gamma_d}$=1.025$\cdot$10$^6$~s$^{-1}$. 

The prompt gamma intensity ($I_{\gamma_p}$) for aluminium and copper is calculated as the product of the neutron absorption intensity and the total prompt gamma yield per absorption ($\gamma_p$) for each material (see Eq.~\ref{eq:promptGamma}).

\begin{equation}
I_{\gamma_p} = I_a\cdot \gamma_p  = I_a\cdot \frac{\sum\limits_i \sigma_{\gamma_i}}{\sigma_a}
\label{eq:promptGamma}
\end{equation}
The number of prompt gammas per absorption is estimated as the ratio of the sum of the gamma line specific cross-sections ($\sigma_{\gamma_i}$) and the absorption cross-section ($\sigma_a$)~\cite{crossSectionDatabase}. 
Due to the gamma cascades, this ratio is not unity, the resulting yields are 
$\gamma_{p,Al}$=1.978 %$\pm$0.028  
and $\gamma_{p,Cu}$=2.665. %$\pm$0.024. 
The prompt gamma intensities are I$_{Al}$=1.493$\cdot$10$^6$~s$^{-1}$, 
I$_{Cu}$=2.345$\cdot$10$^6$~s$^{-1}$ and the total prompt gamma intensity is I$_{\gamma_p}$=3.839$\cdot$10$^6$~s$^{-1}$.

%The total prompt and decay gamma intensity of aluminium and copper is I$_\gamma$ = 4.611$\cdot$10$^6$~s$^{-1}$.
The total prompt and decay gamma intensity of aluminium and copper is I$_\gamma$=4.864$\cdot$10$^6$~s$^{-1}$. 
This intensity is so low that it is overruled by the intensity of photons from the conversion process. In about 94\% of the neutron conversions in $^{10}$B, a 0.48~MeV gamma is emitted~\cite{Knoll}.  
Using the relative absorption ratio of B$_4$C for 3~{\AA} neutrons from the absorption section,
neglecting the neutron absorption of carbon and $^{11}$B, 25.31\% of the incident neutrons are converted.  
This translates to a gamma intensity of I$_{\gamma_B}$=1.190$\cdot$10$^7$~s$^{-1}$ from the converter layer. 
As this is the key source of photons, the total gamma intensity of the detector system for the observed neutron flux is on the order of 10$^{7}$ for all relevant wavelengths.
With a gamma efficiency of the BCS detectors on the order of 10$^{-7}$, the gamma background coming from the detector itself is negligible compared to the detected neutron intensity on the order of 10$^7$~n/s.
% 3\AA detection ratio: 59.59, 
%Absorption is $^{10}$B I$_B$ = $5\cdot10^7 \cdot$  0.2531 $\cdot$0.94 = 1.190$\cdot$10$^7$~1/s

In terms of dose rate after the irradiation, the detector system can be approximated as a surface source of 100$\times$100~cm$^2$, emitting photons with the source intensity and energy corresponding to the decay gamma intensity and energy of the activated isotopes. The resulting fluxes at the end of the irradiation time are $\Phi_{Al}$=37.75~cm$^{-2}$s$^{-1}$, $\Phi_{Cu-64}$=12.78~cm$^{-2}$s$^{-1}$, $\Phi_{Cu-66}$=0.72~cm$^{-2}$s$^{-1}$. The fluxes are dominated by the 1778.969~keV, 511~keV and 1039.231~keV energy photons, respectively, so the fluxes are turned into dose rate with the corresponding flux to dose rate conversion factors presented in Table~\ref{tab:decay}. 
The resulting dose rates are $\dot{D}_{Al}$=0.934~$\mu$Svh$^{-1}$,  $\dot{D}_{Cu-64}$=0.116~$\mu$Svh$^{-1}$ and $\dot{D}_{Cu-66}$=0.012~$\mu$Svh$^{-1}$ with the total value of $\dot{D}$=1.062~$\mu$Svh$^{-1}$. The dose rate after certain cooling time can be calculated in the same way and the results can be scaled for similar detector setups.

As already mentioned, the obtained results correspond to pure unalloyed materials; alloyed materials and impurities may significantly increase the activity and dose rate due to isotopes with high cross-section or long half-life. 
This investigation assumed cold neutrons and single neutron activation, but for fast neutrons
other reaction, like $^{63}$Cu(n,$\alpha$)$^{60}$Co or $^{63}$Cu(n,p)$^{63}$Ni must be taken into consideration.

\FloatBarrier
\section{Scattering}
Another side effect of the increased detector material budget due to the multi panel layout is
the scattering of neutrons. In contrast to absorption, scattering can
degrade the detector performance by producing intrinsic background,
which in turn can impact the signal to background ratio. The latter is
a driving requirement in particular for inelastic neutron instruments
and has to be carefully considered in the detector design process. 

The definition of signal and background depends on the detector
application, as different quantities might be relevant for the
respective measurement technique. Moreover, there are different ways
to quantify scattering in a detector, each giving a different but still
valid insight. A series of studies have already quantified the effect
for the Multi-Blade and the Multi-Grid detectors~\cite{multiBlade2018, DIAN2018173, Dian_2019}. 
In the study that follows, rather than only look at the SANS application initially foreseen for BCS;~all quantities of interest are looked at, so that the results may be considered in the context of all techniques, i.e. reflectometry, diffraction and spectroscopy.

%The more panels of detectors and consequently more amount of material increases not only the unwanted absorption but also the scattering on the detector's materials and therefore the scattered background. Higher detection efficiency doesn't necessarily means a scientific improvement because the additional detection events increase both the signal and the background. 
%so it's vital to well define all the figures. 
%csak a detektorból származó háttérrel foglalkozunk ebben a cikkben (no instrument effects, scattering patern from the sample and no cosmo.. background)

%In this section we aspire to present information about deviation of the following base and derived quantities due to the scattering in the detector materials:
\subsection{Quantities of interest}

The following raw (in boldface font) and derived quantities are of
interest for the scattering study:
\begin{itemize}
\item \textbf{X}: position along the straws (along the wire).
\item \textbf{Y}: position perpendicular to the straws.
\item \textbf{TOF}: neutron time of flight from the source until the detection.
\item $\Theta$: polar angle calculated from the source and the detection event X and Y positions.
\item $\Phi$: azimuthal angle calculated from the source and the detection event X and Y positions.
\item $\lambda$: neutron wavelength calculated from the
  distance between the source and the detection event position (SDD)
  and the TOF, using Eq.~\ref{eq:lambda} as
\begin{equation}
\lambda=\frac{h}{m\cdot \upsilon}= \frac{h}{m}\cdot \frac{TOF}{SDD},
\label{eq:lambda}
\end{equation}
where $h$ is the Planck constant, $m$ is the neutron mass and $\upsilon$ the velocity.
\item E: neutron energy calculated from $\lambda$.
\item Q: scattering vector calculated from
  $\Theta$ and $\lambda$, using Eq.~\ref{eq:Q} (SANS definition):
\end{itemize}

\begin{equation}
Q=\frac{4\pi}{\lambda}\sin\left(\frac{\Theta}{2} \right).
\label{eq:Q}
\end{equation}

\begin{figure}[!h]  
  \centering
  \begin{subfigure}{.5\textwidth} %{8cm}
      \includegraphics[width=\textwidth]{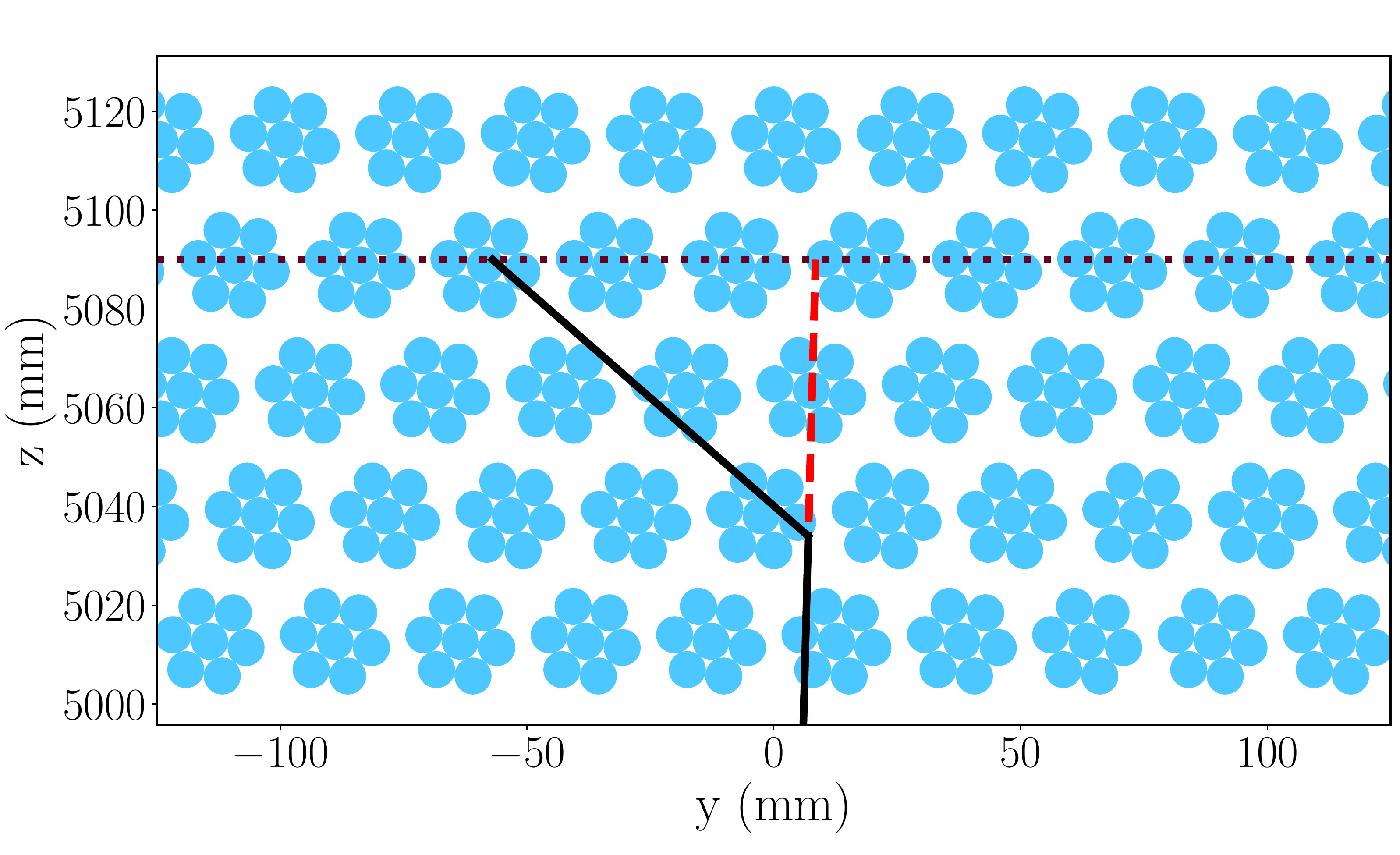}
     \end{subfigure}
   \caption{\footnotesize Demonstration of the ideal detection point
     for a scattered neutron. The black line represents the track of a
     neutron that is scattered in the second panel before being
     converted and detected in the fourth panel. The red dashed line
     shows the track of the ideal, non scattered neutron, that is the
     continuation of the simulated neutron's initial direction until
     the Z plane of its detection point, indicated with the dotted line.}
  \label{fig:projectedDetectionPoint}
\end{figure}

Scattering, elastic and inelastic, leads to a
change in the detection coordinates and subsequently in all derived
quantities in almost every case. To quantify this change, the ideal
values of these quantities must be defined for each neutron. These
values correspond to a non-scattered neutron generated with the same
initial parameters. Without scattering, the neutron $\lambda$
and direction ($\Theta$ and $\Phi$) remain unaltered. The ideal values
of these parameters are easy to define. Q is derived
from $\Theta$ and $\lambda$ so its ideal value is also straightforward
to estimate. For the remaining quantities an ideal detection point and
time have to be defined. This could be done in many ways but for this study it is
defined as the position extrapolated from the source's position and
the neutron's initial $\Theta$ and $\Phi$ to the Z plane of the actual
detection event, as demonstrated in
Fig.\ref{fig:projectedDetectionPoint}. From the ideal detection
coordinates the ideal X, Y and TOF values are calculated. The
difference of the simulated and ideal quantities are referred to
hereinafter as $\delta$X (=$\delta$X$_{sim}$-$\delta$X$_{ideal}$), $\delta$Y, $\delta\Theta$ and so on.
%NOTE: $\delta$ = Simulated - Projected

%{\color{green}Mostly }
Due to the statistical behavior of the detection
process, 
and discrete Y and Z detection coordinates defined by the anode wires, 
the $\delta$ quantities are not zero even for the
non-scattered neutrons. Therefore, in order to decide what is signal and what
is background, an upper and a lower limit are selected for all
quantities. The detection events with a $\delta$ value within the
limits are counted as signal for the respective quantity. 

\begin{figure}[!h]  
  \centering
  \begin{subfigure}{\textwidth} %{8cm}
      \includegraphics[width=\textwidth]{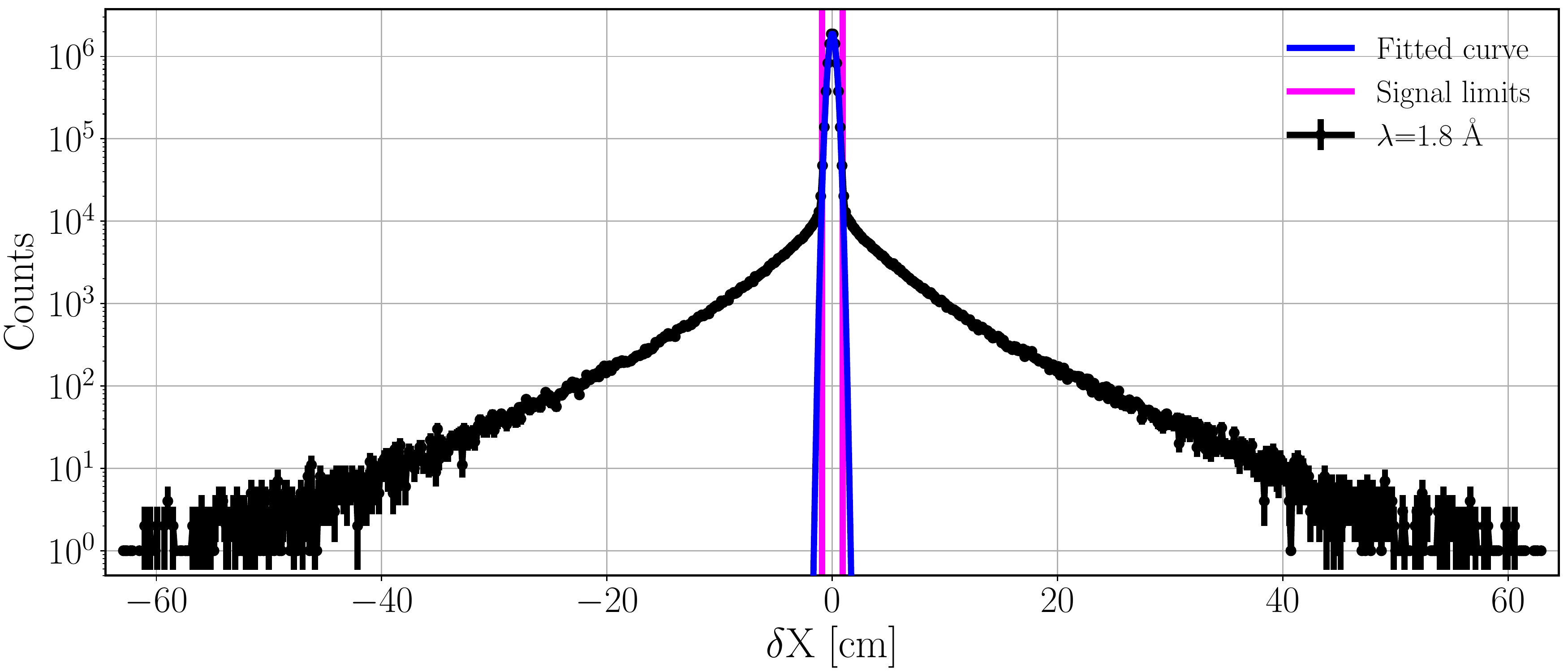}
     \end{subfigure}
   \caption{\footnotesize Finding the limits for signal and background
     separation for $\delta$X using the results from simulation
     with 1.8~{\AA} monoenergetic beam. The Gaussian function (blue)
     is fitted on the data (black) with the 3$\sigma$ limits appearing
     in red. }
  \label{fig:Gaussian_example}
\end{figure}

Fig.~\ref{fig:Gaussian_example} demonstrates the process of finding the signal limits for $\delta$X.
A Gaussian function is fitted to the $\delta$ distribution, and the limits are defined as the mean value $\pm$3$\sigma$. 
% The same method is applied to $\delta\Theta$, $\delta\Phi$, $\delta$TOF and $\delta$Q (see Fig.~\ref{fig:4gaussianQuantity}), but $\delta$Y and $\delta\lambda$ need a different approach.
The same method cannot be applied to $\delta$Y, as
Fig.~\ref{fig:dY_example} shows clearly that a
Gaussian fit is not appropriate. The shape in the centre is the result
of using the position of the wire in the centre of the straw instead
of the exact coordinates of the detection event. 
The maximum difference in Y caused by using the wire coordinate is the outer radius of the converter layer.
The peak corresponds to this difference so all neutrons which follow a straight line until their conversion are contained in it.
This means that the straightforward limit for the separation of signal and background is the outer radius of the converter layer, which is the same as the inner radius of the straw.

%The drop of the curve
%on both the positive and negative side corresponds to the outer radius
%of the converter layer. 
%If the absolute value of $\delta$Y is less
%than this radius, it means that the neutron is detected in a straw
%that has an intersection with the line of the initial direction of the
%neutron. {\color{red} I don't follow, you can just write that the
%  peaks correspond to the straw radius or whatever and contain
%  neutron which followed straight line until their conversion} For neutrons that do not scatter before the conversion, $\delta$Y gives the difference of the wire's and the detection events coordinate in the Y direction. As most of the detection events are close to the converter layer, $\delta$Y is essentially the 1-dimensional projection of a circle, that gives the U-shape in the center.
%This all means that the straightforward limits for the separation of signal and background for $\delta$Y is $\pm$3.725~mm, the outer radius of the converter layer, that is the same as the inner radius of the straw. 
{%\color{red}It would be more clear to say that a non-scattered neutron }

\begin{figure}[!h]  
  \centering
  \begin{subfigure}{\textwidth}
     \centering
      \includegraphics[width=\textwidth]{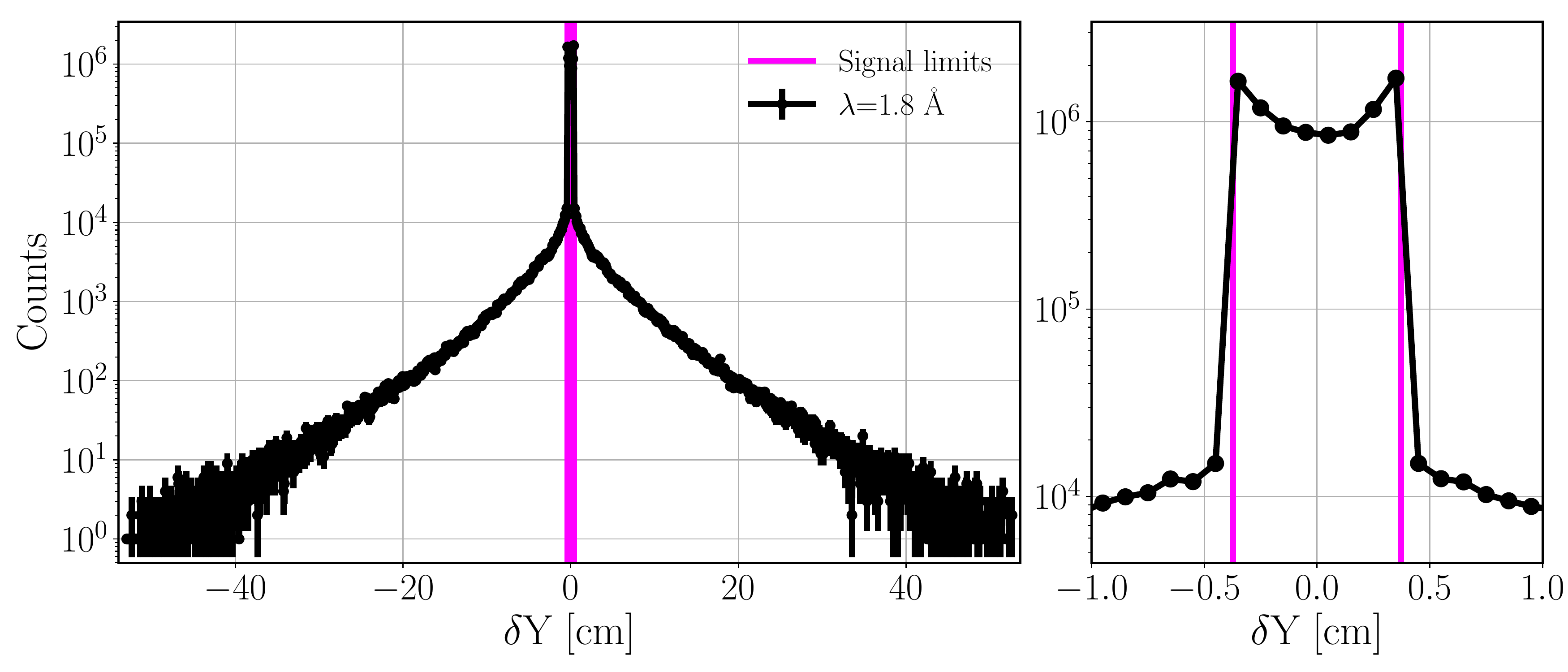}
  \end{subfigure}
       \caption{\footnotesize Finding the limits for signal and
         background separation for $\delta$Y with a 1.8~{\AA} monoenergetic beam. 
         The limits given by the straw inner radius are $\pm$3.725~mm. 
         The figure on the right shows an enlarged view of the center part of the figure on the left. The lines are only joining the points.}
  \label{fig:dY_example}
\end{figure}

%{\color{red} NOPE thats not what dTOF is..
%The maximum difference in TOF comes from the maximum neutron path length difference, that is the same as for Y, the outer radius of the converter layer.
%TODO: express $\Delta$TOF with equation...and explain it (it is the difference of the simulated tof and the TOF calculated from the initial velocity to the wire detection coordinate)the difference of the measured TOF and what TOF would be for the detection point for a non-scatted neutron.
%The difference in TOF depends on the velocity of the neutrons, so the $\delta$TOF limits a calculated for the relevant wavelengths.}

$\delta$TOF is the difference of the simulated TOF and the ideal TOF. 
The former is the time until the neutron reaches its conversion point from the source plus the time until the conversion products deposit enough energy to overcome the applied detection threshold. 
The TOF of the conversion products is negligible compared to the TOF of the neutrons for any relevant distance because they have a much higher velocity due to their initial kinetic energy, so the simulated TOF is practically the TOF of the neutrons until the conversion point.
 The ideal TOF is calculated from the initial wavelength and the distance between the source and the ideal detection point.
For a non-scattered neutron the wavelength, and therefore the velocity does not change, so $\delta$TOF comes from the different distances from the source to the conversion point and to the ideal detection point.
This distance is related to the ion range, that appears to agree with the radius of the straws. 
Therefore, the $\delta$TOF limits to separate signal and background are calculated from the TOF difference caused by this spatial difference, that depends on the neutron velocity so different limits are defined for the wavelengths of interest.

The effect of the discrete Y and Z detection coordinates also appears for $\lambda$.
The ideal $\lambda$ is calculated from the TOF and the source to detection point distance (SDD) according to  Eq.~\ref{eq:lambda}, 
but for the simulated $\lambda$, the distance between the source and the detection event wire coordinates (SWD) is used. 
The maximum difference between source to detection point distance and source to detection event's wire coordinates distance is the inner radius of the straw. 
The resulting $\Delta\lambda$ difference for a non-scattered neutron is calculated using Eq.~\ref{eq:deltaLambda}

%Fig.~\ref{fig:dlambda_example} showing $\delta\lambda$.
%$\delta\lambda$ is the difference of the measured $\lambda$, calculated from the detection events wire coordinates and the measured TOF, and the initial $\lambda$ of the neutron.
 
%This causes a $\Delta\lambda$ difference in the simulated $\lambda$. 
%This $\Delta\lambda$, calculated for the non-scattered neutrons using Eq.~\ref{eq:deltaLambda} derived from Eq.~\ref{eq:lambda}, gives good limits for signal and background separation. 

{
%\begin{equation}
%\Delta\lambda = \frac{h}{m} \cdot \frac{TOF}{(SWD)^2}\cdot {\Delta}SWD =  \frac{h}{m} \cdot \frac{{\Delta}SWD}{v\cdot SWD}
%\label{eq:deltaLambda}
%\end{equation}
\begin{equation}
\Delta\lambda = \lambda_{sim}-\lambda_{ideal} = \frac{h}{m} \cdot \frac{TOF}{SWD} - \frac{h}{m} \cdot \frac{TOF}{SDD}.
\label{eq:deltaLambda}
\end{equation}
%where $h$ is the Planck constant, $m$ is the neutron mass,  and 
%where $SWD$ is the distance between the source and the detection event wire coordinates, and $\Delta$SWD is equal to the inner radius of a straw.}
%$SWD$ differs for all neutrons, that means $\Delta\lambda$ caused by using the wire coordinates differs too, but the difference is rather small because the 5~m source to detector distance gives most of the $SWD$.
%The limits for $\delta\lambda$ were calculated using the average $SWD$ for the detected non scattered neutrons, that is only 10~cm more than the source to detector distance even for the lowest 0.6~{\AA} wavelength. 
$\Delta\lambda$ depends on the SDD and the corresponding TOF. 
The highest difference, that gives good limits for the signal and background separation appears for the shortest SDD, that is 5~m.  
The TOF and therefore the $\delta\lambda$ limits depend on the initial velocity of the neutrons, so different limits are defined for the 5 wavelengths of interest. 
Neutron energy is calculated directly from $\lambda$, so the signal
and background separation for $\delta\lambda$ is valid for $\delta$E, and therefore not repeated.
The limits for $\delta\Theta$, $\delta\Phi$ and $\delta$Q are defined
the same way as for $\delta$X, by fitting a Gaussian function. The $\delta$Q limits are wavelength dependent, so they are defined for each wavelength of interest.

%The same effect appears in Fig.~\ref{fig:dTOF_example} showing $\delta$TOF. 
%%$\delta$TOF is the the measured TOF until the detection event minus the TOF needed for an ideal neutron with the same initial parameters to reach the Z plane of the detection event.
%The maximum neutron path length difference caused by using the wire
%coordinates is the the inner radius of a straw. The corresponding
%time difference, calculated using the neutron's initial velocity,
%gives good limits for signal and background separation.

For all quantities where the limits are defined by the parameters of the fitted Gaussian function, the mean is at least 3 orders of magnitude lower than the standard deviation, so for simplicity zero is used instead. This means that the range for any $\delta$ quantity, within which a detection event is considered as signal is $\pm$ the corresponding limit.
The width of all signal ranges are presented in Tab.~\ref{tab:signalConstantLimits} and Tab.~\ref{tab:signalWavelengthLimits}. The figures showing the limits visually, similarly as for $\delta$X~(Fig.~\ref{fig:Gaussian_example}) and $\delta$Y~(Fig.~\ref{fig:dY_example}), are placed in appendix (see Fig.~\ref{fig:dth_example}--\ref{fig:dQ_example}).
The separation of signal and background using the limits is applied on
the raw data instead of the histograms, to avoid possible distortions
caused by the arbitrary binning choices. 

\begin{table}[htp]
  \begin{center}
    \begin{tabular}{| c | c |}
      \hline
      Quantity & Width of signal range \\
      \hline
      $\delta$X [cm] & 1.75\\
      $\delta$Y [cm]& 0.75\\
      $\delta\Theta$ [degree] & 0.21\\
      $\delta\Phi$ [degree] & 3.25\\
      \hline
    \end{tabular}
  \end{center}
  \caption{\footnotesize The width of the ranges within which detection events are considered as signal for quantities with limits independent of the neutron wavelength. The limits are $\pm$ half of the width presented in this table. }
  \label{tab:signalConstantLimits}
\end{table}

\begin{table}[htp]
  \begin{center}
    \begin{tabular}{| c | c c c c c |}
      \hline
      \multirow{2}{*}{Quantity}  & \multicolumn{5}{c|}{Width of signal ranges for different wavelengths} \\
      & 0.6~{\AA} & 1.8~{\AA} & 3~{\AA} & 5~{\AA} & 11~{\AA}\\
      \hline
      $\delta$TOF [$\mu$s] & 0.57 & 1.70 & 2.83 & 4.71 & 10.63\\
      $\delta\lambda$ [{\AA}]&4.5$\cdot$10$^{-4}$ &1.34$\cdot$10$^{-3}$ & 2.24$\cdot$10$^{-3}$ & 3.73$\cdot$10$^{-3}$ & 8.20$\cdot$10$^{-3}$ \\
      $\delta$Q [1/{\AA}] & 1.95$\cdot$10$^{-2}$ & 6.3$\cdot$10$^{-3}$ & 3.8$\cdot$10$^{-3}$ & 2.2$\cdot$10$^{-3}$ &  1.0$\cdot$10$^{-3}$\\
      \hline
    \end{tabular}
  \end{center}
  \caption{\footnotesize The width of the ranges within which detection events are considered as signal for quantities with wavelength dependent limits. The limits are $\pm$ half of the width presented in this table.}
  \label{tab:signalWavelengthLimits}
\end{table}

\FloatBarrier
\subsection{Impact of scattering on spatial resolution}

The FWHM of the $\delta$X distribution is 0.67~cm, which is a convolution of the detection
coordinate approximation with the weighted average of the deposited
energy by the conversion products and the applied smearing with a FWHM
of 0.6~cm. Small local scatterings could also cause the broadening of
the peak. In order to evaluate this effect, the $\delta$X distribution
is shown in Fig.~\ref{fig:dX_noSmear} consecutively estimated with the neutron conversion coordinates,
the approximated detection coordinates and finally with the application of position smearing
on top of the latter.
%with the smearing, without the smearing and without both the smearing and the detection coordinate approximation. For the last one, $\delta$X of the point of conversion is used that can be non-zero only if the neutron is scattered.
The results demonstrate that the broadening of the peak because of
scattering inside the detector is negligible compared to the other
processes of the detection. This means that scattering inside the
detector does not affect the spatial resolution of the detector.

\begin{figure}[!h]  
  \centering
  \begin{subfigure}{\textwidth}
     \centering
      \includegraphics[width=\textwidth]{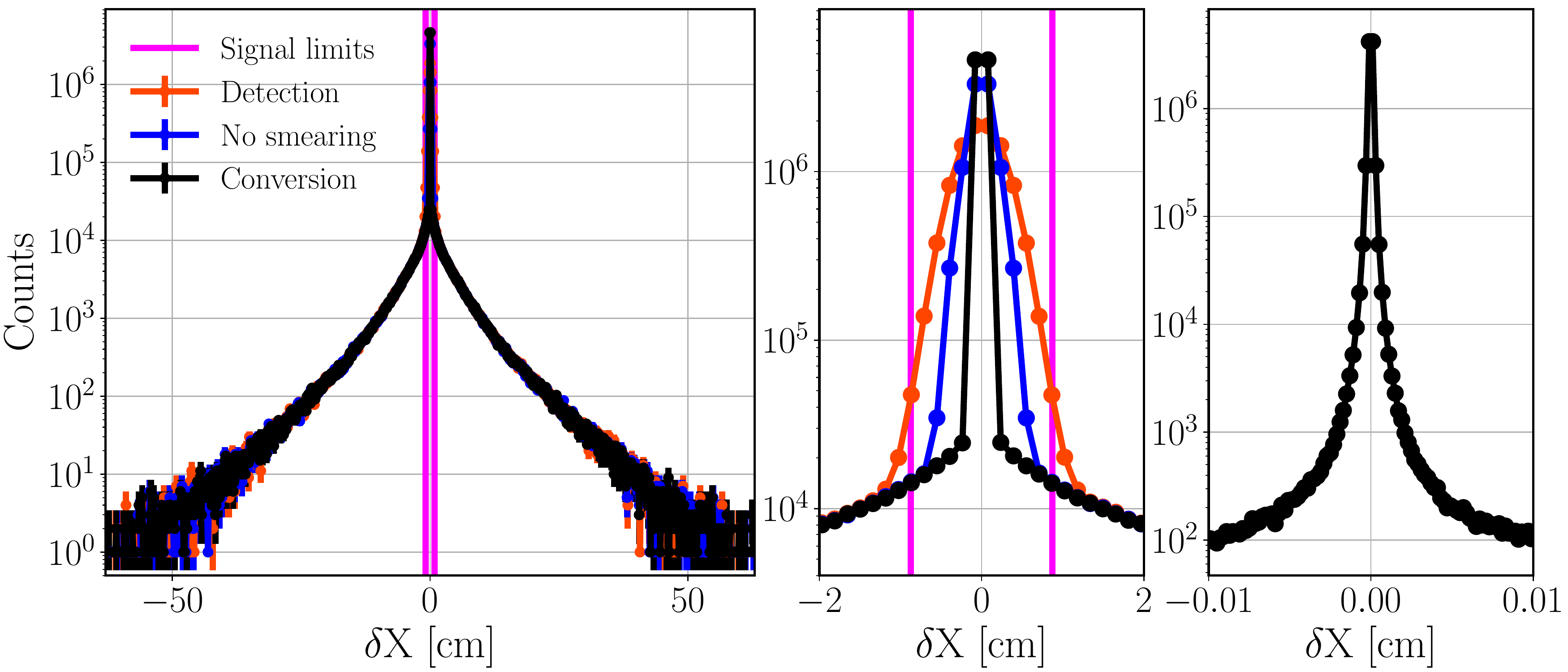}
  \end{subfigure}
     \caption{\footnotesize Decomposition of the peak broadening
       effects for $\lambda$=1.8~{\AA}. $\delta$X is shown for the
       neutron conversion coordinates (black), for the approximated
       detection coordinates (blue) and for the approximated
       detection coordinates with the application of position smearing
       (red). The figure on the right shows only $\delta$X for the conversion point in the
       centermost range with finer binning to reveal the shape of its peak. The lines are only joining the points.}
       %         \caption{\footnotesize Decomposition of the peak broadening
%       effects for $\lambda$ = 1.8~{\AA}. In the figures on the left
%       and the centre, $\delta$X is shown 
%       for normal detection process with detection coordinate approximation and smearing (in orange), 
%       the same without smearing (in blue) 
%       and without both detection coordinate approximation and smearing using the conversion points coordinates (in black). 
%       The figure on the
%       right shows only $\delta$X for the conversion point in the
%       centermost range with finer binning to reveal the shape of its
%       peak. {\color{red} reverse the order of presentation,
%         conversion, detection, smearing as in the text} }
  \label{fig:dX_noSmear}
\end{figure}

\FloatBarrier
\subsection{Impact of panels and material budget}

Fig.~\ref{fig:compare_dth_panel} depicts $\delta\Theta$ for a different number of detector panels for 1.8~{\AA} neutrons. 
It can be seen that the overall fraction of scattered background increases with the number of panels.
The increase of the scattered background with every additional panel
has two components. First, there are scattered neutrons detected in
the downstream panels. Second, there are neutrons backscattered from
the downstream panels that are detected upstream.
\begin{figure}[!h]  
  \centering
  \begin{subfigure}{\textwidth} %{8cm}
    \includegraphics[width=\textwidth]{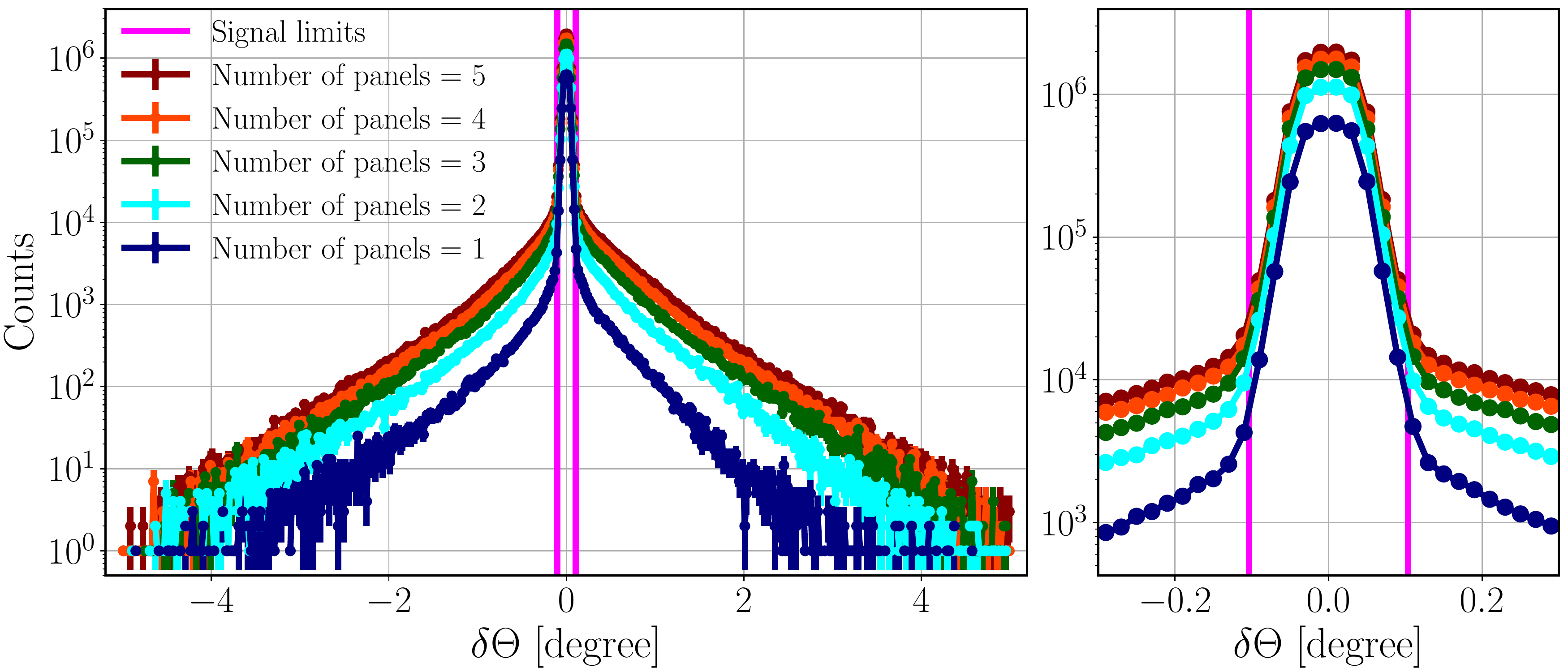}
  \end{subfigure}
  \caption{\footnotesize $\delta\Theta$ for different number of panels from simulations with $\lambda$=1.8~{\AA}. The signal limits are defined from simulation with 5 panels. The figure on the right shows an enlarged view of the center part of the figure on the left. The lines are only joining the points.}
  \label{fig:compare_dth_panel}
\end{figure}

In order to distinguish the scattering effects of the different materials, simulations are performed with 4 different detector models derived from the original one by leaving out some of the materials. All of the models included the Ar/CO$_2$ gas and the B$_4$C layer as they are indispensable for the detection. One of the models had nothing more in it, and two others included either the copper straws or the aluminium tubes. The last one was the original model with both aluminium and copper in place.

Fig~\ref{fig:compare_dth_material}. depicts $\delta\Theta$ for the different detector models with 1.8~{\AA} neutrons. 
Though the quantitative approach to determine the amount of relative scattering is presented in section \ref{subsec:fracScat}, it is already visually clear that the original model that contains all materials produces the highest scattered background.
From the simulations with either aluminium or copper left out, it is
clear that aluminium is the main source of the scattering. Similarly
to absorption, the scattering cross-section of copper is generally
much higher (approximately 5 times higher for 1.8~{\AA}}), but in this case the difference is not high enough to compensate the presence of 17.6 times more aluminium.

For the rest of the quantities the scattered background shows similar
trends as a function of number of panels and detector materials,
therefore only their wavelength dependence is presented using the
original model of 5 detector panels with all materials in
place. 

\begin{figure}[!h]  
  \centering
  \begin{subfigure}{\textwidth} %{8cm}
    \includegraphics[width=\textwidth]{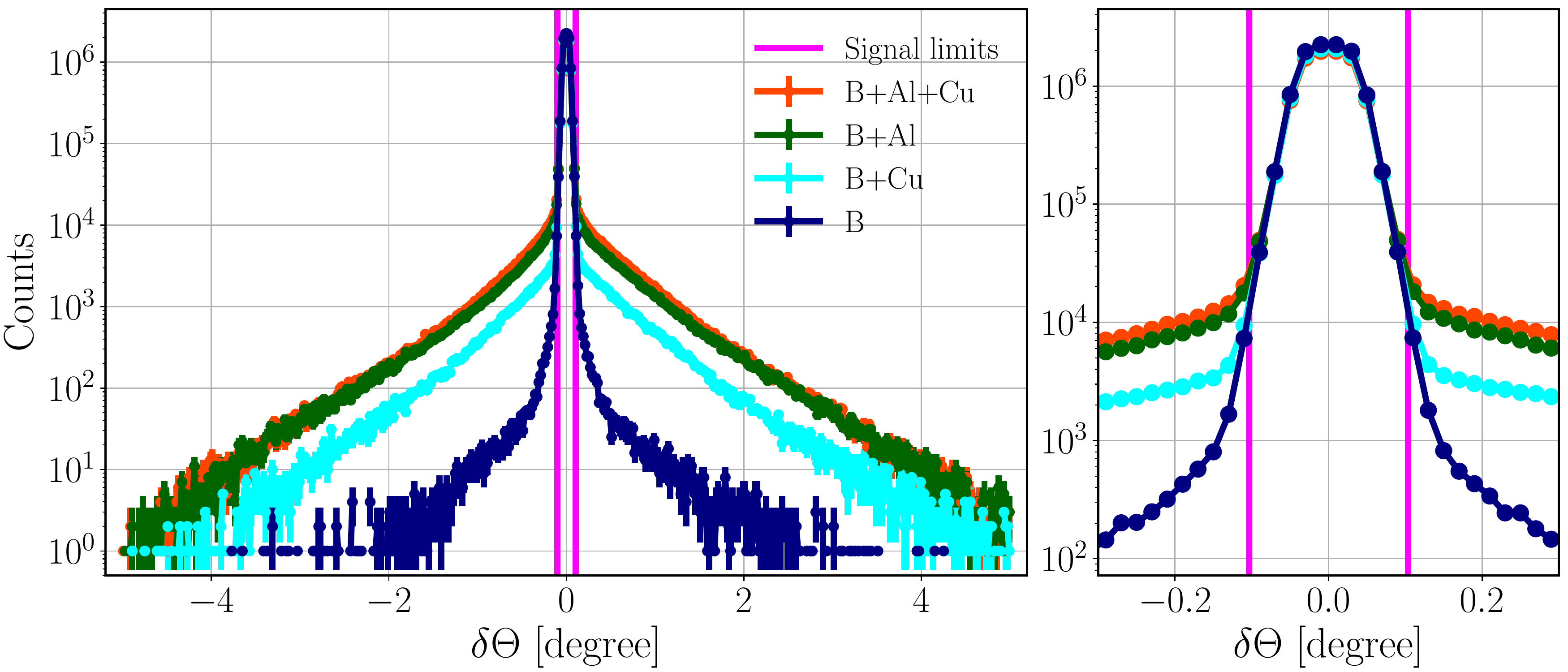}
  \end{subfigure}
  \caption{\footnotesize $\delta\Theta$ for different detector models from simulations with $\lambda$=1.8~{\AA}. The model labeled with `B' contains only Ar/CO$_2$ gas and the B$_4$C converter layer. The models `B+Cu' and `B+Al' contain either the copper straws or the aluminium tubes. `B+Al+Cu' represents the original model with all materials in place.
    The signal limits are defined from simulation with the original model. The figure on the right shows an enlarged view of the center part of the figure on the left. The lines are only joining the points.
}
  \label{fig:compare_dth_material}
\end{figure}

\FloatBarrier
\subsection{Impact of neutron wavelength}

Fig.~\ref{fig:compare_dx_wavelength} depicts $\delta$X for various monoenergetic neutron beams. 
For all wavelengths there is a peak inside the signal range and tails
on both sides outside the limits, so it is visually easy to separate the signal and the scattered background.  
The first and most notable thing is that for higher wavelengths the background gets lower. 
This is not a trivial result, because in some regions the total scattering cross-section of aluminium and copper increases, and with the higher conversion and detection efficiency due to the also increased absorption cross-section of the B$_4$C, more and more of the scattered neutrons are detected. On the other hand, the increased absorption in the converter layers reduces the average path length of the neutrons in aluminium and copper, and therefore the amount of scattering. The results show the latter effect is stronger.
The significant drop in the background between 3~{\AA} and 5~{\AA} is
the result of the Bragg cut-off that is at 4.174~{\AA} for copper and
4.676~{\AA} for aluminium (see Fig.~\ref{ncrystalCrossSections} in
appendix). 
Above these wavelengths, there is no Bragg scattering in
the materials, only incoherent, and coherent inelastic
scattering. 
%{\color{red} I am not sure this is entirely correct, I think Bragg scattering continues to happen but its intensity is dramatically smaller. Ask Thomas or look it up.} 
% Thomas: No, for wavelengths longer than 2 times the maximal d-spacing (with non-zero structure factor), aka "the bragg cutoff", there is indeed no Bragg diffraction at all.
Figures~\ref{fig:compare_dy_wavelength}--\ref{fig:compare_dphi_wavelength}
demonstrate the same effects for $\delta$Y, $\delta\Theta$ and
$\delta\Phi$. Even though $\delta\Phi$ shows non-gaussian shape, the signal defined by the limits from fitting Gaussian function is in good accordance with signal from other quantities.
% {\color{red} isn't it dy, dtheta, dphi? }

 \begin{figure}[!h]  
  \centering
  \begin{subfigure}{\textwidth} %{8cm}
      \includegraphics[width=\textwidth]{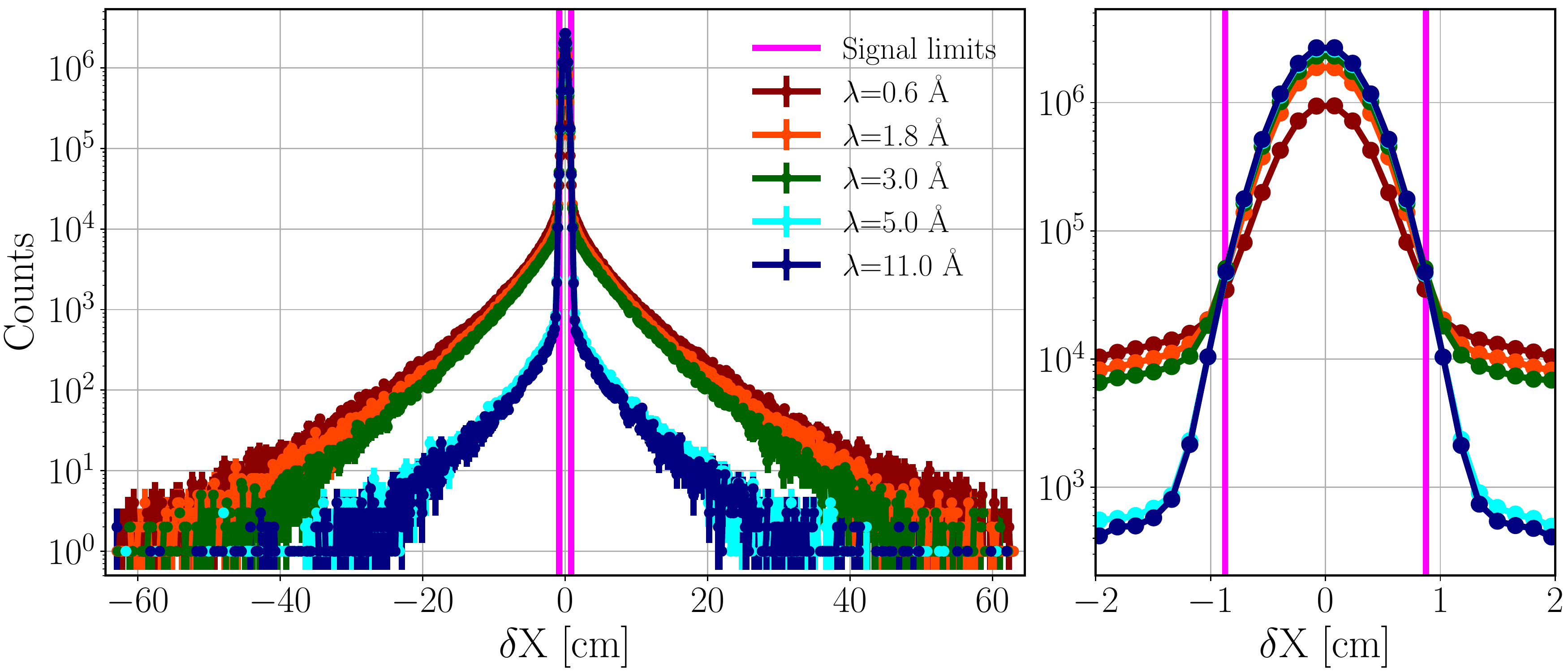}
     \end{subfigure}
      \caption{\footnotesize $\delta$X for different neutron wavelengths. The signal limits are defined based on results from simulation with $\lambda$=1.8~{\AA}. The figure on the right shows an enlarged view of the center part of the figure on the left. The lines are only joining the points.
}
  \label{fig:compare_dx_wavelength}
  \end{figure}
  
%  By the nature of the detector, $\delta$Y has the most
%  distinguishable signal range, that stands for all wavelengths, as
%  opposed to $\delta\Phi$ where no sharp transition is visible.
%   {\color{red} Kelly:Don't know what to do with this sentence.}

\begin{figure}[!h]  
  \centering
  \begin{subfigure}{\textwidth} %{8cm}
    \includegraphics[width=\textwidth]{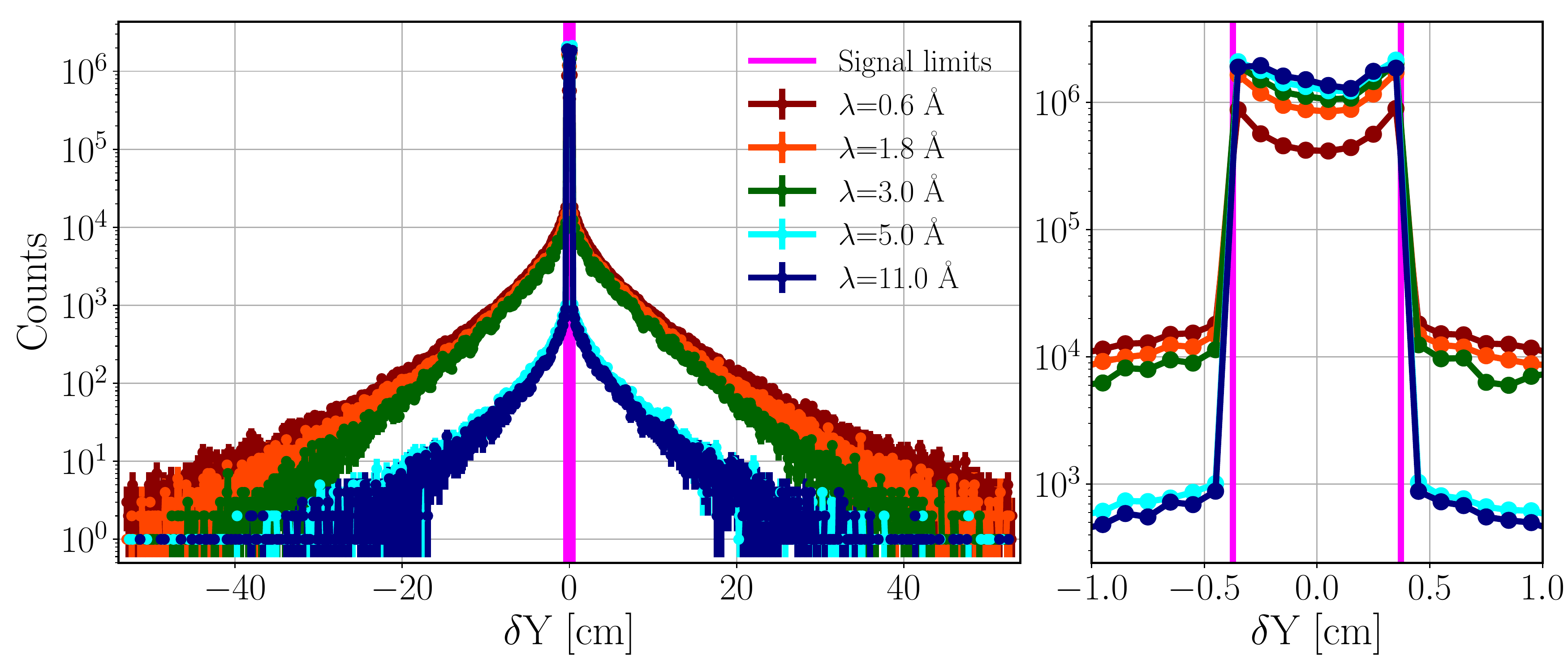}
  \end{subfigure}
  \caption{\footnotesize $\delta$Y for different neutron wavelengths. The signal limits are defined as the inner radius of the copper straws. The figure on the right shows an enlarged view of the center part of the figure on the left. The lines are only joining the points.
}
  \label{fig:compare_dy_wavelength}
\end{figure}

\begin{figure}[!h]  
  \centering
  \begin{subfigure}{\textwidth} %{8cm}
    \includegraphics[width=\textwidth]{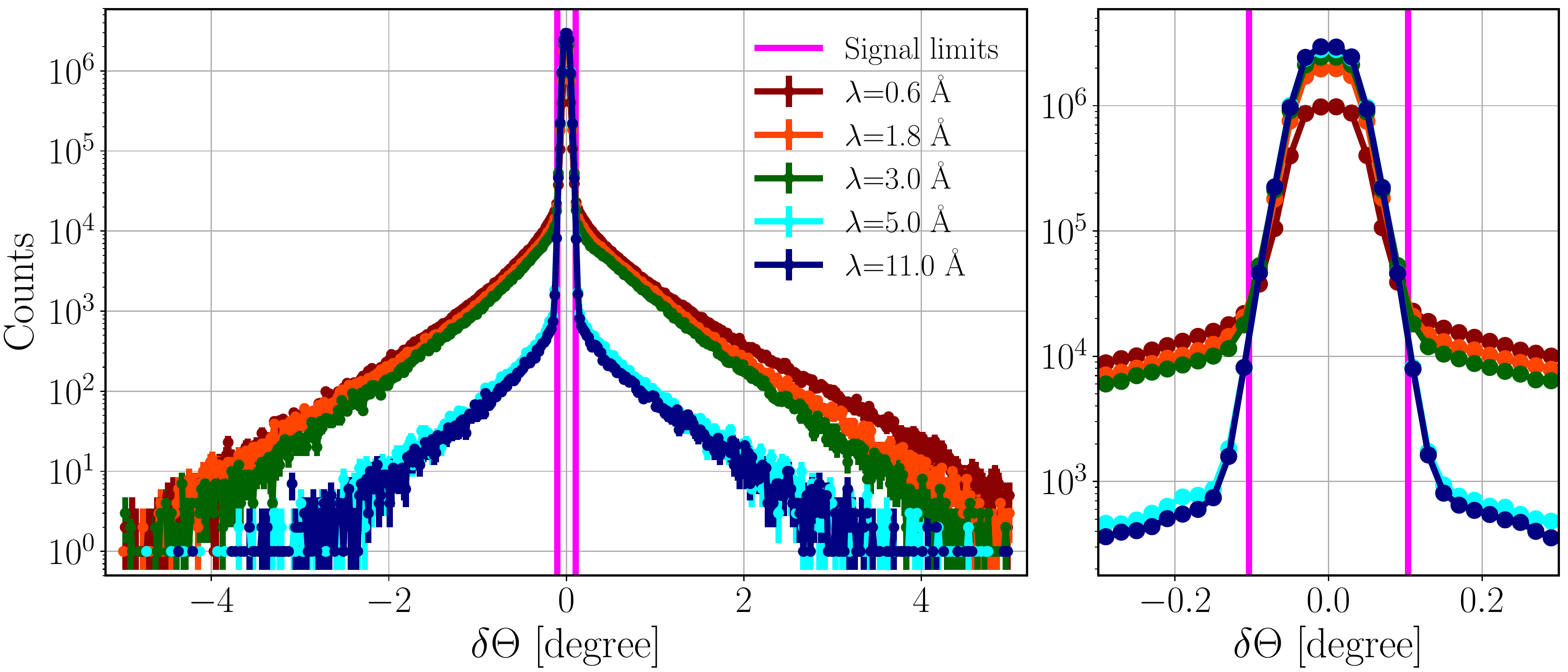}
  \end{subfigure}
  \caption{\footnotesize $\delta\Theta$ for different neutron wavelengths. The range of signal is indicated with pink vertical lines. The signal limits are defined form simulation with $\lambda$=1.8~{\AA}. The figure on the right shows an enlarged view of the center part of the figure on the left. The lines are only joining the points.
}
  \label{fig:compare_dth_wavelength}
\end{figure}

\begin{figure}[!h]  
  \centering
  \begin{subfigure}{\textwidth} %{8cm}
    \includegraphics[width=\textwidth]{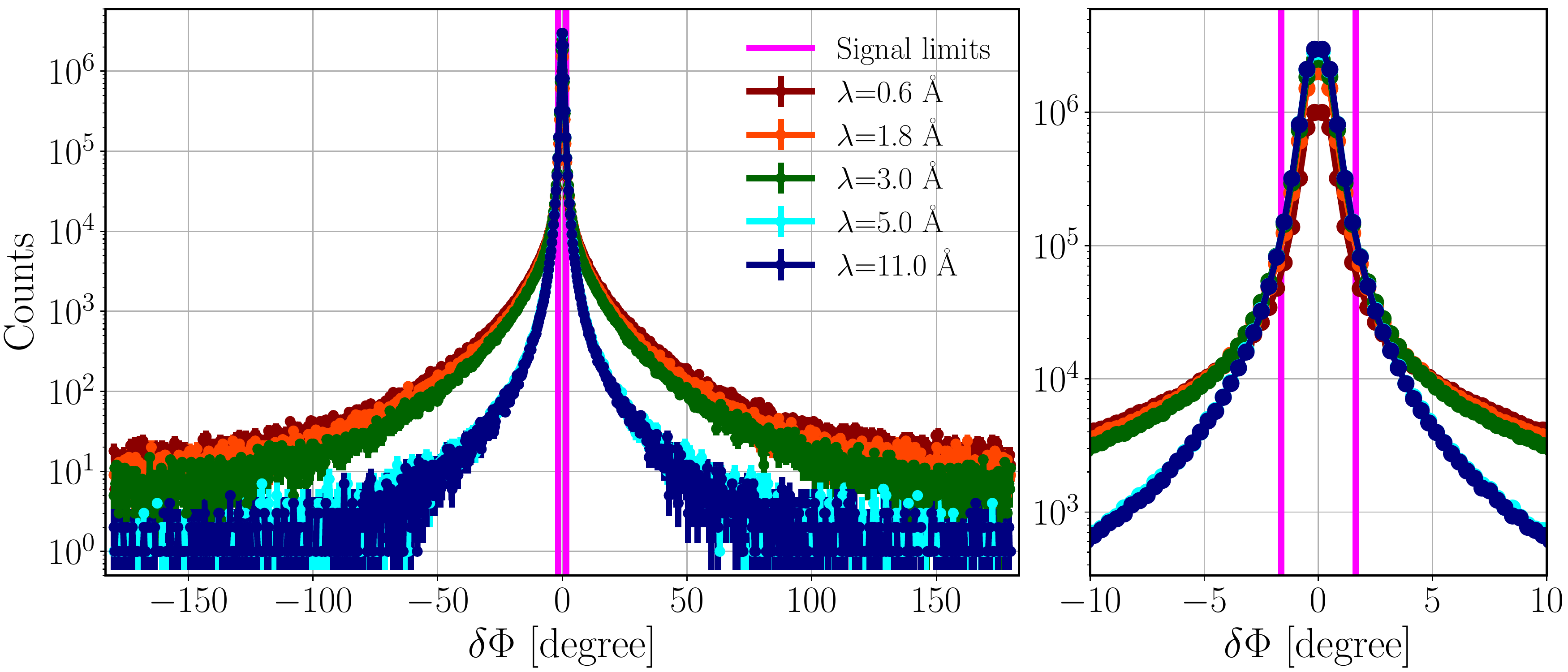}
  \end{subfigure}
  \caption{\footnotesize $\delta\Phi$ for different neutron
    wavelengths. The signal limits are defined based on results from
    simulation with $\lambda$=1.8~{\AA}. The figure on the right shows an enlarged view of the center part of the figure on the left. The lines are only joining the points.
    %{\color{red} not sure this is needed. We have stuffed this section quite a bit.} {\color{green} Honestly, I would drop $\Phi$ from the whole paper.}
    }
  \label{fig:compare_dphi_wavelength}
\end{figure}

%\FloatBarrier
%\subsection{Derived quantities?? Reciprocal space??}

The simulated $\lambda$ is not necessarily equal to the physical wavelength of the detected neutron. It is a measured $\lambda$ that is calculated from TOF with Eq.~\ref{eq:lambda}, hence $\delta\lambda$ and $\delta$TOF are closely connected. 
Scattering inside the detector can change the TOF in two ways: by
changing the direction and the wavelength of the neutron, in case of
inelastic scattering. 
Both effects have a complex impact on TOF,
leading to either increase or decrease thereof. 
A change in the direction can greatly increase the path length of a neutron inside the detector but
it can also decrease it in case the next converter layer is closer in the
new direction.
Thermalisation through scattering generally leads the neutrons toward thermal neutron wavelength but for the wavelength range of interest between 0.6--11~{\AA} this can mean both increase and decrease. 
With increased wavelength the neutrons travel slower inside the
detector, which could result in higher TOF.
On the other hand, higher wavelength also implies a higher
absorption cross-section and therefore possibly a shorter path length,
and the other way around. 
%{\color{red} do we need the rest of the
%  sentence??, and the other way around increased velocity brings
%  higher expected path length.} 
% {\color{red} The thermal neutron
%  wavelength is within the observed range, so thermalisation through
%  scattering can lead the neutrons toward both higher and lower
%  wavelengths, depending on the initial wavelength. Rephrase, I am confused}
The cumulative effect is shown in
Figs.~\ref{fig:compare_dtof_wavelength} and~\ref{fig:compare_dlambda_wavelength}. For shorter wavelengths clearly the positive $\delta$ values are dominant. This means that for these neutrons TOF generally increases due to scattering and so does the resulting $\lambda$. Going for longer initial wavelengths, the negative side becomes more and more significant. This proves that shorter measured TOF and $\lambda$ as a result of scattering can be just as important as longer. 

The change in the measured $\lambda$ due to scattering inside the detector can be expressed in terms of change in the measured neutron energy, that can be important when observing energy transfer in real samples.
Fig.~\ref{fig:compare_dekin_wavelength} depicts the change in the
measured energy for different wavelengths. The scale of $\delta$E depends on the initial $\lambda$, so different ranges are highlighted in the subplots to provide information about the relevant wavelengths.

\begin{figure}[!h]  
  \centering
  \begin{subfigure}{\textwidth} %{8cm}
      \includegraphics[width=\textwidth]{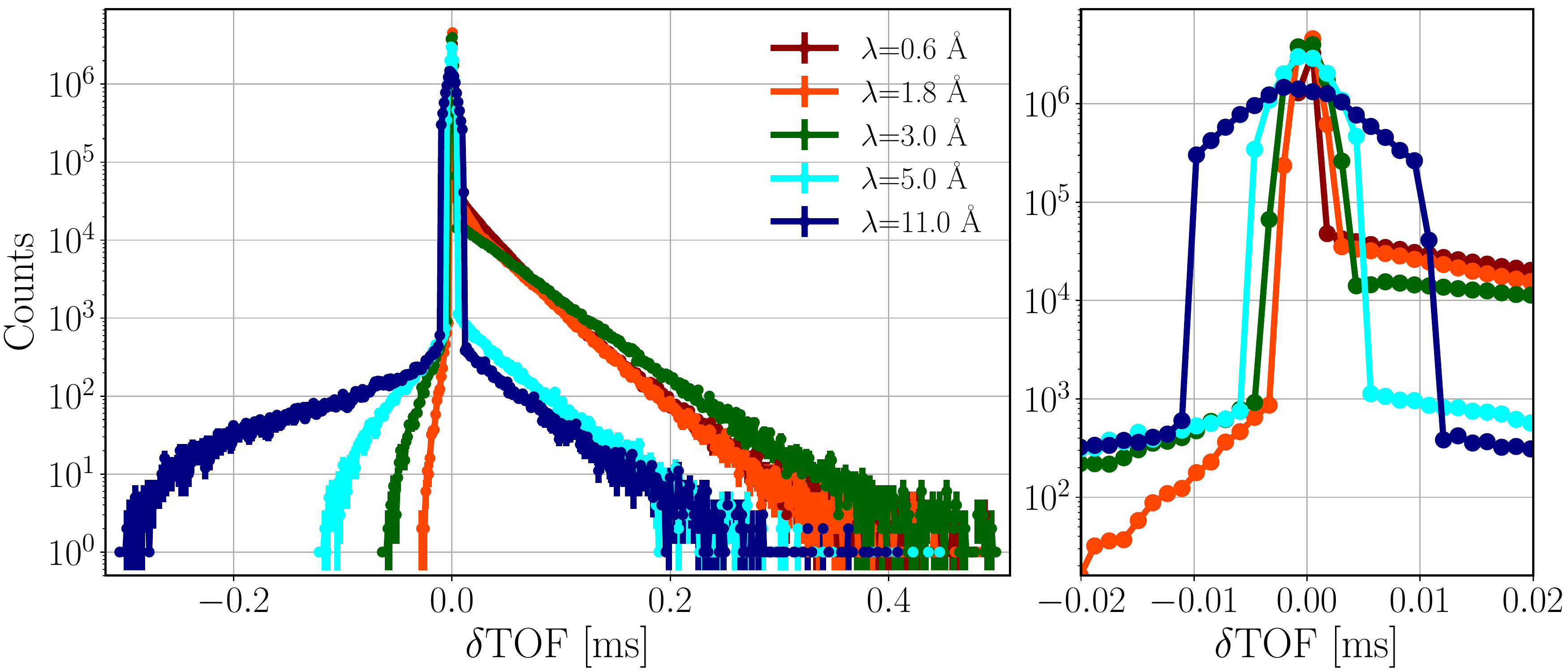}
     \end{subfigure}
       \caption{\footnotesize $\delta$TOF for different neutron wavelengths. The limits are wavelength dependent and not shown. The figure on the right shows an enlarged view of the center part of the figure on the left. The lines are only joining the points.}
  \label{fig:compare_dtof_wavelength}
\end{figure}

\begin{figure}[!h]  
  \centering
  \begin{subfigure}{\textwidth} %{8cm}
      \includegraphics[width=\textwidth]{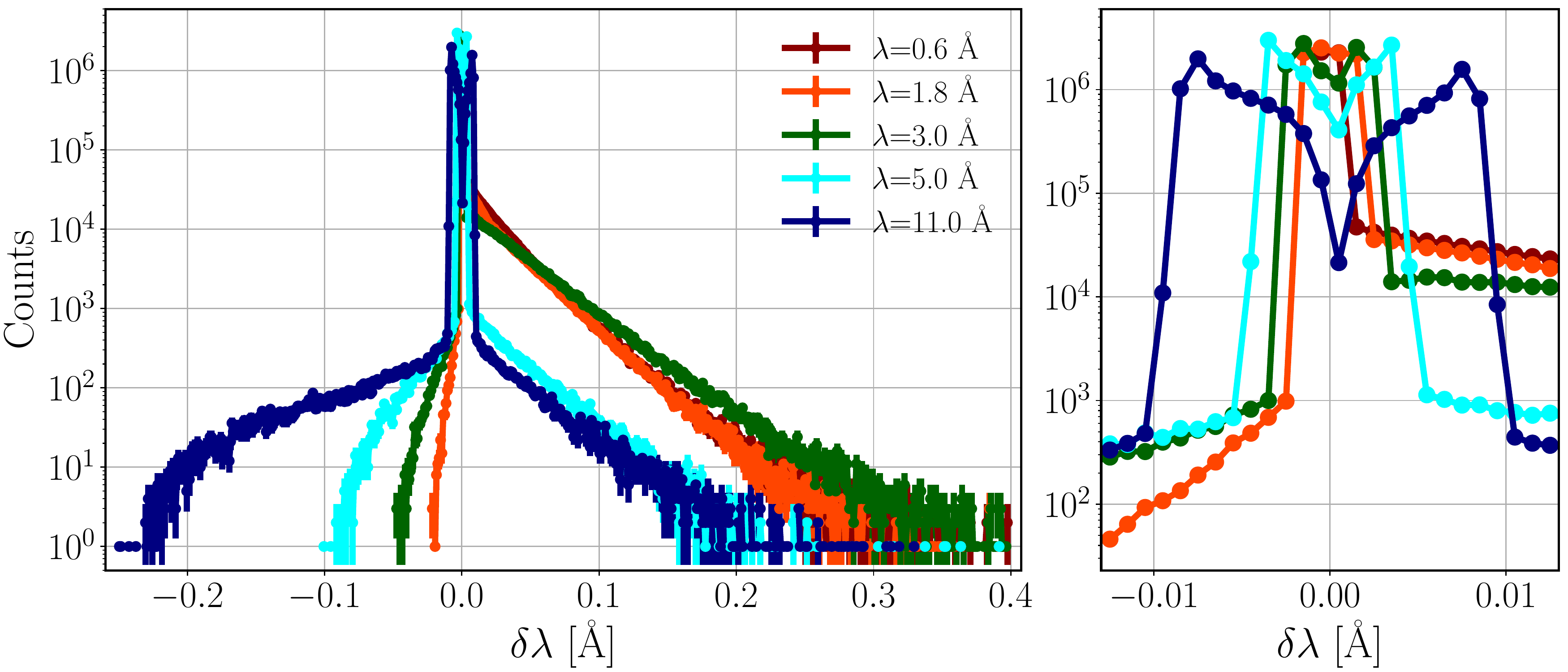}
     \end{subfigure}
         \caption{\footnotesize $\delta\lambda$ for different neutron wavelengths. The limits are strongly wavelength dependent and not shown. The figure on the right shows an enlarged view of the center part of the figure on the left. The lines are only joining the points.
}
  \label{fig:compare_dlambda_wavelength}
\end{figure}

\begin{figure}[!h]  
  \centering
  \begin{subfigure}{\textwidth} %{8cm}
      \includegraphics[width=\textwidth]{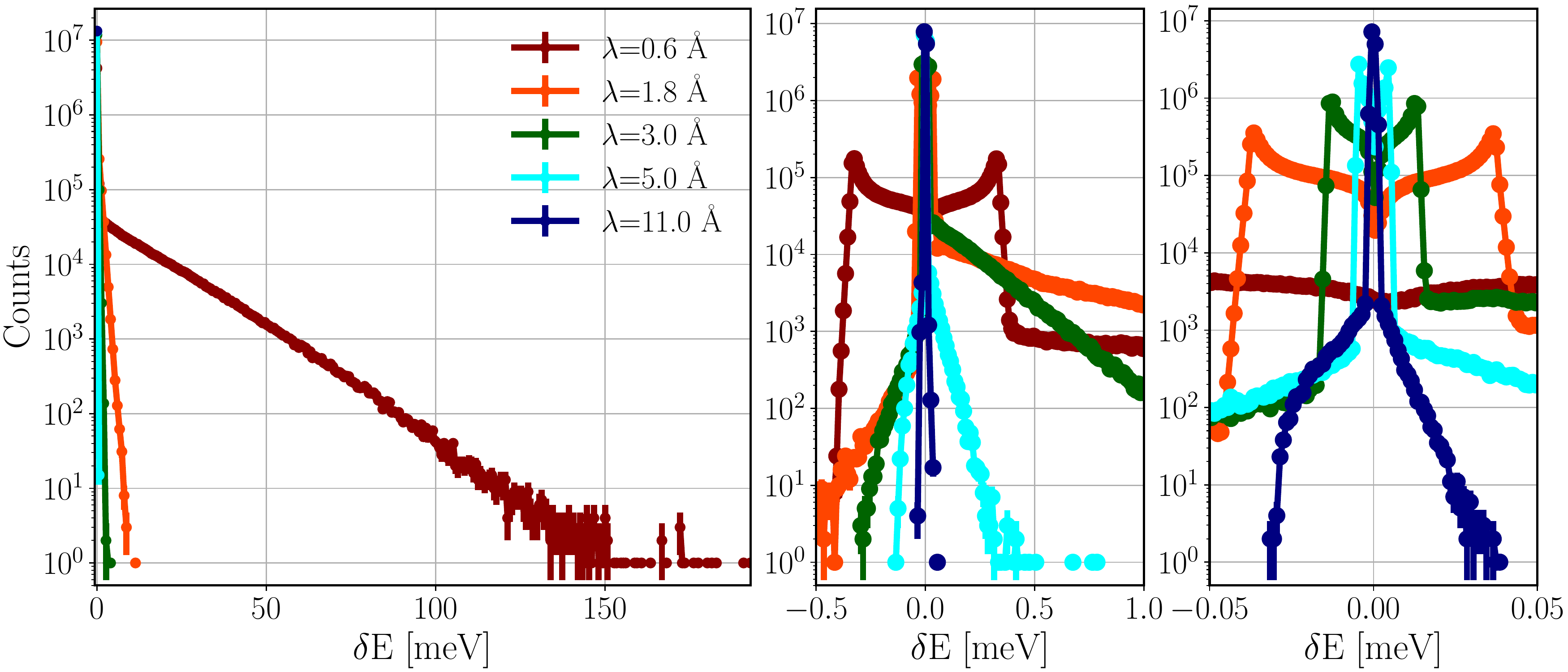}
     \end{subfigure}
         \caption{\footnotesize Change in measured neutron energy for different initial wavelengths. The three figures emphasise different ranges, with different binning. The lines are only joining the points.
}
  \label{fig:compare_dekin_wavelength}
\end{figure}

Q is derived from $\Theta$ and $\lambda$ (see Eq.~\ref{eq:Q}) so $\delta$Q depends on all phenomena mentioned earlier in connection with these quantities. The results are depicted in Fig.~\ref{fig:compare_dQ_wavelength}. The wavelength dependency of both the signal and the background is visible. Simulations with low $\lambda$ values give higher scattered background and broader signal peaks.

\begin{figure}[!h]  
  \centering
  \begin{subfigure}{\textwidth} %{8cm}
      \includegraphics[width=\textwidth]{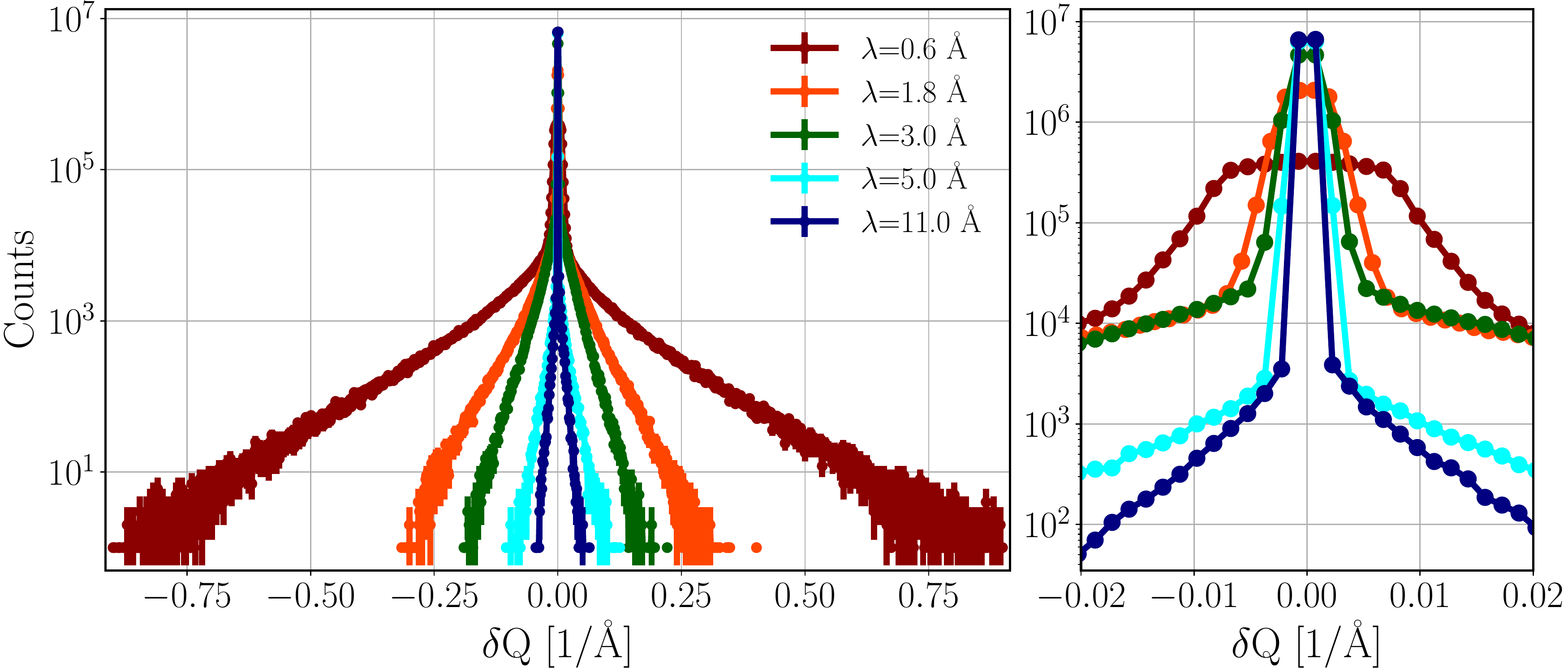}
     \end{subfigure}
      
   \caption{\footnotesize $\delta$Q for different neutron wavelengths. The limits are wavelength dependent and not shown. The figure on the right shows an enlarged view of the center part of the figure on the left. The lines are only joining the points.
}
  \label{fig:compare_dQ_wavelength}
\end{figure}

%{\color{red}TODO: justify where there is no need for new limits for every simulation.\\
%The mean is negligible compared to 3 sigma so 0 is used everywhere.\\}

\FloatBarrier
\subsection{Definition of fractional scattering \label{subsec:fracScat}}

Signal and background relation can be characterised by several
quantities. A common way to express it in neutron scattering is the peak-to-tail ratio, that can be
visually extracted from the presented figures. However, this ratio
does not quantitatively reflect the total amount of signal or
background and in addition it is sensitive to the histogram binning. Instead, a different figure of merit is used that provides
fractional scattering in terms of integrals, as defined by Eq.~\ref{eq:fractionalScattering}:
 \begin{equation}
 \textrm{Fractional Scattering} = \frac{B}{S+B}
 \label{eq:fractionalScattering}
 \end{equation}
 where $S$ and $B$ denote the number of detected neutrons considered
 as signal and background respectively. Although the shape of signal
 and background varies for the observed $\delta$ quantities,
 Fig.~\ref{fig:signalEqual} demonstrates that the integrals of the
 previously defined signal ranges are similar within less than a 3\% range. %The difference between any two of the signal values from the same simulation is less than 3\% for all the presented cases.
This means that any of the $\delta$ quantities and the signal limits
thereof lead to essentially the same fractional scattering
values. The following results are acquired using $\delta\Theta$ for
signal-background separation. 

%{\color{red} I am assuming there are no
% overflows or underflows in the estimate of the total detection
%  events. Otherwise you cannot claim that you end up with the same FS ratio.}
%{\color{green} The total number of detection event does't depend on any quantity and there are no under/overflows because it is from the integral of a histogram ('z hit'  as I remember)) }

\begin{figure}[!h]  
  \centering
  \begin{subfigure}{0.5\textwidth} %{8cm}
      \includegraphics[width=\textwidth]{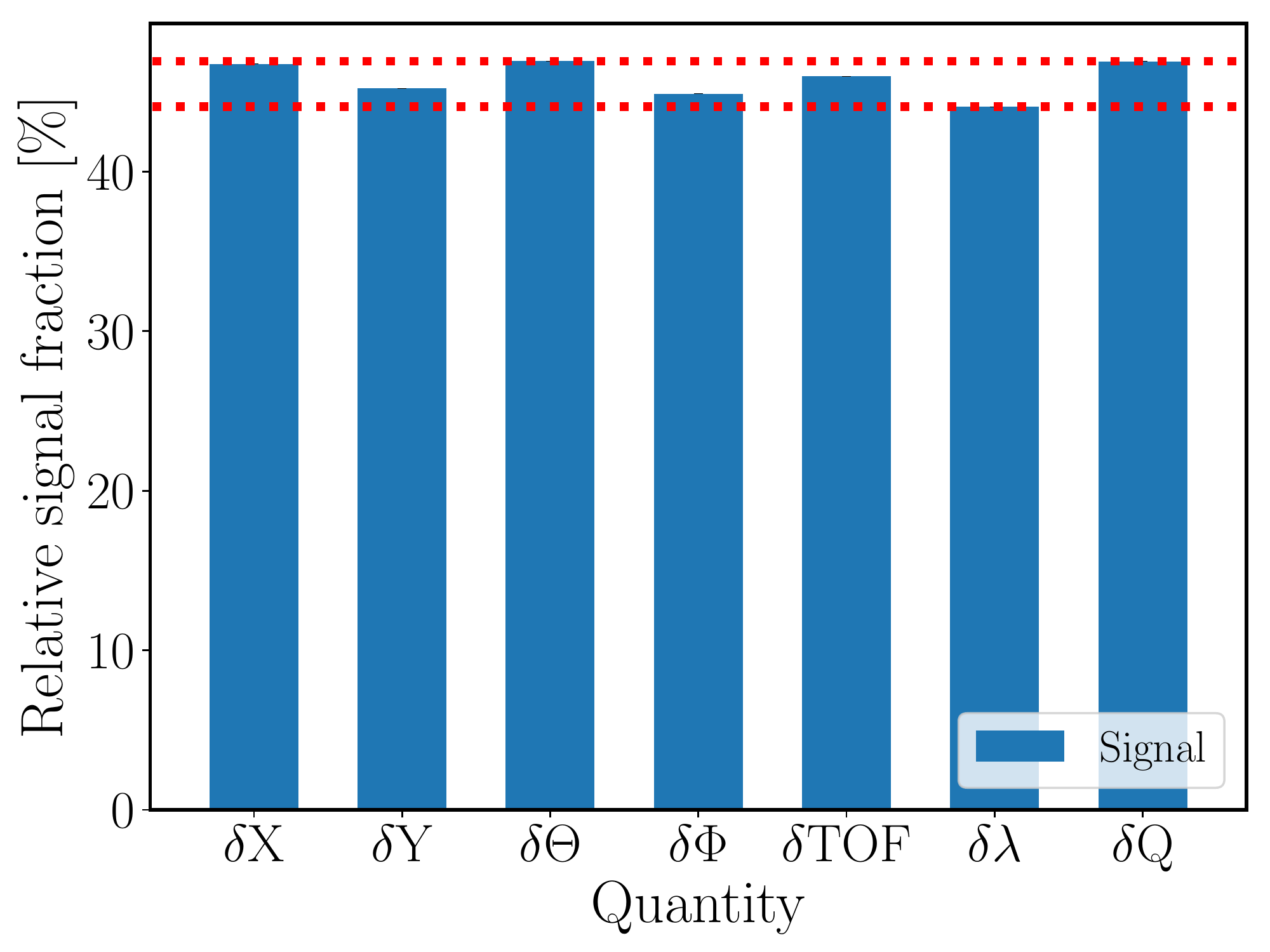}
     \end{subfigure}
   \caption{\footnotesize The portion of neutrons considered as signal of the total number of neutrons,
     based on the limits for different $\delta$ quantities from
     simulation with 1.8~{\AA} neutrons. The dotted lines in red indicate the minimum and maximum value.
    % {\color{red} Ratio is 2 numbers, you only write one. Is this valid for all wavelengths? Say it. Also, half your plot is empty
    %   because of the ymax you chose. Are you plotting FS? Then why call it relative ratio?}
       }
  \label{fig:signalEqual}
\end{figure}

Fig.~\ref{fig:fracScatPanel} depicts the fractional scattering for 3
wavelengths for a different number of panels, as the latter are added
one-by-one to the geometry. It was shown before that additional panels 
not only increase the signal, but the scattered background as well as, via the
detection of more scattered neutrons and the back-scattering of
neutrons to upstream panels. This result shows that the ratio of
signal to background degrades with the additional panels, because the fractional scattering increases monotonously. 
This is more notable for low wavelengths, where the differences are higher, but the tendency is the same for 11.0 {\AA}.
For the same number of panels, fractional scattering is always higher for neutrons with shorter initial $\lambda$.
\begin{figure}[!h]  
  \centering
  \begin{subfigure}{0.5\textwidth} %{8cm}
      \includegraphics[width=\textwidth]{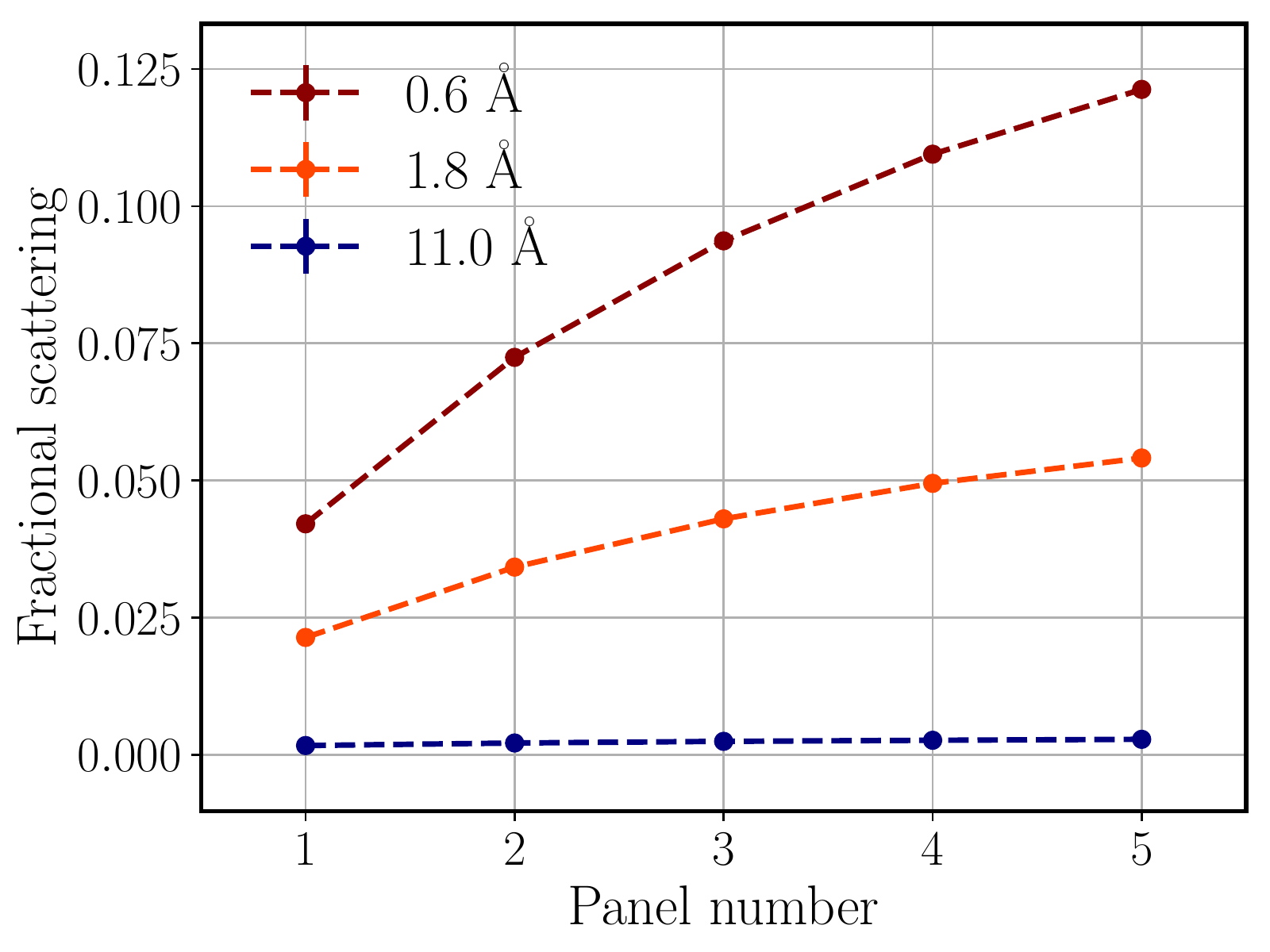}
     \end{subfigure}
   \caption{\footnotesize The change of the fractional scattering for
     different number of detector panels with different neutron
     wavelengths.}
  \label{fig:fracScatPanel}
\end{figure}

Fig.~\ref{fig:fracScatMaterial} demonstrates the fractional scattering
for different wavelengths with 5 panels of detectors, using the 4
models with reduced materials, mentioned earlier in this
section. Additional materials increase the fractional scattering for
all $\lambda$ and in agreement with previous results, aluminium has a
higher impact on scattering than copper, which manifests itself in higher fractional scattering values too.
It can be seen that scattering is a very important background effect to be minimized below the Bragg cut-off.
\begin{figure}[!h]  
  \centering
  \begin{subfigure}{0.5\textwidth} %{8cm}
      \includegraphics[width=\textwidth]{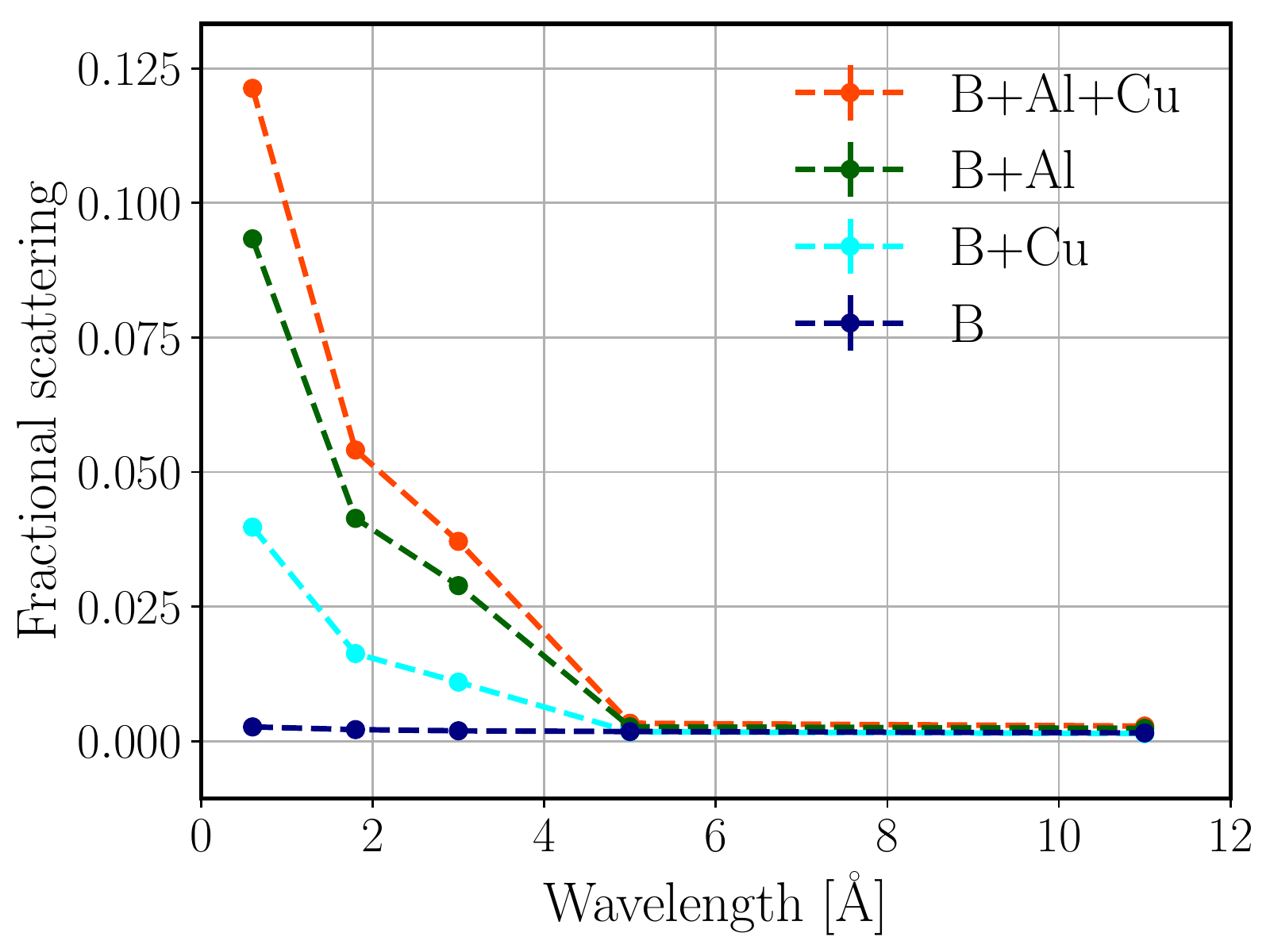}
     \end{subfigure}
   \caption{\footnotesize The effect of the detector material
     components on the fractional scattering based on $\delta\theta$
     for different neutron wavelengths between 0.6--11~{\AA}. `B' (dark blue) represents the geometry where only gas and converter are in place. `B+Cu' (cyan) has enabled copper in the straw
     volumes. `B+Al' (green) has enabled aluminium in the tube
     volumes. Finally `B+Al+Cu' (orange) has all materials in place.}
  \label{fig:fracScatMaterial}
\end{figure}
%\begin{comment}
%\begin{table}[htp]
%\begin{center}
%\begin{tabular}{| c c c c c |}
%\hline
% \multirow{2}{*}{$\lambda$ [{\AA}]}  & \multirow{2}{*}{Absorption} & \multirow{2}{*}{Transmission} & \multicolumn{2}{c|}{Detection}    \\
%& & & Background & Signal  \\\hline\hline
%0.6	&	12.97	&	59.76	&	3.29	&	24.00 \\
%1.8	&	23.62	&	26.82	&	2.66	&	49.56 \\
%3	&	28.58	&	11.83	&	2.20	&	59.64 \\
%5	&	31.38	&	3.57		&	0.20	&	64.84 \\
%11	&	33.29	&	0.33		&	0.17	&	66.12 \\
%\hline
%\end{tabular}
%\end{center}
%\caption{{\color{red}TODO proper name: The fate of neutrons with all material in place. Relative to all neutrons entering the detector. (8e6 or 8e5 neutrons)}}
%\label{tab:todo}
%\end{table}
%\end{comment}
Fig.~\ref{fig:barChartSignal} presents similar information to
Fig.~\ref{fig:absTransDet} with two changes. First, the absorption in
different materials is merged in a single color (purple). Second, the detection is
separated into signal and background using $\delta\Theta$ limits once
again.
\begin{figure}[!h]  
  \centering
  \begin{subfigure}{0.5\textwidth} %{8cm}
      \includegraphics[width=\textwidth]{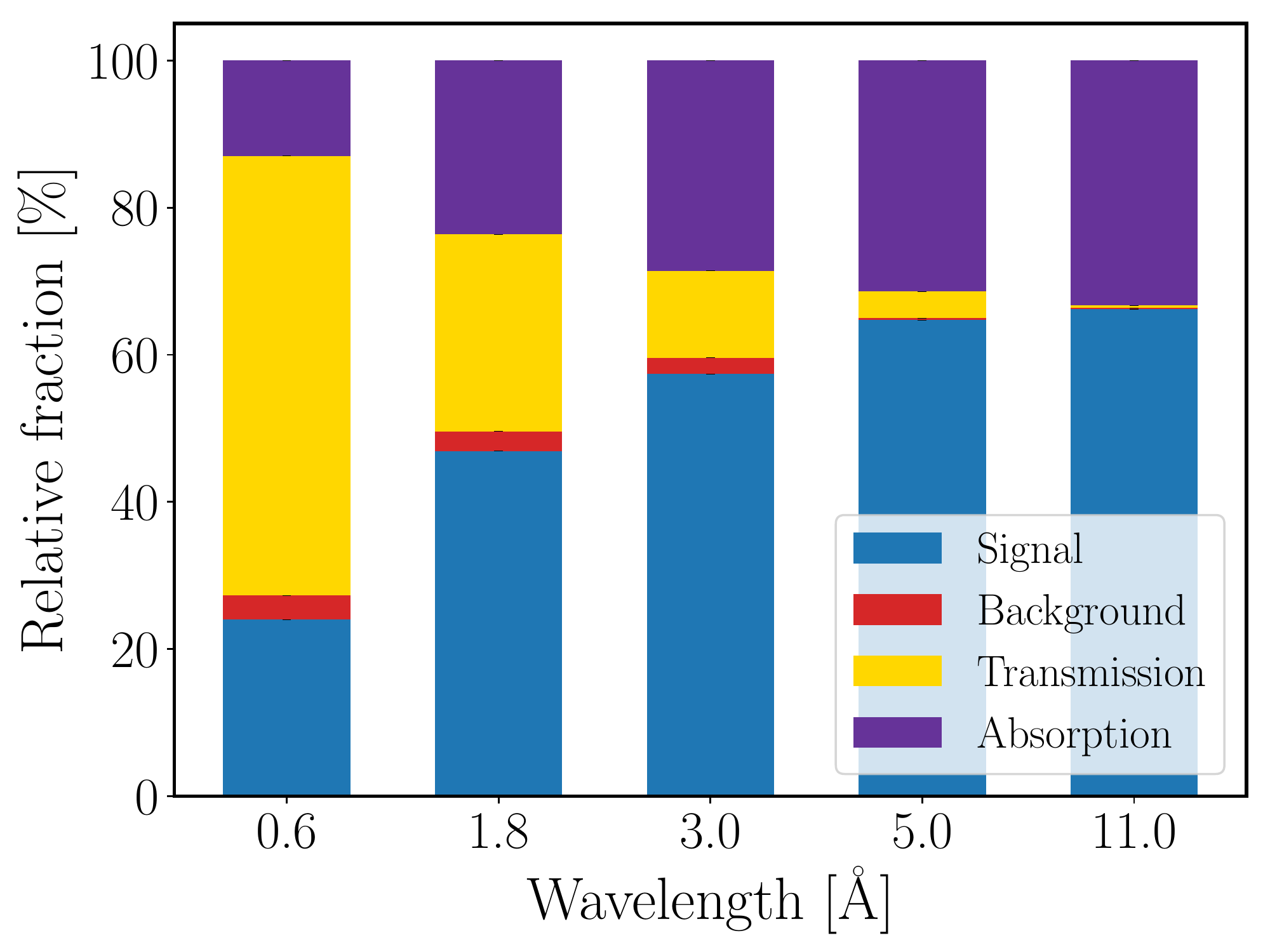}
     \end{subfigure}
   \caption{\footnotesize Proportion of absorption, transmission and detection with the separation of signal and background from simulations with monoenergetic neutrons.}
  \label{fig:barChartSignal}
\end{figure}

%{\color{red}
%COHERENT (run with G4 and encristaled copper, to see if coherent)\\
%TODO: get only incoherent scattering in det with noBragg option (for low wavelength)\\
%}

% {\color{green} Do we want to do that?} {\color{blue} Well, perhaps not for
%   the paper but it would be nice to know what Efinal looks
%   like... and count the percentage of inelastically scattered
%   neutrons. I think you already have the histos. Just see what the
%   bin content is at the maximum wrt to the integral.}
% {\color{red}
%energy of the neutron at the conversion point! to see if the detector scatters inelastically because it will be a problem for the chopper spectrometert as this detector is a candidate for
%(SANS Choppoerspectrometer- how mutch inelasticscatter the material will the detector itself! if too much)
%}

%\begin{comment}
%\FloatBarrier
%\section{Express absorption and scattering in terms of ``radiation'' length}
%{\color{red}
%Adding this more as a reminder. Kelly might work on this in parallel
%at least to check how much work it is. To be discussed with Milan in person.
%}
%\end{comment}

\FloatBarrier
\section{Polyethylene {\it``afterburner''}  block behind the detector }

As previously demonstrated, for short wavelengths such as 0.6--1.8~{\AA}
respectively 60--27\% of the neutrons can escape even 5 panels of
detectors without being absorbed in any of the materials. The
detector performance could be possibly improved by applying additional
panels of detectors but that might not be the most cost-efficient
solution, taking into account that there is little to gain for longer
wavelengths. However, there is a cheap and easy option to increase
detection efficiency by placing a strongly scattering material at the
backside of the detectors~\cite{TK_Backscattering}. The principle is
to back-scatter any transmitted neutron forcing it to enter the
detectors again, giving it a further opportunity for conversion and
detection. In the study cited above it was shown that the detection
efficiencies can be enhanced by applying a single layer of
polyethylene (PE) behind the detectors. 
It is emphasised, however, that as a side effect the scattering in the
back-scatterer layer has a negative impact on the resolution of the
detector so the combined effect should be carefully examined. 

The definition of signal and background from the previous section
offers a practical way to decide whether this technique is
advantageous or not for the BCS detector. To investigate this, a PE layer
 is placed closely behind the fifth panel of
detectors as illustrated in Fig.~\ref{fig:geomPE}. The Geant4 physics
list used for these simulation is
ESS\_QGSP\_BIC\_HP\_TS~\cite{TK_Backscattering}.
\begin{figure}[!h]  
  \centering
  \begin{subfigure}{\textwidth} %{8cm}
    \includegraphics[width=\textwidth]{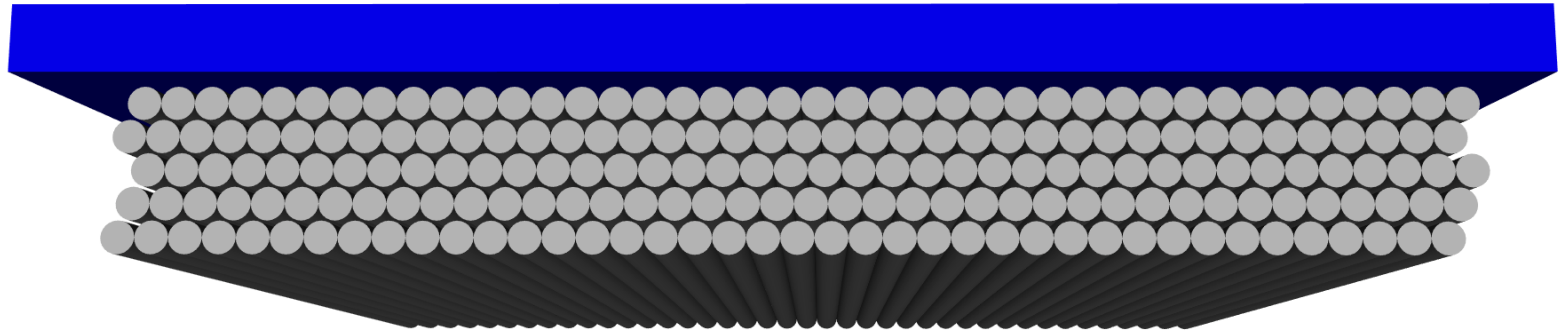}
  \end{subfigure}
    \caption{\footnotesize A layer of PE (blue) is placed behind the 5 panels of BCS detector tubes (silver). The thickness of the PE layer is 50~mm.}
  \label{fig:geomPE}
\end{figure}

The highest gain from the back-scattering of transmitted neutrons is
achieved for low wavelengths. Fig.~\ref{fig:compare_dth_PE}
demonstrates the impact of the additional PE layer on $\delta\Theta$
for the lowest observed wavelength, $\lambda$=0.6~{\AA}. This result
shows a significantly increased background with a slightly increased signal.
 \begin{figure}[!h]  
  \centering
  \begin{subfigure}{\textwidth} %{8cm}
      \includegraphics[width=\textwidth]{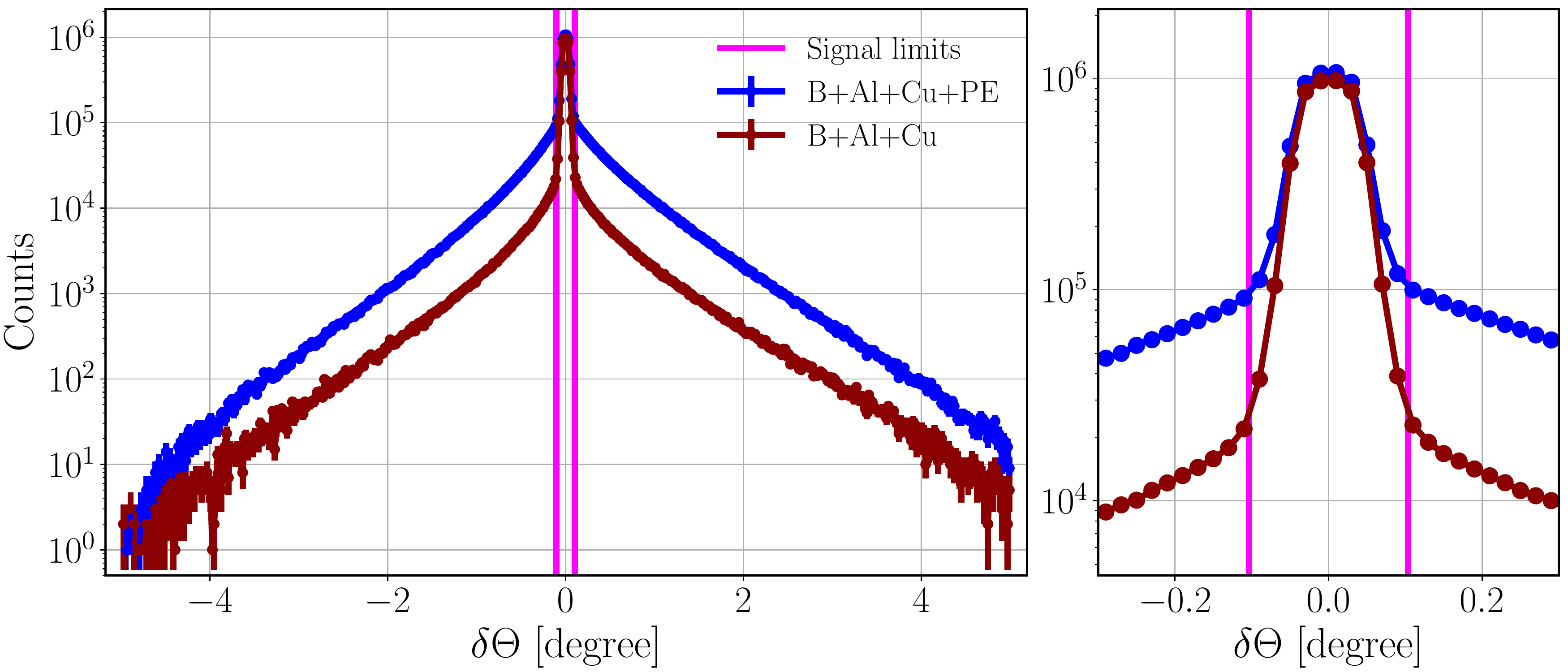}
     \end{subfigure}     
   \caption{\footnotesize $\delta\Theta$ with and without the PE layer from simulations with $\lambda$=0.6~{\AA}. The signal limits are the same as defined in the previous section. The figure on the right shows an enlarged view of the center part of the figure on the left. The lines are only joining the points.
}
  \label{fig:compare_dth_PE}
\end{figure}

Fig.~\ref{fig:fracScatPE} demonstrates that the PE layer has a 
different effect on the different $\delta$ quantities. 
%{\color{red} unless you change the ymax I cannot see it well, same comments as for the previous similar plot, unclear what ratio you are depicting}. 
The signal defined
by the limits for $\delta\Theta$ is 6.3\% higher than for
$\delta\lambda$. This difference is more than double of any previously
experienced. This implies that the fractional scattering depends much
more on the signal definition. 

The focus here is on the case with the highest gain, so the fractional
scattering results presented in Fig.~\ref{fig:fracScatPE} are
calculated with the highest signal from $\delta\Theta$. Even with this
favourable definition, the ratio of the scattered background appears
to be higher with the back-scattering layer than without it. For the
highest wavelength there is no difference but going to lower
$\lambda$ where the PE should help, the fractional scattering becomes
significantly worse.
\begin{figure}[!h]  
  \centering
  \begin{subfigure}{0.49\textwidth} %{8cm}
    \includegraphics[width=\textwidth]{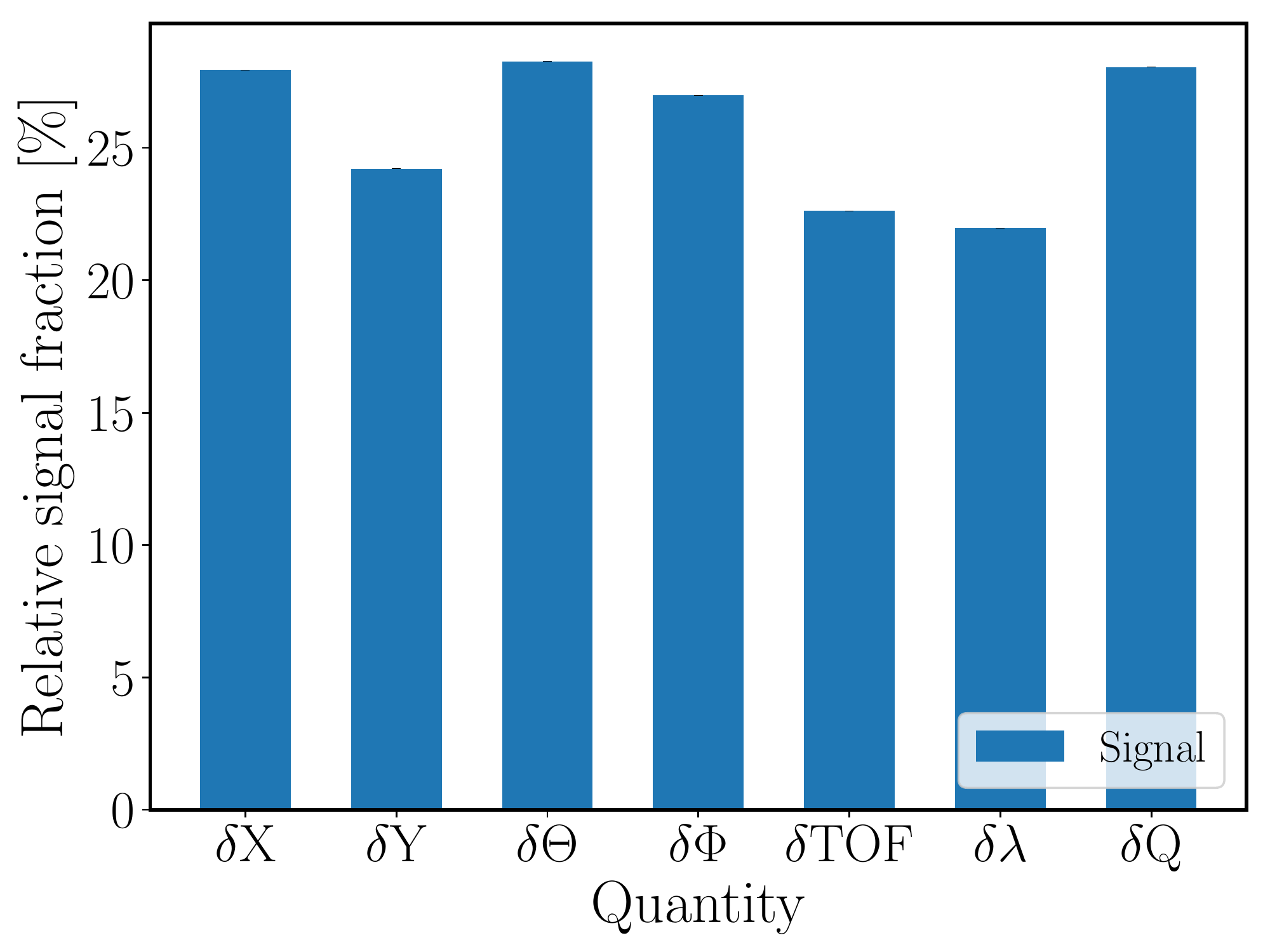}
    %\label{fig:signalEqualPE}
  \end{subfigure}
  \begin{subfigure}{0.49\textwidth} %{8cm}
    \includegraphics[width=\textwidth]{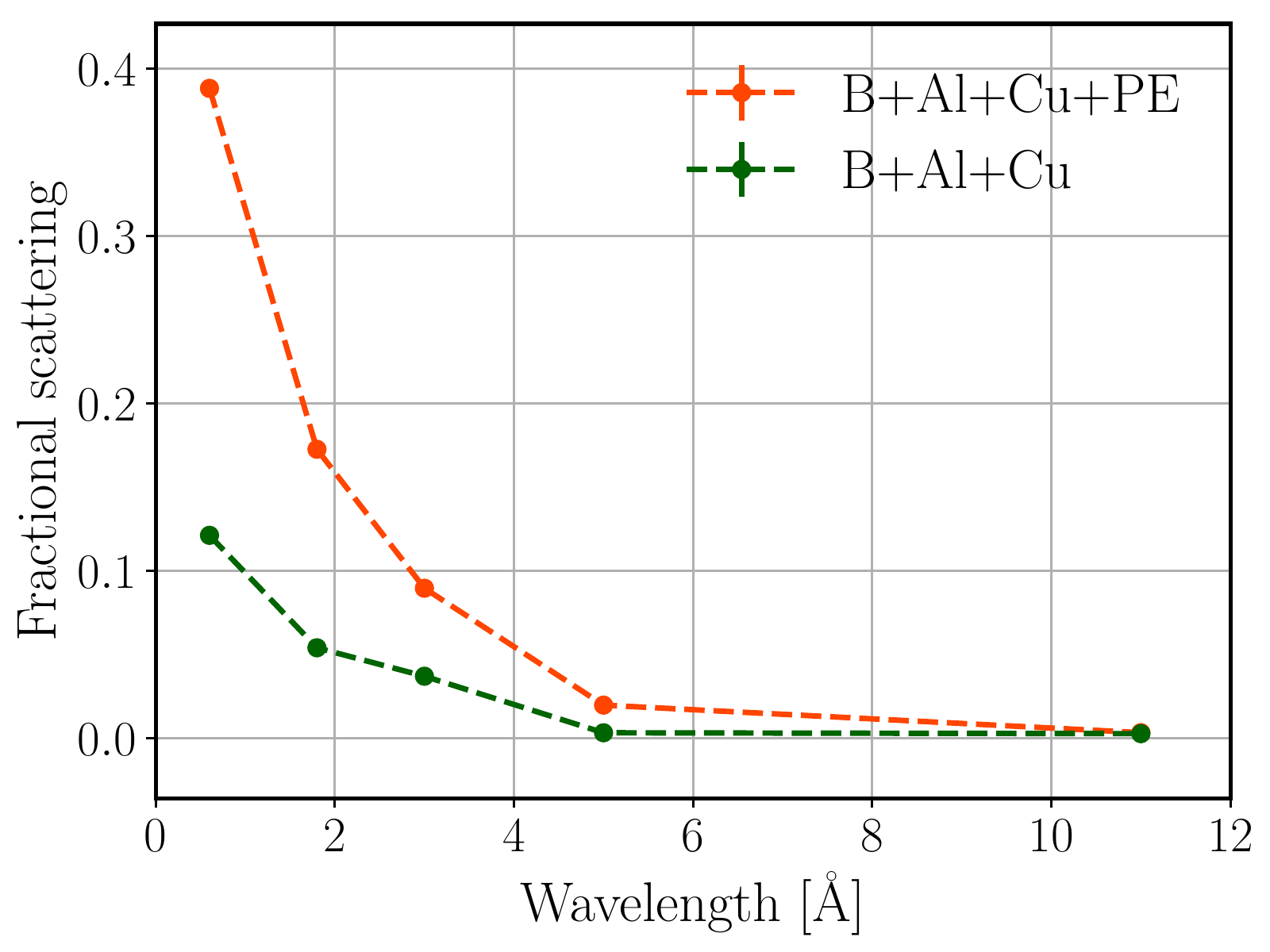}
    %\label{fig:fracScatPE}
  \end{subfigure}
  \caption{\footnotesize 
  The portion of neutrons considered as signal of the total number of neutrons
    based on the limits for different $\delta$ quantities from
    simulation with $\lambda$=0.6~{\AA} (left) and the fractional
    scattering using signal from $\delta\Theta$ for different
    wavelengths (right).}
    \label{fig:fracScatPE}
\end{figure}

%\begin{figure}[!h]  
%  \centering
%  \begin{subfigure}{0.7\textwidth} %{8cm}
%      \includegraphics[width=\textwidth]{fracScatDTheta_Panel_PE}
%     \end{subfigure}
%   \caption{\footnotesize {\color{red}TODO }}
%  \label{fig:fracScatPE}
%\end{figure}

This is in agreement with what can be derived from
Fig.~\ref{fig:barChartPE} showing the effects of PE from various
aspects. For higher wavelengths, where the transmission is negligible, there is
practically no change but for lower wavelengths the significant drop
in the transmission leads to increase in all other areas. Due to the
longer path length in copper and aluminium there is more absorption in
these materials, nevertheless it remains a minor effect. The vast
majority of the back-scattered and not transmitted neutrons are
absorbed in the converter layer with the usual detected/not detected
ratio. This means that the detection efficiency is effectively
enhanced by the PE layer. On the other hand the separation of
detection events into signal and background shows that the
additionally detected neutrons mostly increase the background, not the
signal. Where most gain is expected, for $\lambda$=0.6~{\AA}, the
4.3\% increase in the ratio of the signal is followed by a 14.6\%
increase in the background.
This implies that the only application where a PE afterburner may be beneficial is homeland security, where position resolution is not a concern. 
These results re-emphasise that it is vital to ensure that whilst polyethylene or other hydrogen containing materials are ubiquitous in neutron shielding, next to the detector, the shielding material must have no albedo effect from scattering of thermal neutrons.
\begin{figure}[!h]  
  \centering
  \begin{subfigure}{0.49\textwidth} %{8cm}
      \includegraphics[width=\textwidth]{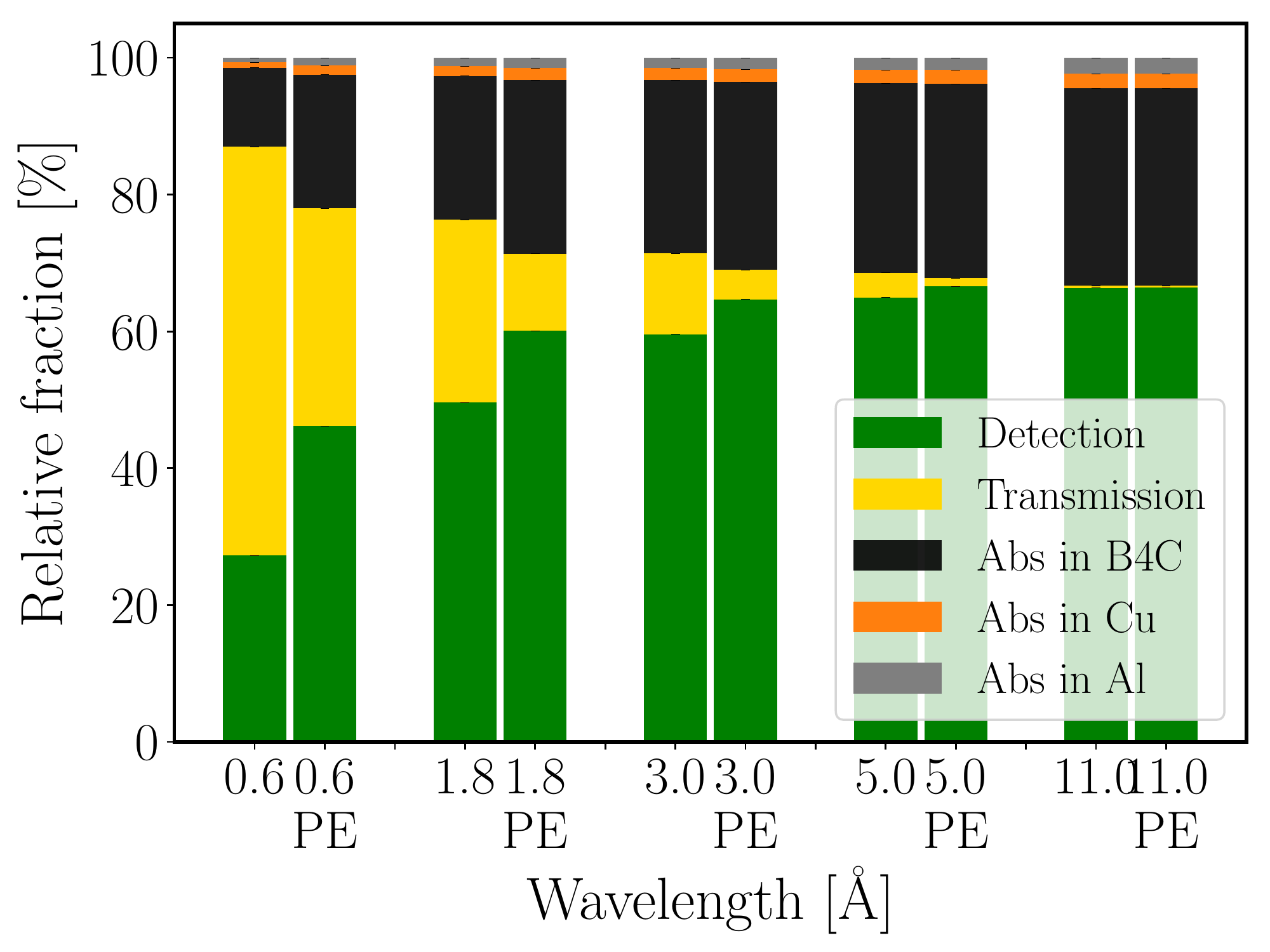}
     \end{subfigure}
  \begin{subfigure}{0.49\textwidth} %{8cm}
      \includegraphics[width=\textwidth]{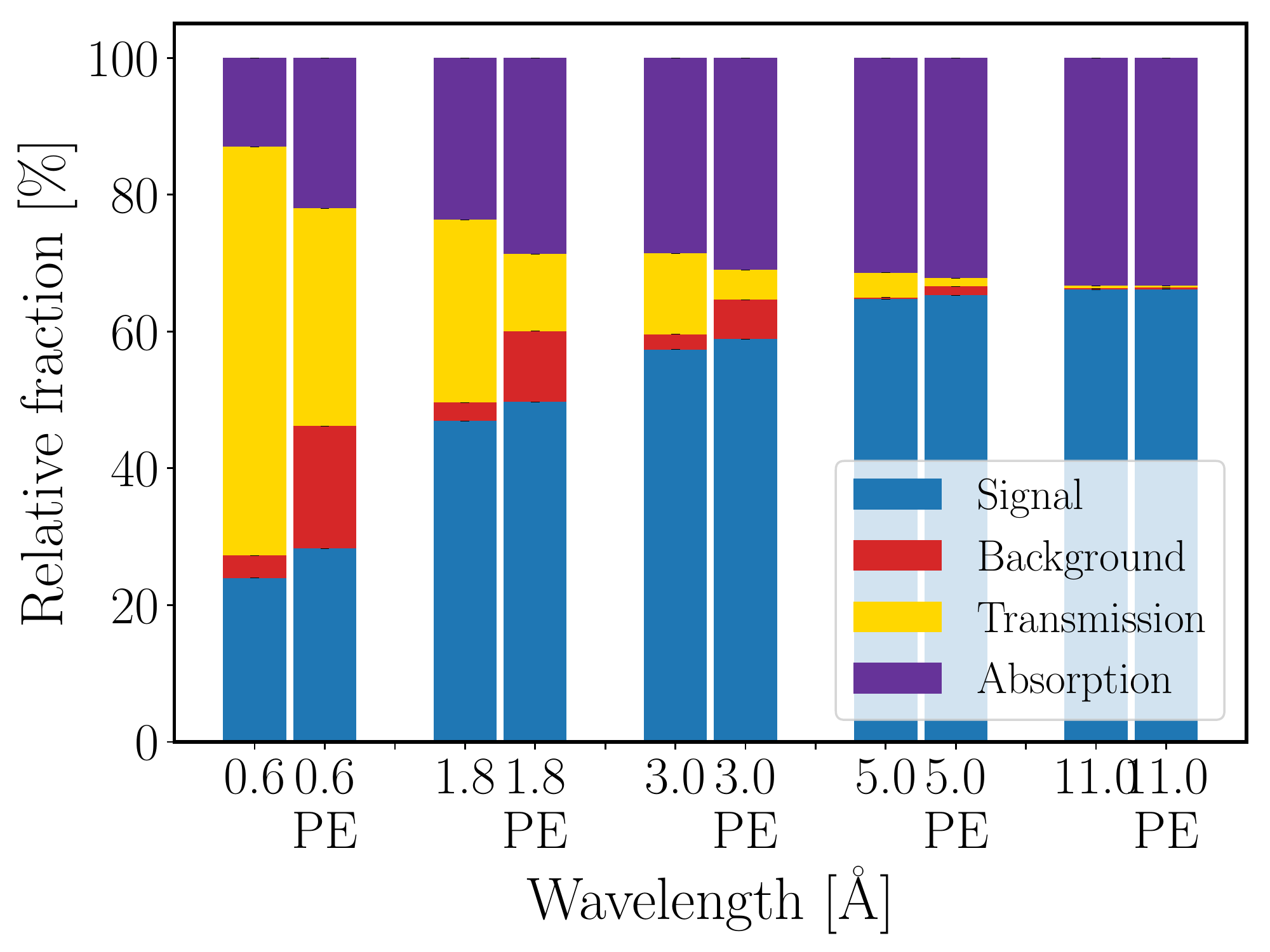}
     \end{subfigure}
   \caption{\footnotesize Comparison of proportion of absorption,
     transmission and detection with and without the PE layer from simulations with monoenergetic neutrons. On the left the absorption is separated for different materials, on the right the detection is separated into signal and background.}
  \label{fig:barChartPE}
\end{figure}

\FloatBarrier
\section{Conclusions}

A generic BCS detector model is implemented for Geant4 simulations.
With this, a complex analysis is carried out in order to evaluate
various aspects of the BCS detector performance. 
This study is made in the context of most realistic applications that
might be envisaged. % not just that of small angle neutron scattering,
                  % which is the initial intended application. 
The aim is to have a complete set of generally applicable results.

The detection efficiency of a single straw, and even of complete
detector tubes with seven straws are shown to be low, as expected. Therefore,
overlapping layers (panels) of detectors are needed to achieve a decent efficiency.
The cost-efficient number of panels depends on the application and the relevant neutron wavelength range.
%The efficiency could also be increased with higher detection to conversion ratio, which is found to be wavelength independent, but the pulse height threshold affects it. For higher efficiency, the threshold should be as low as the background allows it. %Not shown in the paper so it is removed

The absorption (not resulting in conversion to detectable particles) in B$_4$C is 6.5--8 times more than in Al and Cu combined.
The absorption from these two mechanical materials in the detector is in the range of 1.5--4.5\% of the incident neutrons depending on the wavelength. 
Pure unalloyed material was modeled in the study; alloyed materials and impurities may significantly increase this and need to be considered.
At smaller wavelengths the fraction of neutrons transmitted through the detector is high (50--60\% at 0.6 {\AA}) and therefore absorbant shielding behind the detector is a must for applications below 5{\AA}.

Activation analysis of such a detector has been
implemented. The activation is dominated by copper, as expected, with
a cooling time of a few days. 
The radiation background from activated materials will not interfere with the data acquisition. The activation during operation at ESS is not expected to be a limitation for maintenance.
The calculated numbers have been
presented in a fashion that could be scaled to real
applications. 
%{\color{red} You are missing one of the punch lines. Should I worry during daq about background?}

The scattering has been studied in detail, namely its effect on $\delta$X, $\delta$Y, $\delta$TOF, $\delta\Theta$, $\delta\Phi$, $\delta\lambda$, $\delta$E and $\delta$Q in terms of the fraction of neutrons that end up as signal, scattered background, transmission through the detector or absorbed and non-detected. The effect of the detector geometry on the natural shape of the resolution function is shown.
Scattering is highest at low wavelengths and is significant below the Bragg cut-off. It can be considered to be at acceptable levels for applications such as SANS and diffraction, however, may be considerable for applications which are highly sensitive to it such as spectroscopy. Any application for spectroscopy would need detailed consideration of its effect on performance.

A polyethylene {\it``afterburner''} block placed behind the detector was investigated and found to increase signal by up to 4\%, however, background correspondingly increased up to 15\%. Therefore this is not a good solution for most applications.
It also re-emphasises the need for the layer of shielding closest to the detector to be made of materials with very low neutron albedo.

\FloatBarrier
\section*{Acknowledgements}

This work has been supported by the In-Kind collaboration between
ESS ERIC (contract number: NIK5.4 \#10 [ESS]) and the Hungarian
Academy of Sciences, Centre for Energy Research (MTA EK).
The authors would like to thank the DMSC Computing Centre~\cite{dmscURL} which provided the computing resources for the simulations.
Richard Hall-Wilton and Kalliopi Kanaki would like to acknowledge 
support from BrightnESS [EU Horizon2020 grant 676548].

\FloatBarrier
\section*{Appendix}

The capabilities of Geant4 are enhanced by the NCrystal library~\cite{ncrystalArxiv,ncrystal} in order to treat the thermal neutron transport correctly in crystalline materials by taking into account the material structure and effects of inter-atomic bindings. With this tool aluminium and copper are treated as crystalline materials with the cross-sections presented in Fig.~\ref{ncrystalCrossSections}.

\begin{figure}[!h]  
  \centering
  \begin{subfigure}{0.5\textwidth}
    \centering
    \includegraphics[width=\textwidth]{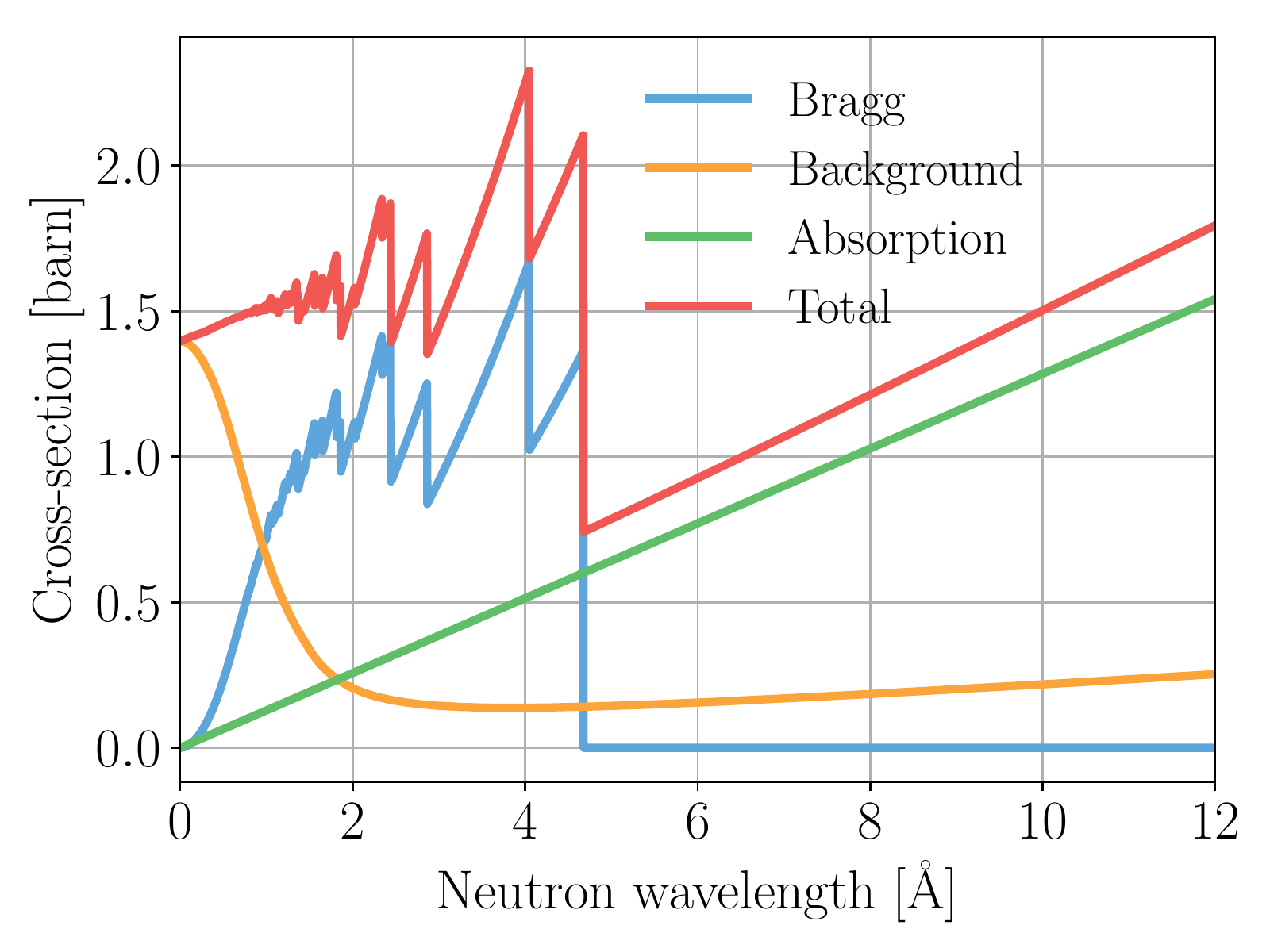}
  \end{subfigure}%
  \begin{subfigure}{0.5\textwidth}
    \centering
    \includegraphics[width=\textwidth]{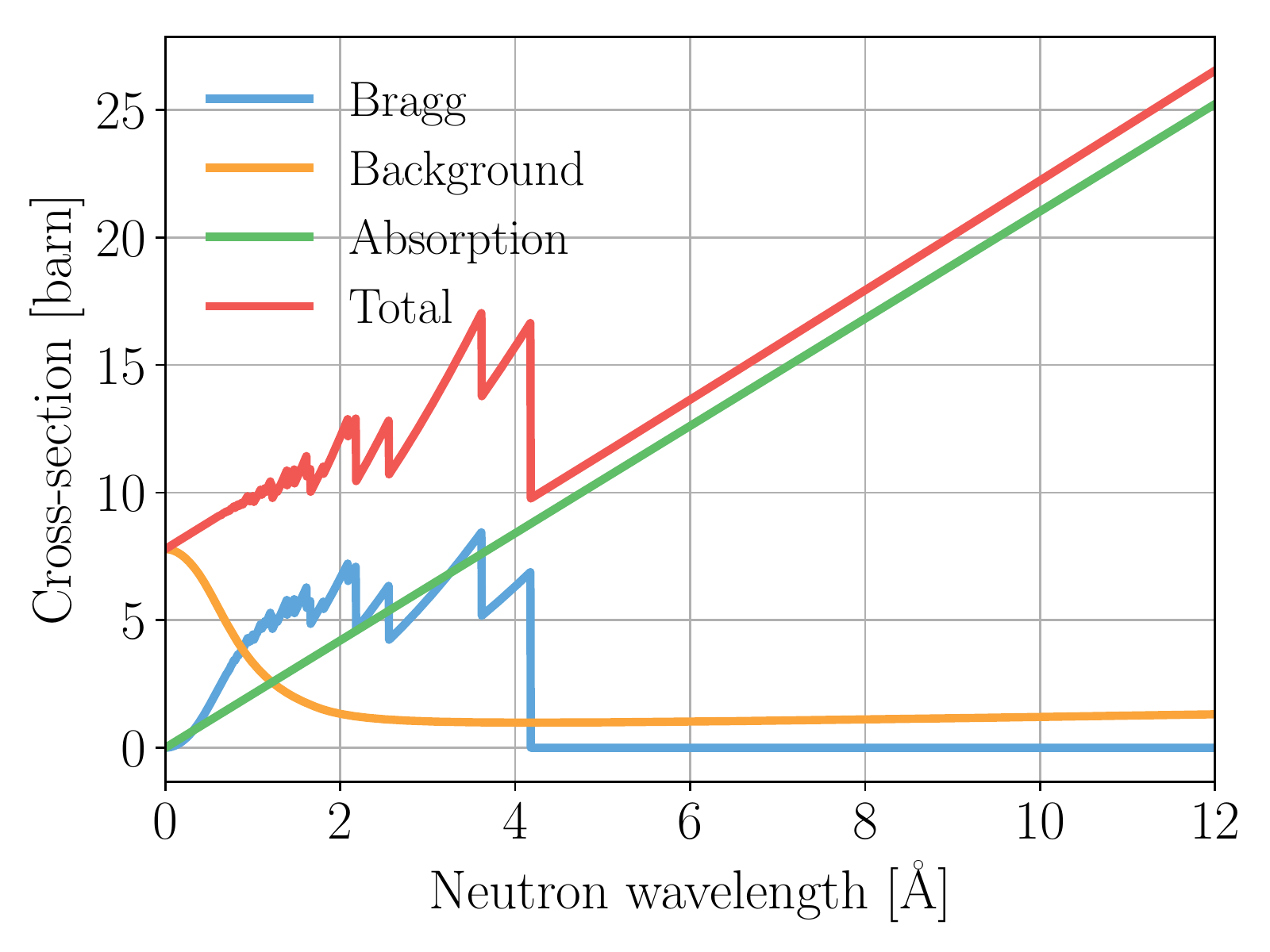}
  \end{subfigure}
  \caption{\footnotesize  Cross-sections of aluminium (left) and copper (right) from the NCrystal library~\cite{ncrystalArxiv,ncrystal}.}
  \label{ncrystalCrossSections}
\end{figure}

Figures~\ref{fig:dth_example}--\ref{fig:dphi_example} demonstrate the signal limits defined for $\delta\Theta$ and $\delta\Phi$ (similarly as Figures~\ref{fig:Gaussian_example}--\ref{fig:dY_example} demonstrated the limits for $\delta$X and $\delta$Y).
In addition, as an example, Figures~\ref{fig:tof_example}--\ref{fig:dQ_example} depict the signal limits for $\delta$TOF, $\delta\lambda$ and $\delta$Q for one of the wavelengths of interest.

\begin{figure}[!h]  
  \centering
  \begin{subfigure}{\textwidth} %{8cm}
      \includegraphics[width=\textwidth]{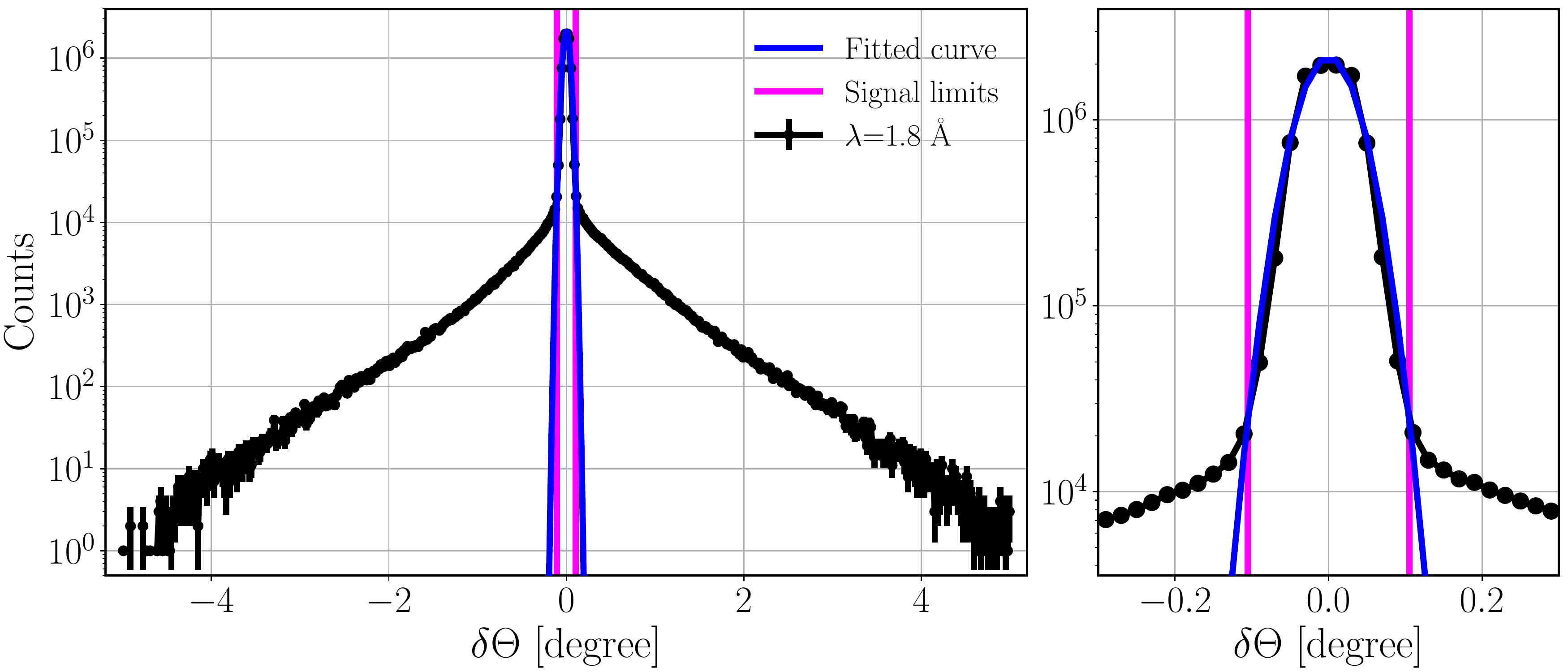}
     \end{subfigure} 
   \caption{\footnotesize Finding the limits for signal and background
     separation for $\delta\Theta$ with a 1.8~{\AA} monoenergetic beam. The figure on the right shows an enlarged view of the center part of the figure on the left. The lines are only joining the points.
}
  \label{fig:dth_example}
\end{figure} 

\begin{figure}[!h]  
  \centering
  \begin{subfigure}{\textwidth} %{8cm}
      \includegraphics[width=\textwidth]{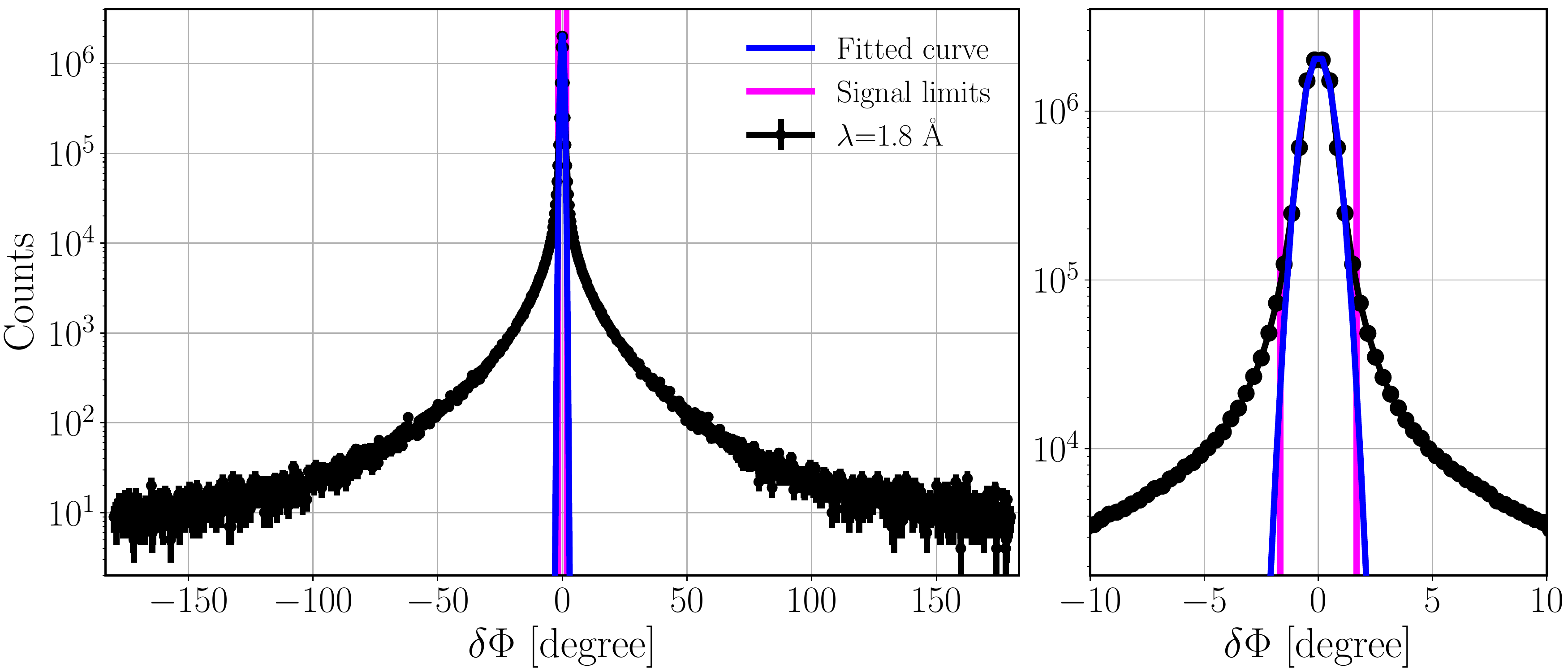}
     \end{subfigure} 
   \caption{\footnotesize Finding the limits for signal and background
     separation for $\delta\Phi$ with a 1.8~{\AA} monoenergetic beam. The figure on the right shows an enlarged view of the center part of the figure on the left. The lines are only joining the points.
}
  \label{fig:dphi_example}
\end{figure} 

\begin{figure}[!h]  
  \centering
  \begin{subfigure}{\textwidth} %{8cm}
      \includegraphics[width=\textwidth]{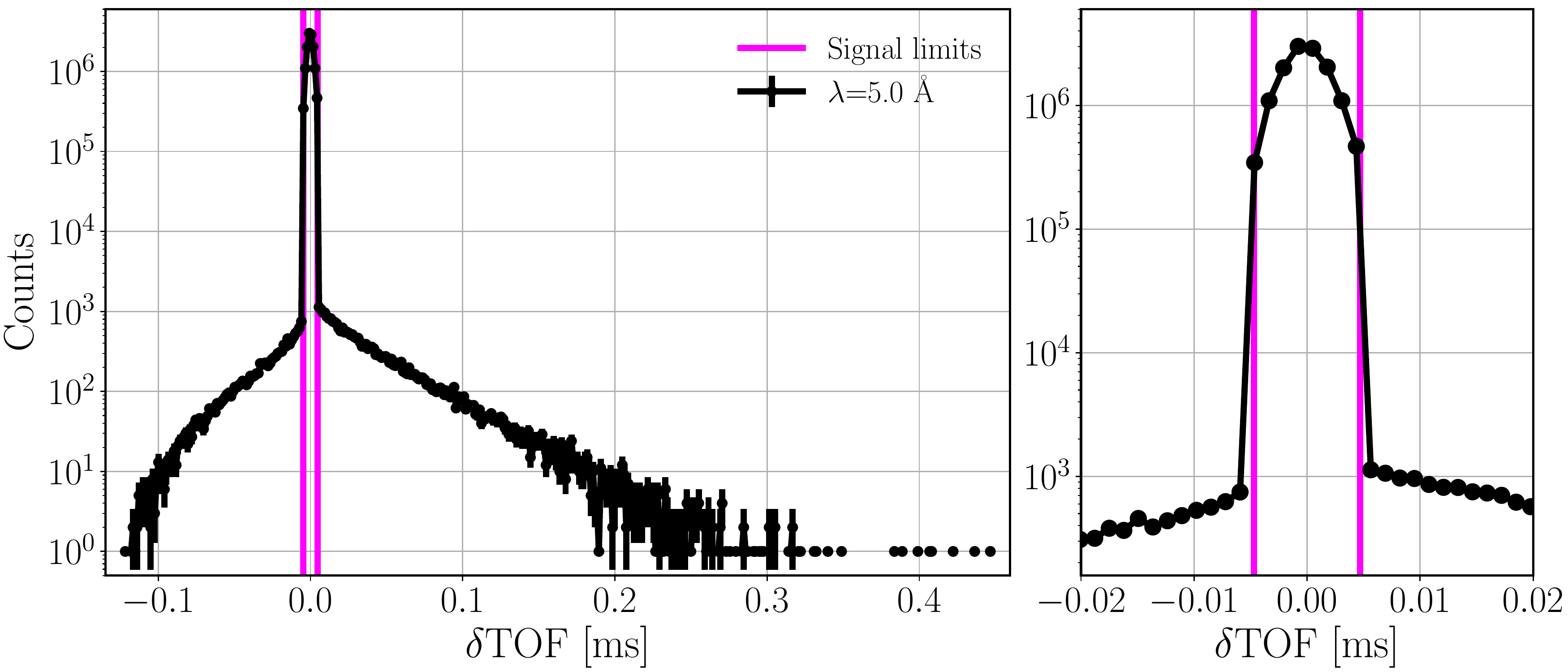}
     \end{subfigure} 
   \caption{\footnotesize Finding the limits for signal and background
     separation for $\delta$TOF with a 5~{\AA} monoenergetic beam. The figure on the right shows an enlarged view of the center part of the figure on the left. The lines are only joining the points.
}
  \label{fig:tof_example}
\end{figure} 

\begin{figure}[!h]  
  \centering
  \begin{subfigure}{\textwidth} %{8cm}
      \includegraphics[width=\textwidth]{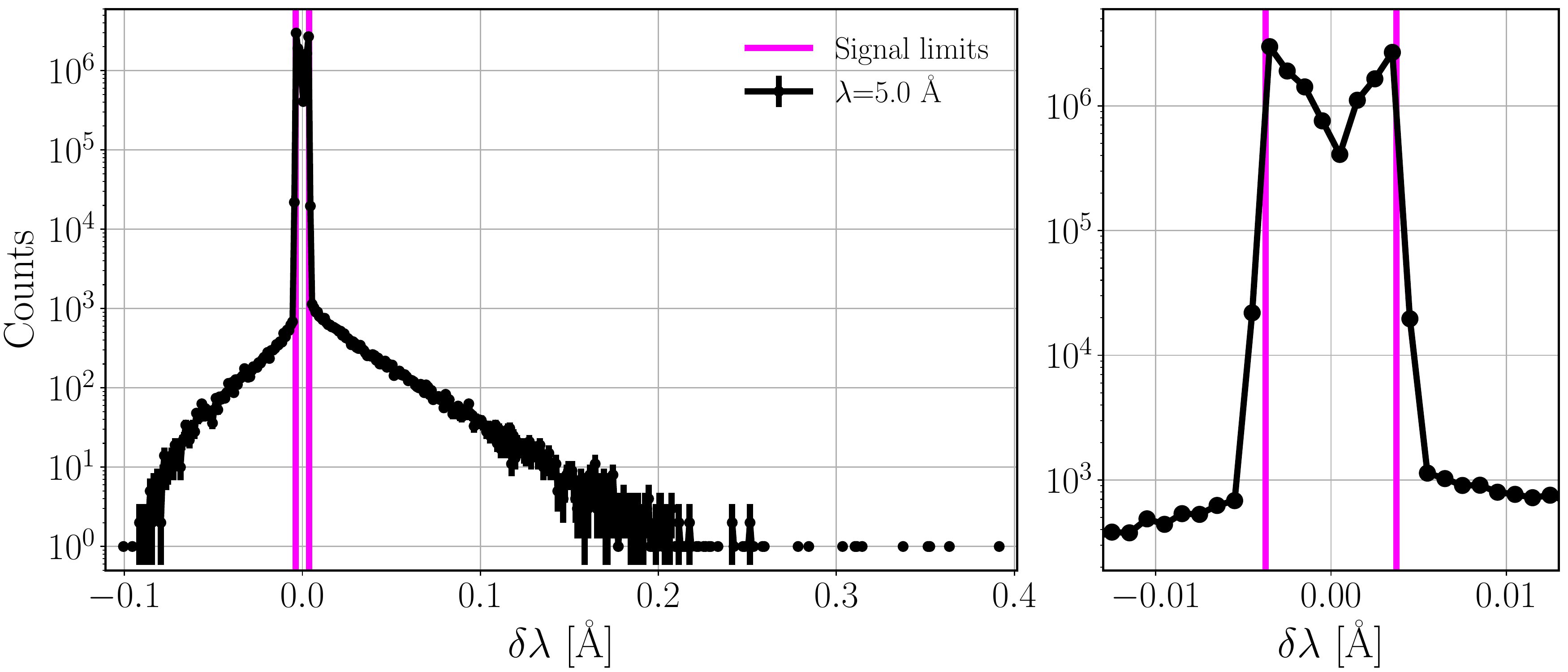}
     \end{subfigure} 
   \caption{\footnotesize Finding the limits for signal and background
     separation for $\delta\lambda$ with a 5~{\AA} monoenergetic beam. The limits are derived from the straw inner radius, using Eq.~\ref{eq:deltaLambda}. The figure on the right shows an enlarged view of the center part of the figure on the left. The lines are only joining the points.
}
  \label{fig:dlambda_example}
\end{figure} 

\begin{figure}[!h]  
  \centering
  \begin{subfigure}{\textwidth} %{8cm}
      \includegraphics[width=\textwidth]{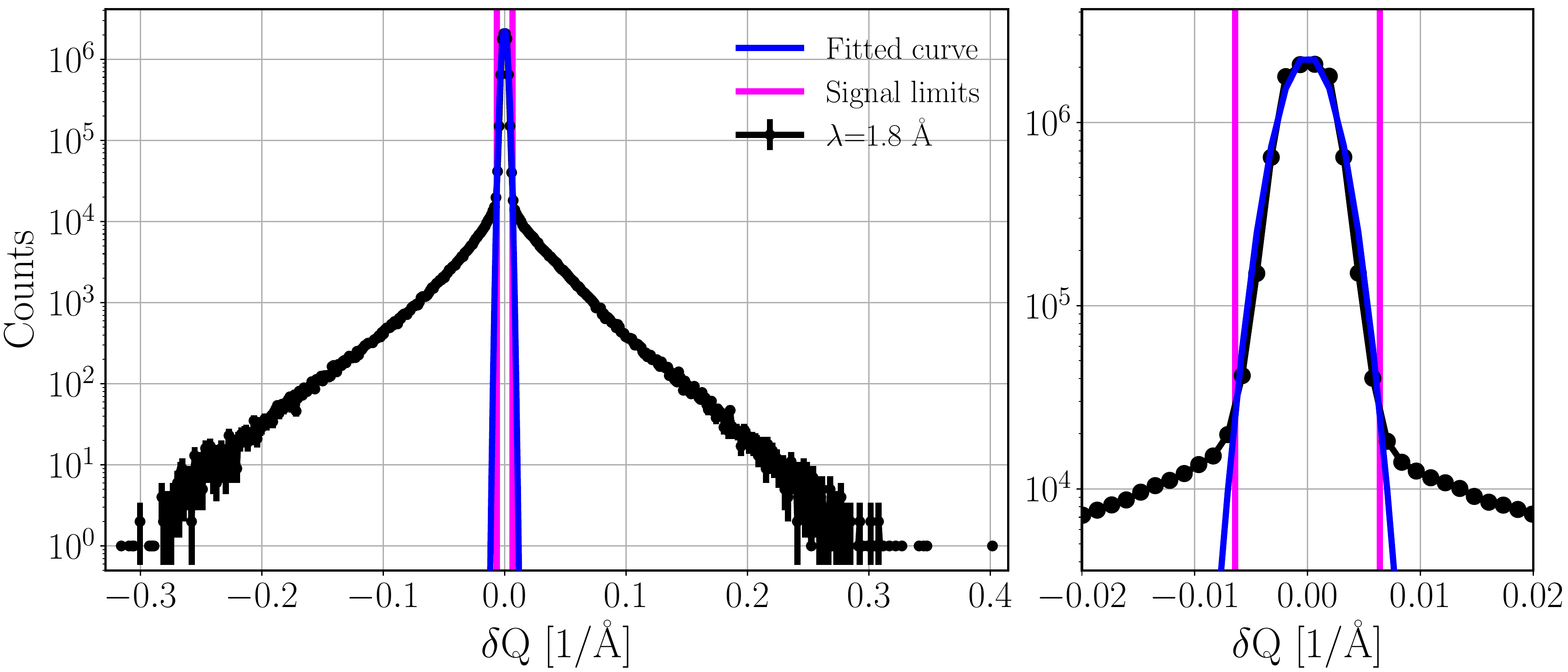}
     \end{subfigure} 
   \caption{\footnotesize Finding the limits for signal and background
     separation for $\delta$Q with a 1.8~{\AA} monoenergetic beam. The figure on the right shows an enlarged view of the center part of the figure on the left. The lines are only joining the points.
}
  \label{fig:dQ_example}
\end{figure}

%\begin{figure}[!h]  
%  \centering
%  \begin{subfigure}{0.49\textwidth} %{8cm}
%      \centering
%      \includegraphics[width=\textwidth]{dth_gaussian_ang1p8_AlCu_8e5}
%   \end{subfigure}
%   \begin{subfigure}{0.49\textwidth}
%     \centering      
%     \includegraphics[width=\textwidth]{dPhi_gaussian_ang1p8_AlCu_8e5}
%   \end{subfigure}
%     
%   \begin{subfigure}{0.49\textwidth} %{8cm}
%      \centering
%      \includegraphics[width=\textwidth]{dTOF_gaussian_ang1p8_AlCu_8e5}
%   \end{subfigure}
%   \begin{subfigure}{0.49\textwidth}
%      \centering      
%      \includegraphics[width=\textwidth]{dQ_gaussian_ang1p8_AlCu_8e5}
%   \end{subfigure}
%   \caption{\footnotesize Finding the limits for signal and background
%     separation for $\delta$TOF using the results from simulation with
%     1.8~{\AA} monoenergetic beam. The limits are derived from the
%     straw inner radius and the neutron initial velocity. {\color{red}
%   what do you mean derived? You did a fit right? Unless you mean
%   coincide or shaped or dictated.}}
%  \label{fig:4gaussianQuantity}
%\end{figure}

\FloatBarrier

\end{document}